\newif\ifIEEEcompliantpdf
\IEEEcompliantpdffalse

\ifIEEEcompliantpdf
\RequirePackage{xr}%
\def\href#1#2{#2}%
\else
\RequirePackage{xr}%
\PassOptionsToPackage{pdftex,pdfpagemode=none, pdftoolbar=true,pdffitwindow=true,pdfcenterwindow=true}{hyperref}%
\fi

\PassOptionsToPackage{hang,flushmargin}{footmisc}
\PassOptionsToPackage{russian,english}{babel}

\newif\ifpnastypesetdefaults
\pnastypesetdefaultsfalse
\newif\ifpnastypesetarxiv
\pnastypesetarxivtrue

\ifpnastypesetdefaults
  \documentclass[9pt,twocolumn,twoside,lineno]{pnas-new}%
  \templatetype{pnasresearcharticle}
\else
  \ifpnastypesetarxiv
    \documentclass[10pt,twocolumn,twoside]{pnas-new}%
    \templatetype{pnasresearcharticle}
    \setboolean{displaywatermark}{false} 
    \makeatletter
    \fancypagestyle{firststyle}{
       \fancyfoot[L]{\footerfont \textbf{PNAS}\hspace{0.5pc}{2021}\hspace{0.5pc}Vol.~118\hspace{0.5pc}No.~4\hspace{0.5pc}\href{\THEDOI}{e2011216118}}%
       \fancyfoot[R]{\footerfont\textbf{\THEDOI}\hspace{0.5pc}|\hspace{0.5pc}\textbf{\thepage\ of \pageref{LastPage}}}%
    }
    \makeatother
    \usepackage{filecontents}
  \else
    \documentclass[12pt,onecolumn,twoside,lineno]{pnas-new}%
    \templatetype{pnasresearcharticle}
    \usepackage{setspace}%
  \fi
\fi

\usepackage{flushend}

\makeatletter
\renewcommand{\showacknow}{
\@ifundefined{@acknow}{}{\vskip 3.25ex plus 1ex minus .2ex
\sffamily\small\noindent{\bfseries ACKNOWLEDGMENTS.\hspace{1.5ex plus .2ex}}\@acknow}}

\newcommand{\dataavailability}[1]{\def\@dataavailability{#1}}

\newcommand{\showdataavailability}{
\@ifundefined{@dataavailability}{}{\vskip 3.25ex plus 1ex minus .2ex
\sffamily\small\noindent{\bfseries Data Availability.\hspace{1.5ex plus .2ex}}\@dataavailability}}
\makeatother

\usepackage{bm,amssymb,amsmath,mathrsfs,amsthm}
\usepackage{subcaption}
\usepackage{float}

\usepackage{algorithm}
\usepackage{tabularx,colortbl,makecell}
\usepackage[tableposition=top]{caption}
\usepackage{enumitem}
\setlist{noitemsep} 

\usepackage[space]{grffile}

\usepackage[T2A,T1]{fontenc}
\usepackage{txfonts}
\usepackage{dsfont}

\usepackage[utf8]{inputenc}
\usepackage{substitutefont}

\makeatletter
\DeclareRobustCommand{\cyrins}[1]{%
  \begingroup\hyphenpenalty\@M\exhyphenpenalty\@M\fontfamily{Tempora-TLF}%
  \foreignlanguage{russian}{#1}%
  \endgroup }
\makeatother

\DeclareMathVersion{sans}
\SetSymbolFont{operators}{sans}{OT1}{cmsmf}{m}{n}
\SetSymbolFont{letters}{sans}{OML}{cmssm}{m}{it}
\SetSymbolFont{symbols}{sans}{OMS}{cmsssy}{m}{n}
\SetMathAlphabet{\mathit}{sans}{OT1}{cmssm}{m}{sl}
\SetMathAlphabet{\mathbf}{sans}{OT1}{cmsssy}{bx}{n}
\SetMathAlphabet{\mathtt}{sans}{OT1}{cmssex}{m}{n}
\SetSymbolFont{largesymbols}{sans}{OMX}{cmssex}{m}{n}

\def\mathsffamily{\sffamily\mathversion{sans}}

\usepackage{xcolor}
\usepackage{graphicx}

\definecolor{darkgreen}{rgb}{0,0.5,0}
\definecolor{darkyellow}{rgb}{0.5,0.5,0}
\definecolor{darkred}{rgb}{0.6667,0,0}
\definecolor{light-gray}{gray}{0.5}

\definecolor{pythonC0}{rgb}{0.1216,0.4667,0.7059}  
\definecolor{pythonC1}{rgb}{1.0000,0.4980,0.0549}  
\definecolor{pythonC2}{rgb}{0.1725,0.6275,0.1725}  
\definecolor{pythonC3}{rgb}{0.8392,0.1529,0.1569}  
\definecolor{pythonC4}{rgb}{0.5804,0.4039,0.7412}  
\definecolor{pythonC5}{rgb}{0.5490,0.3373,0.2941}  
\definecolor{pythonC6}{rgb}{0.8902,0.4667,0.7608}  
\definecolor{pythonC7}{rgb}{0.4980,0.4980,0.4980}  
\definecolor{pythonC8}{rgb}{0.7373,0.7412,0.1333}  
\definecolor{pythonC9}{rgb}{0.0902,0.7451,0.8118}  

\colorlet{pythonblue}{pythonC0}
\colorlet{pythonorange}{pythonC1}
\colorlet{pythongreen}{pythonC2}
\colorlet{pythonred}{pythonC3}
\colorlet{pythonpurple}{pythonC4}
\colorlet{pythonbrown}{pythonC5}
\colorlet{pythonpink}{pythonC6}
\colorlet{pythongrey}{pythonC7}
\colorlet{pythonyellow}{pythonC8}
\colorlet{pythoncyan}{pythonC9}

{\catcode`\@=11\obeyspaces\obeylines
\gdef\beginpython{\begingroup\scriptsize\baselineskip=8pt\ttfamily\frenchspacing\bfseries %
  \spaceskip=\z@ \xspaceskip=\z@ %
  \parskip=\z@ \parindent=\z@
  \chardef\other=12  %
  \catcode`\$=\z@    %
  \catcode`\&=\other %
  \catcode`\^=\other %
  \catcode`\_=\other %
  \catcode`\~=\other %
  \catcode`\%=\other %
  \catcode`\#=\other %
  \catcode`\}=\other %
  \catcode`\{=\other %
  \catcode`\"=\other %
  \catcode`\\=\other %
  \obeyspaces\gdef {\ifvmode\leavevmode\fi\nobreak\space}%
  \obeylines\gdef^^M{\ifvmode\penalty-100\vskip6pt\fi\par\penalty9999}}}

\makeatletter
\newcommand{\customlabel}[2]{%
   \protected@write \@auxout {}{\string \newlabel {#1}{{#2}{\thepage}{#2}{#1}{}}}%
   \hypertarget{#1}{\relax}}
\makeatother

\def\twoauthors#1#2#3#4{\gdef\@address{}
   \gdef\@name{\begingroup\normalsize\begin{tabular}{@{}c@{}}
        {\large\em #1}\\\noalign{\vskip 6pt plus 3pt minus 3pt}
        #2\relax
   \end{tabular}\hskip 0.3333in plus.1667in minus.1667in\begin{tabular}{@{}c@{}}
        {\large\em #3}\\\noalign{\vskip 6pt plus 3pt minus 3pt}
        #4\relax
\end{tabular}\endgroup}}

\def\threeauthors#1#2#3#4#5#6{\gdef\@address{}
   \gdef\@name{\begin{tabular}{@{}c@{}}
        {\em #1}\\\noalign{\vskip 6pt plus 3pt minus 3pt}
        #2\relax
   \end{tabular}\hskip .5in plus.5in minus.125in\begin{tabular}{@{}c@{}}
        {\em #3}\\\noalign{\vskip 9pt plus 3pt minus 3pt}
        #4\relax
   \end{tabular}\hskip .5in plus.5in minus.125in\begin{tabular}{@{}c@{}}
        {\em #5}\\\noalign{\vskip 6pt plus 3pt minus 3pt}
        #6\relax
\end{tabular}}}

\makeatother

\mathchardef\colonord="003A

\def\T{{\scriptscriptstyle\rm T}}   
\let\humlaut=\H
\def\H{\ifmmode{\scriptscriptstyle\rm H}\else\humlaut\fi}   

\def\by{\ifmmode $\hbox{-by-}$\else \leavevmode\hbox{-by-}\fi}
\def\sqrtm1{{\sqrt{\!-1}}}
\let\(=\langle
\let\)=\rangle

\DeclareMathOperator{\Poisson}{Poisson}
\DeclareMathOperator{\Normal}{Normal}

{\catcode`\_=12\relax \global\let\underscoreregular=_\relax}%
\def\underscore{{\texttt\underscoreregular}}

\makeatletter
\def\hgfarg#1{\left(\null\vcenter{\normalbaselines\m@th
    \ialign{&$\displaystyle##$\hfil\crcr
      \mathstrut\crcr\noalign{\kern-\baselineskip}
      #1\crcr\mathstrut\crcr\noalign{\kern-\baselineskip}}}\right)}
\def\Heqalign#1{\null\,\vcenter{\openup\jot\m@th
  \ialign{\strut\hfil$##$:\quad&\hfil$\displaystyle{##}$&$\displaystyle
      {{}##}$\hfil&\qquad##\hfil\crcr#1\crcr}}\,}
\def\eqalignno#1{\displ@y \tabskip\@centering
  \halign to\displaywidth{\hfil$\@lign\displaystyle{##}$\tabskip\z@skip
    &$\@lign\displaystyle{{}##}$\hfil\tabskip\@centering
    &\llap{$\@lign##$}\tabskip\z@skip\crcr
    #1\crcr}}
\def\dbleqalignno#1{\displ@y \tabskip\@centering
  \halign to\displaywidth{\hfil$\@lign\displaystyle{##}$\tabskip\z@skip
    &$\@lign\displaystyle{{}##}$\hfil
    &$\@lign\displaystyle{{}##}$\hfil\tabskip\@centering
    &\llap{$\@lign##$}\tabskip\z@skip\crcr
    #1\crcr}}

\makeatother

\newif\ifmathtomb \mathtombfalse

\def\tombstone{\unskip\penalty50   
  \hskip 0pt plus-1fill \null\nobreak\hskip 0pt plus1fill
  \enskip \vrule width.3333em height.7em depth.2em
  \ifmmode \global\mathtombtrue \else \global\mathtombfalse \fi}

\makeatletter
  {\futurelet\next\hpr@oftext}
  {\ifmathtomb \else \tombstone \fi \widowpenalty=10000  
   \par \ifmathtomb \else \addvspace{\medskipamount}\fi \global\mathtombfalse}
\def\hpr@oftext{\ifx\next[\let\temp\ohpr@@ftext\else\let\temp\hpr@@ftext\fi\temp}
\def\hpr@@ftext{\beginhpr@@f{Proof}}
\def\ohpr@@ftext[#1]{\beginhpr@@f{#1}}
\def\beginhpr@@f#1{\par \addvspace{\bigskipamount}%
  \noindent{\bf #1:\enspace}\ignorespaces }
\makeatother

\def\twittersize{\small}
\def\twitter#1{{#1}}
\def\twitterhlt#1{{\color{twitterhashtag}#1}}
\definecolor{twitterblue}{rgb}{0.9025, 0.9363, 0.9625}
\definecolor{twitterhashtag}{rgb}{0.2588, 0.5647, 0.8275}

\usepackage{etoolbox}
\makeatletter
\newlength{\qrr@dimen@}
\expandafter\pretocmd\csname tabular*\endcsname{\setlength{\qrr@dimen@}{#1}}{}{}
\newcommand*{\Rowcolor}[2][\tabcolsep]{%
    \ifx\relax#1\relax\else
        \kern-\the\dimexpr#1\relax
    \fi
    \makebox[0pt][l]{%
        \fboxsep=0pt
        \colorbox{#2}{%
            \strut\kern\qrr@dimen@
        }%
    }%
    \ifx\relax#1\relax\else
        \kern\the\dimexpr#1\relax
    \fi
    \ignorespaces
}

\newsavebox\saved@arstrutbox
\newcommand*{\setarstrut}[1]{%
  \noalign{%
    \begingroup
      \global\setbox\saved@arstrutbox\copy\@arstrutbox
      #1%
      \global\setbox\@arstrutbox\hbox{%
        \vrule \@height\arraystretch\ht\strutbox
               \@depth\arraystretch \dp\strutbox
               \@width\z@
      }%
    \endgroup
  }%
}
\newcommand*{\restorearstrut}{%
  \noalign{%
    \global\setbox\@arstrutbox\copy\saved@arstrutbox
  }%
}

\makeatother

\newcolumntype{b}{>{\columncolor{twitterblue}}l}
\newcolumntype{a}{>{\columncolor{twitterblue}}c}

\makeatletter
\newif\ifretweet
\def\tweettext{\bgroup\catcode`\_=12\relax \retweetfalse \tweettext@ }
\def\retweettext{\bgroup\catcode`\_=12\relax \retweettrue \tweettext@ }
\long\def\tweettext@#1#2#3#4#5#6#7#8{\dimen@=\linewidth \dimen@i=27pt\relax \dimen@ii=6pt\relax
  \advance\dimen@ by-\dimen@i \advance\dimen@ by-\dimen@ii
  \hbox to\linewidth{\normalfont\mathsffamily\fontsize{7}{9}\frenchspacing
    \vbox to0pt{\vskip-0.6667\dimen@ii\hbox to\dimen@i{\includegraphics[width=\dimen@i]{#4}\hss}\vss}\hskip\dimen@ii \hss
      \vtop{\hsize\dimen@ \parskip0pt\relax \parindent0pt\relax
        \vtop{\raggedright\strut\textbf{{\color{darkred}[id #3]} #1}
            {\color{light-gray}@#2 $\;{\cdot}\;$#5\ifretweet$\;{\cdot}\;$Retweeted\fi}}%
        \vskip3pt\relax
        \vbox{\raggedright #6}%
}\egroup}}
\makeatother

\makeatletter
\def\optionalwordifspace#1{\skip@=.3333em plus.1667em minus.1111em\relax
  \setbox\z@\hbox{#1\hskip\skip@}%
  \unskip \hskip\skip@ \penalty\z@
  \null\nobreak \cleaders\hbox{#1\hskip\skip@}\hskip \wd\z@ minus\wd\z@\penalty\z@ }

\makeatother

\def\hquad{\hskip .5em\relax }
\def\thinskip{\hskip .16667em\relax }  

\def\thinthinspace{\kern .041667em\relax }
\def\thinthinskip{\hskip .041667em\relax }  

{\catcode`\@=11 \catcode`|=\active \catcode`\!=\active
\gdef\references{\catcode`|=\active \catcode`\!=\active
  \def\!{\char`\!}%
  \def|##1|{{\sc ##1}}      
  \def!##1!{\emph{##1}}     
  \def\<##1>{{\bf##1}\futurelet\next\number@ptional}  
  \def\number@ptional{\ifx\next(\def\@temp{\n@mber}\else
    \ifx\next:\def\@temp{\p@gesc@l@n}\else\def\@temp{}\fi\fi \@temp}%
  \def\n@mber(##1){\thinspace(##1)\futurelet\next\page@ptional}%
  \def\page@ptional{\ifx\next:\def\@temp{\p@gesc@l@n}\else
    \def\@temp{}\fi \@temp}%
  \def\p@gesc@l@n:{\thinspace:\penalty-20\thinskip}%
  \frenchspacing}}
\let\virgule=/\relax
\let\?=?\relax
\let\mathshift=$\relax
{\catcode`\@=11 \catcode`\|=14  
  \catcode`\/=\active \catcode`\%=\active \catcode`\?=\active \catcode`\&=\active \catcode`\_=\active
  \catcode`\$=12\relax
  \gdef\doicite:{\bgroup \catcode`\/=\active \catcode`\_=\active \catcode`\&=12\relax \catcode`\$=12\relax
    \catcode`\%=12\relax
    \def\@citen@me{doi}\def\@citer@@t{https://doi.org/}\d@i@rxiv }|
  \gdef\arxivcite:{\bgroup \catcode`\/=\active \catcode`\_=\active \catcode`\&=12\relax \catcode`\$=12\relax
    \catcode`\%=12\relax
    \def\@citen@me{arxiv}\def\@citer@@t{https://arxiv.org/abs/}\d@i@rxiv }|
  \gdef\d@i@rxiv#1:{|
    \def/{\unpenalty\penalty\@M\slash }|
    \def_{{\noexpand\texttt\underscoreregular }}|
    \edef\d@inum{#1}|
    \let/\virgule
    \let_\underscoreregular
    \edef\d@ilink{\@citer@@t #1}|
    \def\d@ic@l@n{\thinthinspace:\penalty-20\thinthinskip}|
    \hyphenpenalty\@M\relax
    \frenchspacing
    \@citen@me\d@ic@l@n\href{\d@ilink}{\color{black}\d@inum}\egroup }|
  \gdef\webcite[{\bgroup \catcode`\/=\active \catcode`\%=\active \catcode`\?=\active \catcode`\&=\active
       \catcode`\_=\active \catcode`\$=12\relax \@webcite }|
  \gdef\@webcite#1]{|
    \def/{\unpenalty\penalty\@M\slash }|
    \def?{\discretionary{}{}{}\?}|
    \def&{\discretionary{}{}{}\&}|
    \def
    \def_{{\noexpand\texttt\underscoreregular }}|
    \edef\@linktxt{#1}|
    \let/\virgule \let?\?\relax \let&\&\relax \let
    \edef\@linktgt{#1}|
    \hyphenpenalty\@M\relax
    \frenchspacing
    \href{\@linktgt}{\color{black}\@linktxt}\nobreak\egroup }|
  \gdef\urlcite[{\bgroup \catcode`\/=\active \catcode`\%=\active \catcode`\?=\active \catcode`\&=\active
       \catcode`\_=\active \catcode`\$=12\relax \@urlcite }|
  \gdef\@urlcite#1]{|
    \def/{\unpenalty\penalty\@M\slash }|
    \def?{\discretionary{}{}{}\?}|
    \def&{\discretionary{}{}{}\&}|
    \def
    \def_{{\noexpand\texttt\underscoreregular }}|
    \edef\@linktxt{#1}|
    \let/\virgule \let?\?\relax \let&\&\relax \let
    \edef\@linktgt{#1}|
    \hyphenpenalty\@M\relax
    \frenchspacing
    \@linktxt\egroup }|
}

\newcommand{\N}{\mathcal{N}}
\newcommand{\Graph}{\mathcal{G}}
\newcommand{\VSet}{\mathcal{V}}
\newcommand{\ESet}{\mathcal{E}}

\newcommand{\ASetClosed}{\mathcal{A}}
\newcommand{\ZkSet}{\mathcal{Z}^k}
\newcommand{\ZlSet}{\mathcal{Z}^l}
\newcommand{\ZSetClosed}{\mathcal{Z}}
\newcommand{\ZSetOpen}{\mathcal{Z}_{-i}}
\newcommand{\A}{\bm{A}}
\newcommand{\Z}{\bm{Z}}

\newcommand{\Zopen}{\Z_{-i}}
\newcommand{\YSet}{\mathds{Y}}
\newcommand{\YmisSet}{\mathds{Y}_{\rm mis}}
\newcommand{\YobsSet}{\mathds{Y}_{\rm obs}}
\newcommand{\Y}{\bm{Y}}
\newcommand{\z}{\bm{z}}

\newcommand{\ave}{\text{ave}}

\renewcommand\refname{}

\newcommand{\X}{\bm{X}}

\def\legendbar#1{$\vcenter{\hbox{\vrule width1.5em height#1\relax}}\,$}
\def\legendbarhalf#1{$\vcenter{\hbox{\vrule width.75em height#1\relax}}\,$}
\def\legendbardashed#1{\legendbarhalf{#1}\hskip.25em\legendbarhalf{#1}}

\def\ieeev#1-{{#1}\hbox{-}\nobreak}

\makeatletter
\long\def\XR@test#1#2#3#4\XR@{%
  \let\XR@next\@gobbletwo
  \ifx#1\newlabel
    \let\XR@next\@firstoftwo%
  \else\ifx#1\@input
     \let\XR@next\@secondoftwo
  \fi\fi
   \XR@next{\newlabel{\XR@prefix#2}{#3}}{\edef\XR@list{\XR@list#2\relax}}%
  \ifeof\@inputcheck\expandafter\XR@aux
  \else\expandafter\XR@read\fi}

\def\customexternaldocument[#1]#2#3{\gdef\@customXRprefix{#1}\gdef\@customXRfilename{#2}\gdef\@customXRDOI{#3}%
  \externaldocument[#1]{#2}}

\def\customxrref#1{%
  \begingroup
    \def\@getpdfanchorname{#1}
    \toks@=\expandafter{\csname r@\@customXRprefix#1\endcsname}%
    \expandafter\let\expandafter\getpdfanchor@refmacro\the\toks@
    \ifx\getpdfanchor@refmacro\@undefined
      \def\getpdfanchor@refmacro{{}{}{}{}{}}%
    \else
      \ifx\getpdfanchor@refmacro\relax
        \def\getpdfanchor@refmacro{{}{}{}{}{}}%
      \fi
    \fi
    \edef\@getpdfanchorname{\expandafter\getpdfanchorHy@setref@link\getpdfanchor@refmacro\@empty\@empty}%
    \colorlet{saved}{.}%
    \href{\@customXRDOI}{\color{saved}\ref*{\@customXRprefix#1}}%
  \endgroup
}
\def\getpdfanchorHy@setref@link#1#2#3#4#5#6#7{%
  #4%
}
\let\getpdfanchorname\getpdfanchor@refstar
\makeatother

\def\orcidlink#1{\href{https://orcid.org/#1}{\hbox to 9pt{\hss
  \color{orcid}\put(0,3){\circle*{8}}\put(-2.15,1.1){\hbox{\color{white}\sffamily
    \textmd{\fontsize{6.5pt}{6.5pt}\selectfont i\fontsize{4.65pt}{4.65pt}\selectfont D}}}\hss}}}

\hypersetup{
    citecolor = .,
    linkcolor = .,
    urlcolor = .,
}

\def\mainpaperfile{io_detection_pnas_final}
\def\SIpaperfile{io_detection_pnas_si_final}

\def\mainpaperDOI{https://doi.org/10.1073/pnas.2011216118}
\def\SIpaperDOI{https://www.pnas.org/lookup/suppl/doi:10.1073/pnas.2011216118/-/DCSupplemental}

\customexternaldocument[SI-]{\SIpaperfile}{\SIpaperDOI}

\definecolor{orcid}{rgb}{0.6824,0.7961,0.3294}

\def\THETITLE{Automatic detection of influential actors in disinformation networks}
\title{\THETITLE}

\def\THEDOI{https://doi.org/10.1073/pnas.2011216118}
\def\THESIURL{https://www.pnas.org/lookup/suppl/doi:10.1073/pnas.2011216118/-/DCSupplemental}
\doi{\url{\THEDOI}}

\author[a,1,2,\orcidlink{0000-0003-0730-0535}]{Steven T. Smith}
\author[a,1,\orcidlink{0000-0001-8880-0691}]{Edward K. Kao}
\author[a,1,\orcidlink{0000-0002-2928-0806}]{Erika D. Mackin}
\author[a,\orcidlink{0000-0002-0821-6661}]{Danelle C. Shah}
\author[a,\orcidlink{0000-0003-4847-3010}]{Olga Simek}
\author[b,c,d,1,2,\orcidlink{0000-0001-7127-9262}]{Donald B. Rubin}

\affil[a]{MIT Lincoln Laboratory, Lexington MA 02421 USA}
\affil[b]{Fox School of Business, Temple University, Philadelphia, PA 19122}
\affil[c]{Yau Mathematical Sciences Center, Tsinghua University, Beijing 100084, China}
\affil[d]{Department of Statistics, Harvard University, Cambridge, MA 02138}

\dates{Contributed by Donald B. Rubin, November 10, 2020
(sent for review July 21, 2020; reviewed by Michael Sobel, Kate
Starbird, and Stefan Wager)}

\leadauthor{Smith}

\significancestatement{\bfseries Hostile influence operations (IOs) that weaponize
digital communications and social media pose a rising threat to open
democracies.  This paper presents a system framework to automate
detection of disinformation narratives, networks, and influential
actors. The framework integrates natural language processing, machine
learning, graph analytics, and network causal inference to quantify
the impact of individual actors in spreading the IO narrative. We
present a classifier that detects reported IO~accounts with 96\%
precision, 79\% recall, and 96\%~AUPRC, demonstrated on real social
media data collected for the 2017 French presidential election and
known IO~accounts disclosed by Twitter. Our system also discovers
salient network communities and high-impact accounts that are
independently corroborated by US Congressional reports and
investigative journalism.}

\authorcontributions{S.T.S., E.K.K., E.D.M., D.C.S., O.S., and~D.B.R. designed
  research; S.T.S., E.K.K., and~E.D.M. performed research; and~S.T.S.,
  E.K.K., E.D.M., and~D.B.R. wrote the paper.\par \vskip5pt\relax
  Reviewers: M.S., Columbia University; K.S., University of Washington; and S.W., Stanford University.}
\authordeclaration{The authors declare no competing interest.\par \vskip5pt\relax
This open access article is distributed
under \href{https://creativecommons.org/licenses/by-nc-nd/4.0/}{Creative
Commons Attribution-NonCommercial-NoDerivatives License 4.0 (CC
BY-NC-ND)}.}

\equalauthors{\textsuperscript{1}S.T.S., E.K.K., E.D.M., and D.B.R. contributed equally to this work.}

\correspondingauthor{\sffamily\textsuperscript{2}To whom correspondence may be
  addressed. Email: \href{mailto:stsmith@ll.mit.edu}{\color{black}stsmith@\discretionary{}{}{}ll.mit.edu}
  or~\href{mailto:rubin@stat.harvard.edu}{\color{black}rubin@\discretionary{}{}{}stat.harvard.edu}.\par\vskip5pt\relax
  This article contains supporting information online at
  {\url{\THESIURL}.}}

\dataavailability{Comma-separated value (CSV) data of the narrative
networks analyzed in this paper have been deposited in GitHub
({\url{https://github.com/Influence-Disinformation-Networks/PNAS-Narrative-Networks}})
and Zenodo
({\url{https://doi.org/10.5281/zenodo.4361708}}).}

\acknow{This material is based upon work supported by the Under
Secretary of Defense for Research and Engineering under Air~Force
Contract No.\ FA8702-15-D-0001. Any opinions, findings, conclusions or
recommendations expressed in this material are those of the
author(s) and do not necessarily reflect the views of the Under
Secretary of Defense for Research and Engineering.}


\begin{abstract}
The weaponization of digital communications and social media to
conduct disinformation campaigns at immense scale, speed, and reach
presents new challenges to identify and counter hostile influence
operations (IOs). This paper presents an end-to-end framework to
automate detection of disinformation narratives, networks, and
influential actors. The framework integrates natural language
processing, machine learning, graph analytics, and a network
causal inference approach to quantify the impact of individual actors
in spreading IO narratives. We demonstrate its capability on
real-world hostile IO campaigns with Twitter datasets collected during
the 2017 French presidential elections, and known IO~accounts
disclosed by Twitter over a broad range of IO campaigns
(May~2007 to February~2020), over 50,000 accounts, 17~countries,
and different account types including both trolls and bots.  Our
system detects IO~accounts with 96\% precision, 79\% recall, and
96\%~area-under-the-PR-curve, maps out salient network communities,
and discovers high-impact accounts that escape the lens of traditional
impact statistics based on activity counts and network
centrality. Results are corroborated with independent sources of known
IO~accounts from US Congressional reports, investigative journalism,
and IO datasets provided by Twitter.
\end{abstract}

\begin{document}

\ifpnastypesetdefaults\else\ifpnastypesetarxiv
\makeatletter
\@input{io_detection_pnas_si_final_aux.tex}%
\makeatother
\fi\fi

\maketitle
\thispagestyle{firststyle}
\ifthenelse{\boolean{shortarticle}}{\ifthenelse{\boolean{singlecolumn}}{\abscontentformatted}{\abscontent}}{}



\dropcap{A}lthough propaganda is an ancient mode of statecraft, the
weaponization of digital communications and social media to conduct
disinformation campaigns at previously unobtainable scales, speeds,
and reach presents new challenges to identify and counter hostile
influence operations~\cite{king2017, stella2018, vosoughi2018,
starbird2019, rid2020, schmidt2020}. Before the internet, the tools
used to conduct such campaigns adopted longstanding---but
effective---technologies. For example, Mao's guerrilla strategy
emphasizes---``[p]ropaganda materials are very important. Every large
guerrilla unit should have a printing press and a mimeograph stone''
(ref.~\citenum{mao1937}, p.~85).  Today, many powers have exploited the
internet to spread propaganda and disinformation to weaken their
competitors. For example, Russia's official military doctrine calls to
``[e]xert simultaneous pressure on the enemy throughout the enemy's
territory in the global information space''~(ref.~\citenum{putin2014},
section~II).



Online influence operations (IOs) are enabled by the low cost,
scalability, automation, and speed provided by social media platforms
on which a variety of automated and semiautomated innovations are
used to spread disinformation~\cite{king2017, stella2018,
starbird2019}. Situational awareness of semiautomated IOs at speed and
scale requires a semiautomated response capable of detecting and
characterizing IO narratives and networks, and estimating their impact
either directly within the communications medium, or more broadly in
the actions and attitudes of the target audience. This arena presents
a challenging, fluid problem whose measured data is comprised of large
volumes of human- and machine-generated multimedia
content~\cite{gadde2018}, many hybrid interactions within a social
media network~\cite{smith2018b}, and actions or consequences resulting
from the IO campaign~\cite{birnbaum2019}. These characteristics of
modern IOs can be addressed by recent advances in machine learning in
several relevant fields: natural language processing (NLP),
semisupervised learning, and network causal inference.

This paper presents a framework to automate detection and
characterization of IO campaigns. The contributions of the paper are:
1)~an end-to-end system to perform narrative detection, IO account
classification, network discovery, and estimation of IO causal impact;
2)~a robust semisupervised approach to IO account classification;
3)~a method for detection and quantification of causal influence on a
network~\cite{smith2018b}; and 4)~application of this approach to
genuine hostile IO campaigns and datasets, with classifier and impact
estimation performance curves evaluated on confirmed IO networks.  Our
system discovers salient network communities and high-impact accounts
in spreading propaganda.  The framework integrates natural language
processing, machine learning, graph analytics, and network causal
inference to quantify the impact of individual actors in spreading IO
narratives.  Our general dataset was collected over numerous IO
scenarios during 2017 and contains nearly $800$~million tweets and
$13$~million accounts.  IO account classification is performed using a
semisupervised ensemble-tree classifier that uses both semantic and
behavioral features, and is trained and tested on accounts from our
general dataset labeled using Twitter's election integrity dataset
that contains over $50{,}000$ known IO accounts posting between
May~2007 to February~2020 from $17$ countries, including both trolls and
bots~\cite{gadde2018}.  To the extent possible, classifier performance
is compared to other online account classifiers. The impact of each
account is inferred by its causal contribution to the overall
narrative propagation over the entire network, which is not accurately
captured by traditional activity- and topology-based impact
statistics. The identity of several high-impact accounts are
corroborated to be agents of a foreign influence operations or
influential participants in known IO campaigns using Twitter's
election integrity dataset and reports from the US Congress and
investigative journalists~\cite{gadde2018, USHPSCI2017a, USHPSCI2017b, marantz2017,
kessler2018, birnbaum2019}.


\section*{Framework}
\customlabel{sec:framework}{Framework}

\begin{figure}[t]
\centering
\includegraphics[width=1\linewidth]{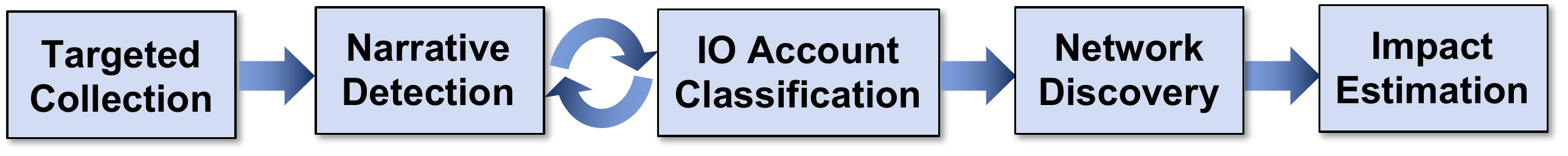}
\caption{Framework block diagram of end-to-end IO detection and characterization.}\label{fig:block_diagram}
\end{figure}

\begin{figure*}[b]
\centering
\includegraphics[width=0.667\linewidth]{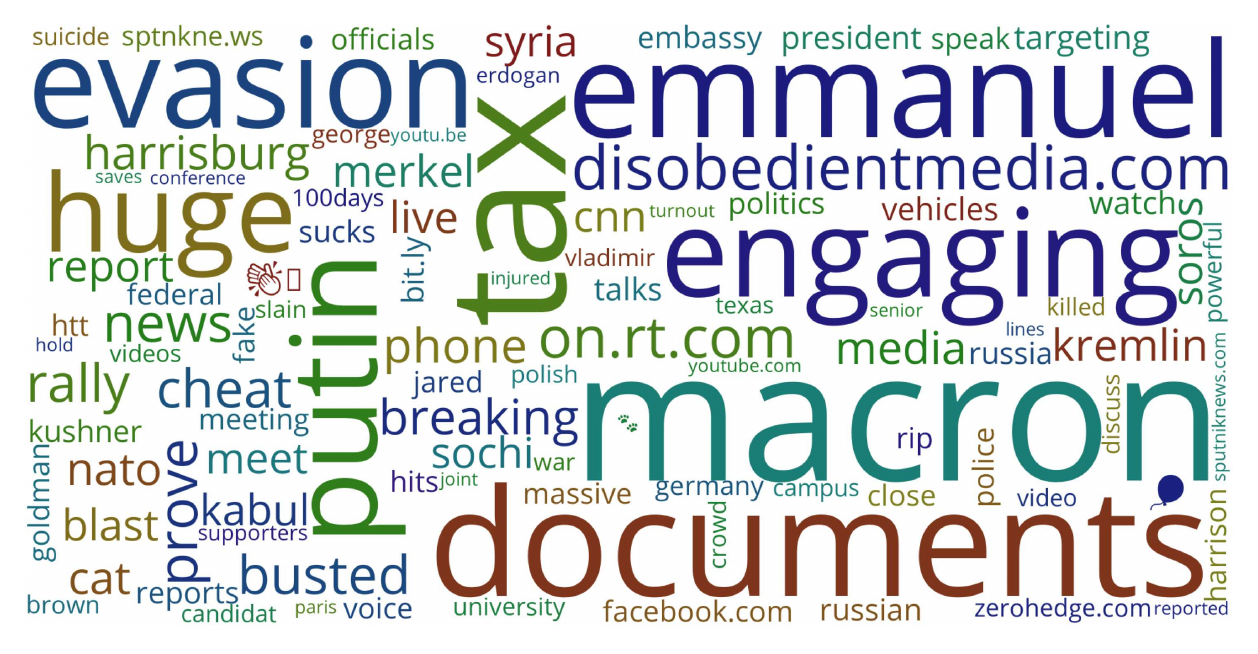}
\includegraphics[width=0.667\linewidth]{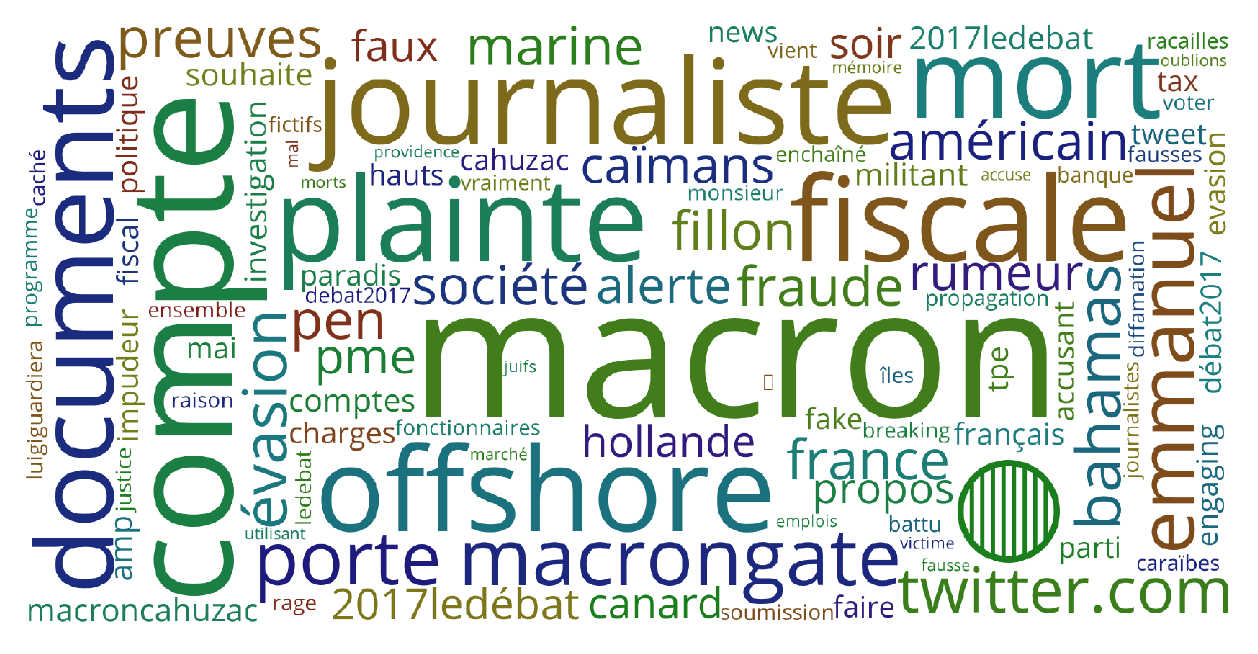}
\caption{Word~clouds associated with the `offshore accounts' topic in
English (above) and French (below). Topics are selected from those
generated from the English corpus \mathsffamily (${N=152{,}203}$) and
the French corpus (${N=1{,}070{,}158}$) as described in
{\color{blue}\textit{\href{\THESIURL}{SI
Appendix}}}.}\label{fig:wordcloud_enfr}
\end{figure*}

The end-to-end system framework collects contextually relevant data,
identifies potential IO narratives, classifies accounts based on their
behavior and content, constructs a narrative network, and estimates
the impact of accounts or networks in spreading specific narratives
(Fig.~\ref{fig:block_diagram}). First, potentially relevant social
media content is collected using the Twitter public application
programming interface (API) based on keywords,
accounts, languages, and spatiotemporal ranges. Second, distinct
narratives are identified using topic modeling, from which narratives
of interest are identified by analysts. In general, more sophisticated
NLP techniques that exploit semantic similarity, e.g., transformer
models~\cite{reimers2019}, can be used to identify salient narratives.
Third, accounts participating in the selected narrative receive an IO
classifier score based on their behavioral, linguistic, and content
features. The second and third steps may be repeated to provide a more
focused corpus for IO narrative detection. Fourth, the social network
of accounts participating in the IO narrative is constructed using
their pattern of interactions. Fifth, the unique impact of each
account---measured using its contribution to the narrative spread over
the network---is quantified using a network causal inference
methodology.  The end product of this framework is a mapping of the IO
narrative network where IO~accounts of high impact are identified.

\showmatmethods{} 

\section*{Methodology}
\customlabel{sec:methodology}{Methodology}

\subsection*{Targeted Collection}
\customlabel{sec:target}{Targeted Collection}

Contextually relevant Twitter data are collected using the Twitter API
based on keywords, accounts, languages, and spatiotemporal filters
specified by the authors.  For example, during the 2017 French
presidential election, keywords include the leading
candidates, \twitter{\#Macron} and \twitter{\#LePen}, French
election-related issues, including hostile narratives expected to harm
specific candidates, e.g., voter abstention~\cite{RT2017} and
unsubstantiated allegations~\cite{borger2017,schmidt2020}.  Because
specific narratives and influential IO~accounts are discovered
subsequently, they offer additional cues to either broaden or refocus
the collection. In the analysis described in the preceding subsection,
$28$~million Twitter posts and nearly $1$~million accounts potentially
relevant to the 2017 French presidential election were collected over
a $30$-d period preceding the election, and a total of nearly
$800$~million tweets and information on $13$~million distinct accounts
were collected.

\begin{figure}[t]
\centering
\includegraphics[width=\linewidth]{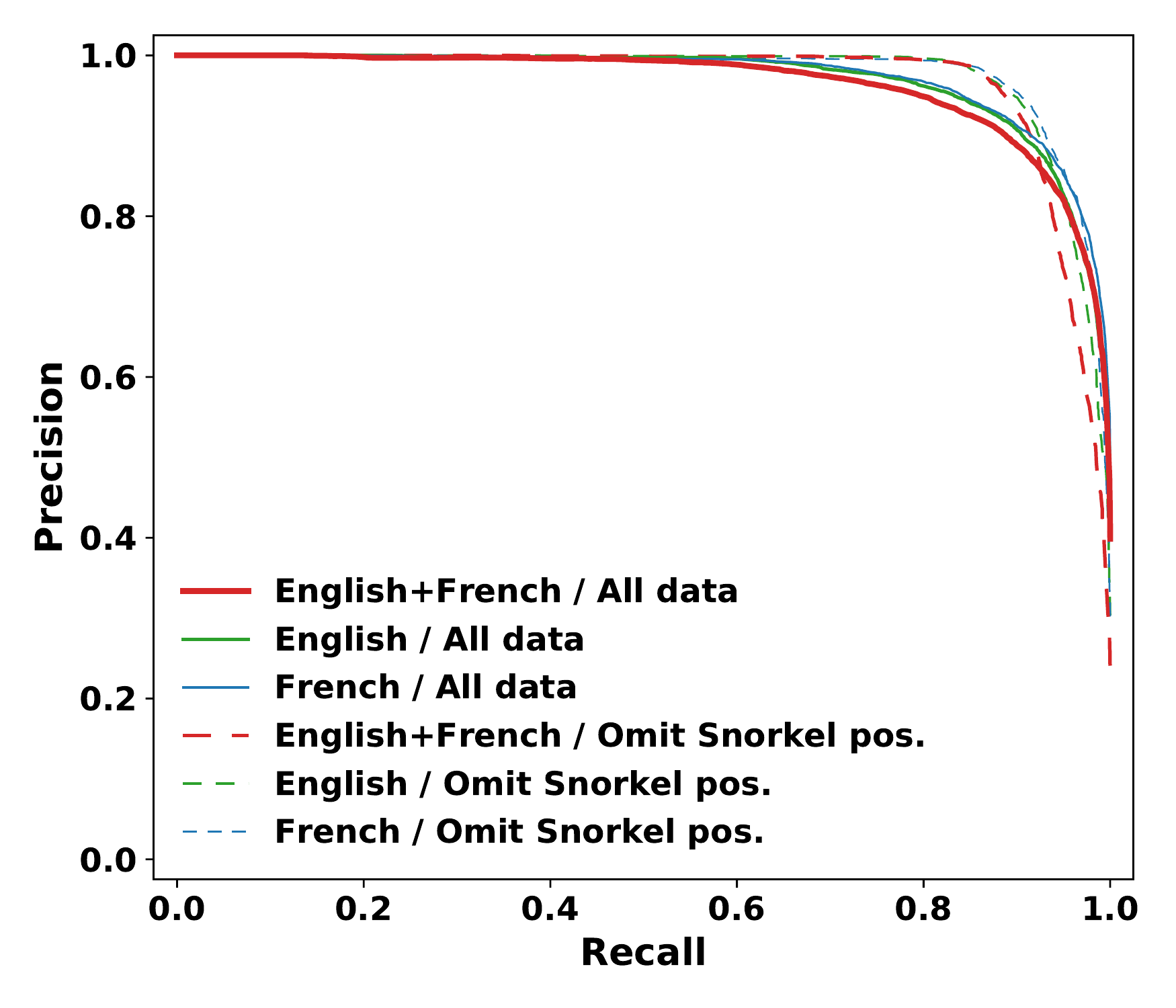}
\caption{P\hbox{-}R classifier performance with \mathsffamily $3{,}151$ known IO
accounts~\cite{gadde2018} and language-specific training data:
combined English and French (red curves, ${N=17{,}999}$); English only
(green curves, ${N=13{,}155}$); French only (blue curves,
${N=13{,}159}$). Cross-validated using a~${90:10}$ split and all data
(solid curves), and Snorkel positives omitted (dashed
curves).}\label{fig:classPRCompare}
\end{figure}


\begin{figure*}[b]
\centering
\includegraphics[width=0.667\linewidth]{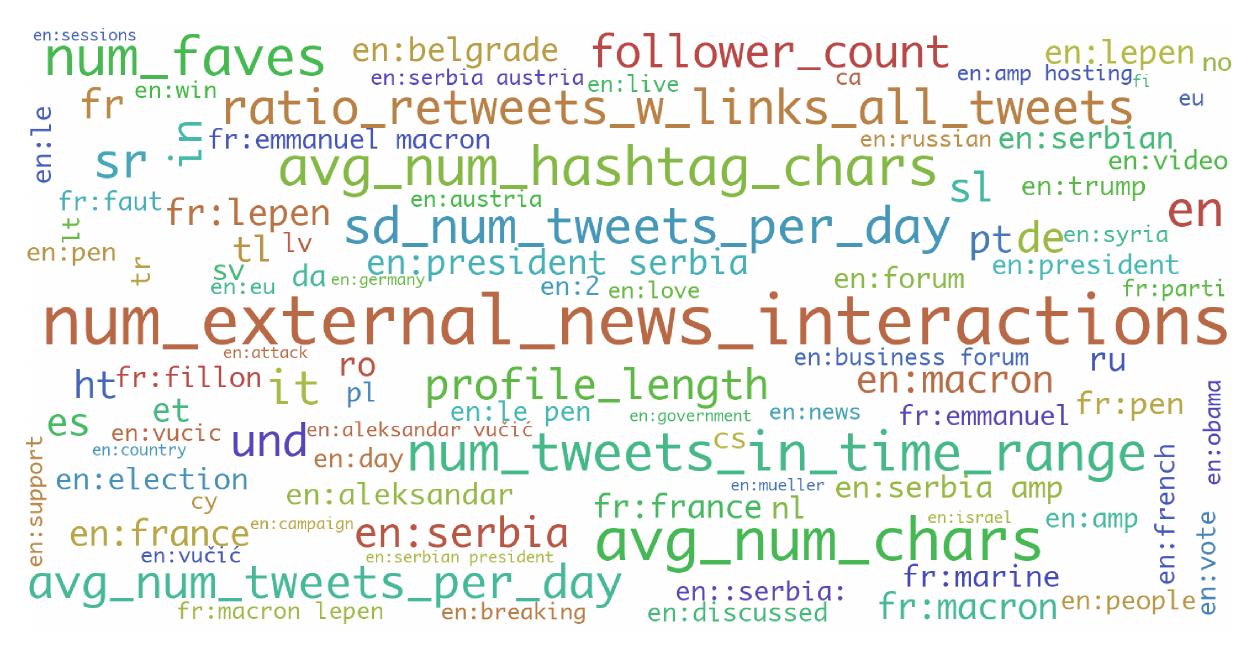}
\caption{The 100 most important features in the French and English
combined classifier, represented by relative size in a word~cloud.}
\label{fig:feature_imp}
\end{figure*}

\subsection*{Narrative Detection}
\customlabel{sec:narratives}{Narrative Detection}

Narratives are automatically generated from the targeted Twitter data
using a topic modeling algorithm~\cite{mccallum2002}. First, accounts
whose tweets contain keywords relevant to the subject, or exhibit
pre-defined, heuristic behavioral patterns within a relevant time
period are identified. Second, content from these accounts are passed
to a topic modeling algorithm, and all topics are represented by a
collection, or bag of words. Third, interesting topics are identified
manually. Fourth, tweets that match these topics above a pre-defined
threshold are selected. Fifth, a narrative network is constructed with
vertices defined by accounts whose content matches the selected
narrative, and edges defined by retweets between these accounts.  In
the case of the 2017 French elections, the relevant keywords are:
``Macron,'' ``leaks,'' ``election,'' and ``France''; the languages used in
topic modeling are English and French.

\subsection*{IO Account Classification}
\customlabel{sec:methodology_io_account_classification}{IO Account Classification}

Developing an automated IO classifier is challenging because the
number of actual examples is necessarily small, and the behavior and
content of these accounts can change over both time and scenario.
Semisupervised classifiers trained using heuristic rules can address
these challenges by augmenting limited truth data with additional
accounts that match IO heuristics within a target narrative that does
not necessarily appear in the training data.  This approach has the
additional intrinsic benefits of preventing the classifier from being
overfit to a specific training set or narrative, and provides a path
to adapt the classifier to future narratives.  IO~account classifier
design is implemented using this semisupervised machine-learning
approach built with the open~source libraries scikit-learn and
Snorkel~\cite{pedregosa2011, ratner2017}, and soft labeling functions
based on heuristics of IO~account metadata, content, and~behavior.
The feature space comprises behavioral characteristics, profile
characteristics, languages used, and the $1$- and $2$\hbox{-}grams
used in tweets; full details are provided in {\color{blue}\textit{SI
Appendix}}, {\color{blue}section~\customxrref{app:ioclassifier}}, and
feature importances are illustrated in~\textit{\ref{sec:results}}
(see Fig.~\ref{fig:feature_imp}). Our design approach to semisupervised
learning is to develop labeling functions only for behavioral and
profile characteristics, not narrative- or content-specific features,
with the expectation that such labeling functions are transferable
across multiple scenarios.

The classifier is trained and tested using sampled accounts
representing both IO-related and general narratives. The approach is
designed to prevent overfitting by labeling these sampled accounts
using semisupervised Snorkel heuristics.  These accounts are
collected as described in \textit{\ref{sec:target}}. There are four
categories of training and testing data: known IO accounts (known
positives from the publicly available Twitter elections integrity
dataset)~\cite{gadde2018}, known non-IO accounts (known negatives
comprised of mainstream media accounts), Snorkel-labeled positive
accounts (heuristic positives), and Snorkel-labeled negative accounts
(heuristic negatives).  All known positives are eligible to be in the
training set, whether they are manually operated troll IO accounts or
automated bots.  Because we include samples that represent both
accounts engaging in the IO~campaign and general accounts, our
training approach is weakly dependent upon the IO~campaign in two
ways.  First, only known IO accounts whose content includes languages
relevant to the IO~campaign are used.  Second, a significant fraction
(e.g., $67\%$) of Snorkel-labeled accounts (either positive or
negative) must be participants in the narrative. Finally, we establish
confidence that we are not overfitting by using cross-validation to
compute classifier performance. Additionally, overfitting is observed
without using Snorkel heuristics (see~{\color{blue}\textit{SI
Appendix}}, {\color{blue}section~\customxrref{app:class_comp}}),
supporting the claim that semisupervised learning is a necessary
component of our design.  Dimensionality reduction and classifier
algorithm selection is performed by optimizing precision--recall
performance over a broad set of dimensionality reduction approaches,
classifiers, and parameters (see~{\color{blue}\textit{SI Appendix}},
{\color{blue}section~\customxrref{app:ioclassifier}}). In the results section,
dimensionality reduction is performed with Extra-Trees
(ET)~\cite{pedregosa2011} and the classifier is the Random Forest (RF)
algorithm~\cite{pedregosa2011}.  Ensemble tree classifiers learn the
complex concepts of IO~account behaviors and characteristics without
over-fitting to the training data through a collection of decision
trees, each representing a simple concept using only a few features.

\begin{figure*}[t]
\centering
\hbox{\raise248pt\vbox to0pt{\vss\hbox to0pt{\kern0pt\footnotesize\sf
Macron allegations\hss}}%
\raise256pt\vbox to0pt{\vss\hbox to0pt{\kern220pt\footnotesize\sf Pro-Abstention\hss}}%
\raise48pt\vbox to0pt{\vss\hbox to0pt{\kern360pt\footnotesize\sf Pro-Macron\hss}}%
\includegraphics[width=\linewidth]{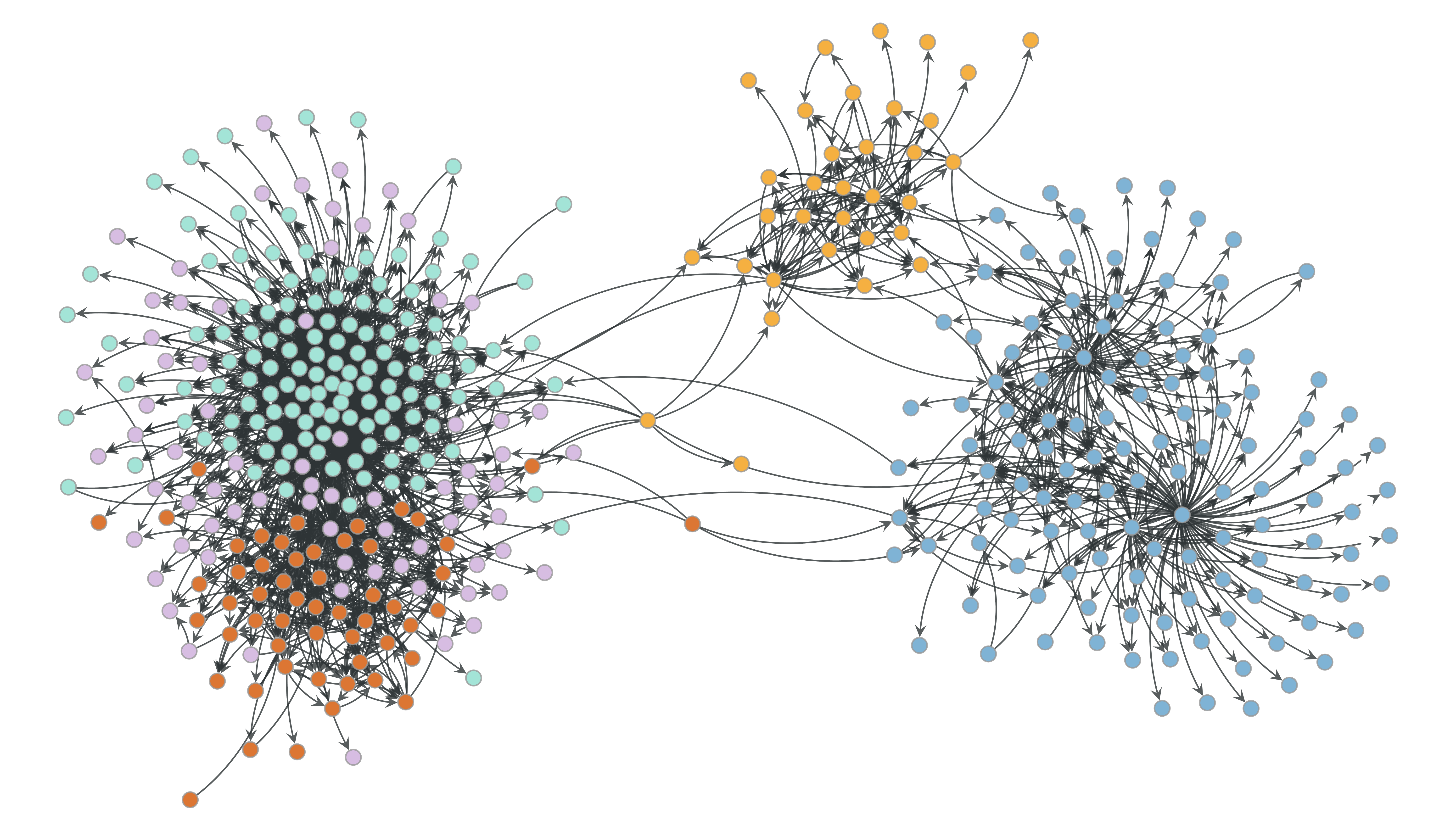}}
\caption{Community membership in the French narrative network
(Fig.~\ref{fig:wordcloud_enfr}). Colors show inferred membership from
a blockmodel~\cite{peixoto2014b}.}\label{fig:fr_communities}
\end{figure*}

\subsection*{Network Discovery}
\customlabel{sec:methodology_network_discovery}{Network Discovery}

The narrative network---a social network of participants involved in
discussing and propagating a specific narrative---is constructed from
their observed pattern of interactions. In \textit{\ref{sec:results}},
narrative networks are constructed using retweets. Narrative networks
and their pattern of influence are represented as graphs whose edges
represent strength of interactions.  The (directed) influence from
(account) vertex~$v_i$ to vertex~$v_j$ is denoted by the weighted
edge~$a_{ij}$. For simplicity in the sequel, network vertex~$v_i$ is
also referred to as~$i$.  The influence network is represented by the
adjacency matrix $\bm{A}=(a_{ij})$. Because actual influence is not
directly observable, the influence network is modeled as a random
variable with Poisson distribution parameterized by the observed
evidence of influence. Specifically, influence~$a_{ij}$ is modeled
with prior distribution $a_{ij}\sim\Poisson(\text{frequency of
interactions from $i$ to $j$})$, as counts of interactive influence in
real-world networks. Observations of past interactions or influence on
a subset of edges can be used to estimate the rates on the missing
edges through inference on a network model that captures realistic
characteristics such as sparsity, varying vertex degrees, and
community structure~\cite{kao2019}.

\subsection*{Impact Estimation}
\customlabel{sec:methodology_impact_estimation}{Impact Estimation}

Impact estimation is based on a method that quantifies each
account's unique causal contribution to the overall narrative
propagation over the entire network. It accounts for social
confounders (e.g., community membership, popularity) and disentangles
their effects from the causal estimation. This approach is based on
the network potential outcome framework~\cite{kao2017}, itself based
upon Rubin's causal framework~\cite{imbensrubin2015}. Mathematical
details are provided in {\color{blue}\textit{SI Appendix}},
{\color{blue}section~\customxrref{SI_causalImpact}}.

The fundamental quantity is the network potential outcome of each
vertex, denoted $Y_i(\bm{Z},\bm{A})$, under exposure to the narrative
from the source vector $\bm{Z}$ via the influence network
$\bm{A}$. Precisely, $\bm{Z}$ is a binary $N$-vector of narrative
sources (a.k.a.\ treatment vector). In this study, vertices are user
accounts, edges represent influence as described
in \textit{\ref{sec:methodology_network_discovery}}, and the potential
outcomes are the number of tweets in the narrative. The influence
network is an important part of the treatment exposure mechanism. An
account's exposure to the narrative is determined by both the sources
as well as exposures to them delivered through the influence
network. The impact $\zeta_j$ of each vertex $j$ on the overall
narrative propagation is defined using network potential outcome
differentials averaged over the entire network:
\begin{equation}
\label{eq:individual_impact_causal_estimand}
	\zeta_j(\bm{z}) \buildrel{\rm def}\over= \frac{1}{N} \sum_{i =
          1}^N \bigl(Y_i(\bm{Z}=\bm{z}_{j+}, \bm{A})
        -Y_i(\bm{Z}=\bm{z}_{j-}, \bm{A})\bigr).
\end{equation}
This causal estimand is the average difference between the individual
outcomes with $v_j$ as a source such~that $\bm{z}_{j+}:=
(z_1,\ldots,{z_j:=1},\discretionary{}{}{}\ldots,z_N)^\T$, versus $v_j$
not a source, $\bm{z}_{j-}:=
(z_1,\ldots,{z_j:=0},\discretionary{}{}{}\ldots,z_N)^\T$.  This impact
is the average (per~vertex) number of additional tweets generated by
an user's participation in the narrative. The source is said to be
uniquely impactful if it is the only source.

\begin{figure*}[t]
\centering
\hbox{\raise248pt\vbox to0pt{\vss\hbox to0pt{\kern0pt\footnotesize\sf
Macron allegations\hss}}%
\raise256pt\vbox to0pt{\vss\hbox to0pt{\kern220pt\footnotesize\sf Pro-Abstension\hss}}%
\raise48pt\vbox to0pt{\vss\hbox to0pt{\kern360pt\footnotesize\sf Pro-Macron\hss}}%
\raise36pt\vbox to0pt{\vss\hbox to0pt{\kern184pt\includegraphics[width=.25\linewidth]{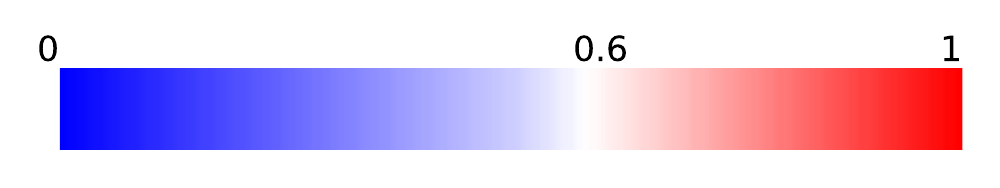}\hss}}%
\includegraphics[width=\linewidth]{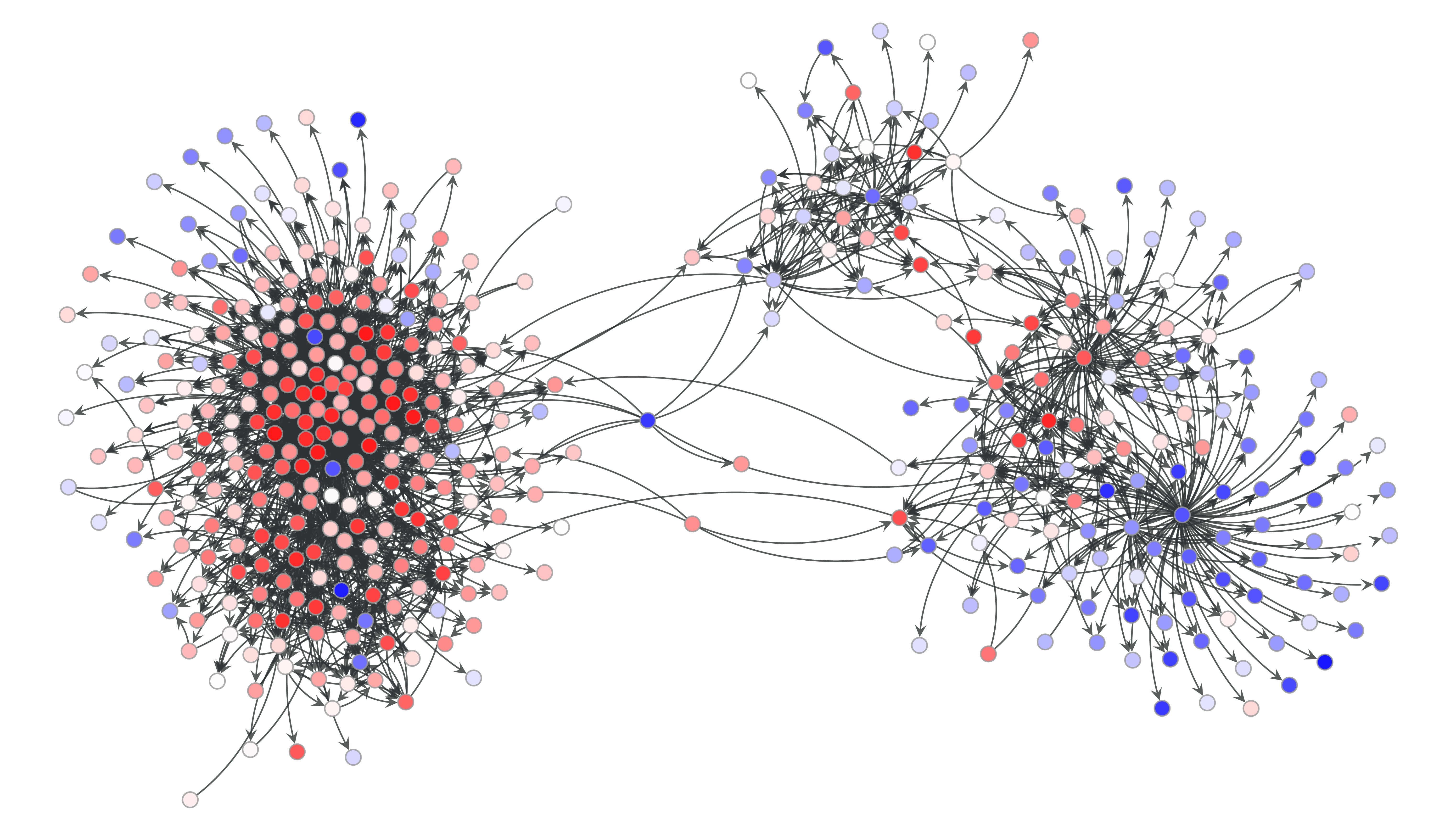}}
\caption{Classifier scores over the French narrative
network (Fig.~\ref{fig:wordcloud_enfr}). The \mathsffamily $0$--$1$
score range indicates increasing similarity to known IO~accounts.}\label{fig:fr_classifier}
\end{figure*}

\begin{figure}[b]
\centering
\includegraphics[width=0.96\linewidth]{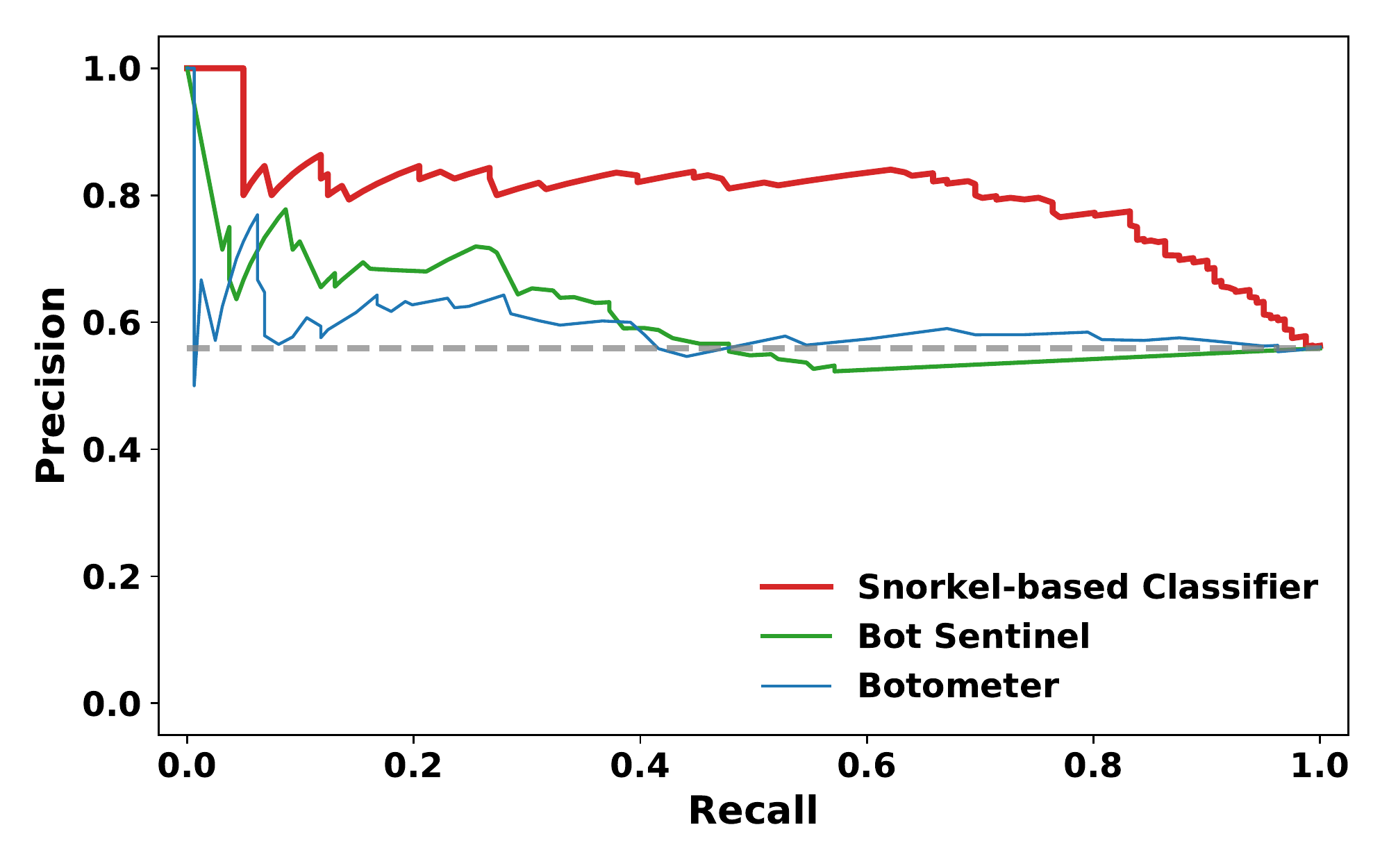}
\caption{Classifier performance comparison: Snorkel-based classifier
(red curve, \mathsffamily ${N=415}$) vs.\ Botometer (green curve,
${N=289}$) vs.\ Bot~Sentinel (blue curve, ${N=288}$), given proxy,
community-based truth. The dashed gray horizontal line at $63$\% is
the fraction of presumptively true examples in the community-based
proxy for known IO~accounts, and therefore represents random chance
precision performance.}\label{fig:fr_PRcomp}
\end{figure}


\begin{figure*}[b]
\centering
\includegraphics[width=0.96\linewidth]{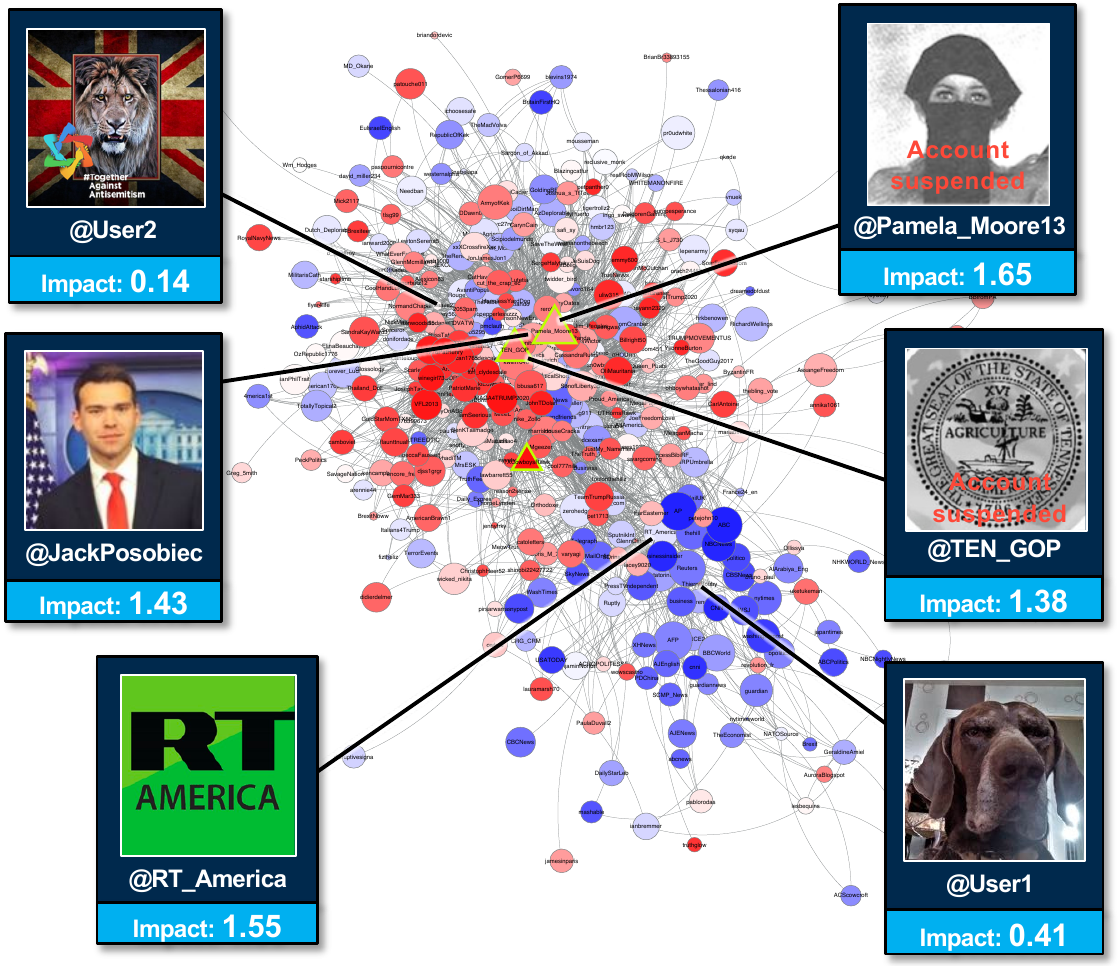}
\caption{Impact network (accounts sized by impact) colored by IO
classifier score on the English narrative network
Fig.~\ref{fig:wordcloud_enfr}). Known IO~accounts are highlighted in
triangles. Image credits: Twitter\slash JackPosobiec, Twitter\slash
RT\underscore America, Twitter\slash Pamela\underscore Moore13,
Twitter\slash TEN\underscore GOP.}
\label{fig:en_network}
\end{figure*}

It is impossible to observe the outcomes at each vertex with both
exposure conditions under source vectors $\bm{z}_{j+}$ and
$\bm{z}_{j-}$; therefore, the missing potential outcomes must be
estimated, which can be accomplished using a model.  After estimating
the model parameters from the observed outcomes and vertex covariates,
missing potential outcomes in the causal estimand $\zeta_j$ can be
imputed using the fitted model. Potential outcomes are modeled using a
Poisson generalized linear mixed model (GLMM) with the canonical log
link function and linear predictor coefficients
$(\tau,\bm{\gamma},\bm{\beta},\mu)$ corresponding to the source
indicator $Z_i$, $n$-hop exposure vector $s^{(n)}_i$, the covariate
vector $\bm{x}_i$, and the baseline outcome. The covariate vector
$\bm{x}_i$ includes the potential social confounders such as
popularity and community membership. These confounders are accounted
for through covariate adjustment in order to disentangle actual causal
impact from effects of homophily (birds of a feather flock together)
and vertex degree on outcomes, by meeting the key unconfounded
influence network
assumption~\customxrref{as:unconfoundedInfluenceNetworkUnderNetworkInterference}
in {\color{blue}\textit{SI Appendix}},
{\color{blue}\textit{Assumption~\customxrref{as:unconfoundedInfluenceNetworkUnderNetworkInterference}}}.
For correctness and rigor, imputation of the missing potential
outcomes are designed to meet the unconfoundedness assumptions that
lead to ignorable treatment exposure mechanism under network
interference, detailed in {\color{blue}\textit{SI Appendix}},
{\color{blue}section~\customxrref{sec:imputation_and_unconfoundedness_assumption}}.
The GLMM model for the potential outcomes is
\begin{align}
\label{eq:PO_PoissonGLM}
Y_i(\bm{Z},\bm{A}) & \sim \Poisson(\lambda_i), \nonumber \\
\log\lambda_i & = \tau Z_i + \biggl(\sum_{n=1}^{N_{\text{hop}}}
                   {s^{(n)}_i \tau \prod_{k=1}^n{\gamma_k}}\biggr) +
                   \bm{\beta}^\mathrm{T} \bm{x}_i + \mu + \epsilon_i,
\end{align}
where $\tau Z_i$ represents the primary effect from the source,
$\sum_{n=1}^{N_{\text{hop}}} {s^{(n)}_i \tau \prod_{k=1}^n{\gamma_k}}$
represents the accumulative social influence effect from $n$-hop
exposures $s^{(n)}_i$ to the source, $\gamma_k$ (between $0$ and $1$)
represents how quickly the effect decays over each additional $k$th hop, 
$\bm{\beta}^\T\bm{x}_i$ is the effect of the unit covariates $\bm{x}_i$ including potential
social confounders such as popularity and community membership, $\mu$
is the baseline effect on the entire population, and
$\epsilon_i \sim \Normal(0,\sigma_{\!\epsilon}^2)$ provides independent
and identically distributed variation for heterogeneity between the
units.  The amounts of social exposure at the $n$th hop are determined
by $(\bm{A}^\T)^n\bm{Z}$. This captures narrative propagation via all
exposure to sources within the narrative network.  Diminishing return
of additional exposures is modeled using (elementwise) log-exposure,
$\bm{s}^{(n)} =\log\bigl((\bm{A}^\T)^n\bm{Z} + 1\bigr)$. The influence matrix $\bm{A}$,
with prior distribution specified in~\textit{\ref{sec:methodology_network_discovery}}, is
jointly estimated with the model parameters $\tau$, $\bm{\gamma}$, $\bm{\beta}$,
$\mu$, through Markov Chain Monte~Carlo (MCMC) and Bayesian regression.

\section*{Results}
\customlabel{sec:results}{Results}

\subsection*{Targeted Collection}

The targeted collection for the 2017 French presidential election
includes $28{,}896{,}185$ potentially relevant tweets and $999{,}883$
distinct accounts, all collected over a $30$-d period preceding the
election on 7~May 2017. Targeted collections for several other IO
scenarios were also collected during 2017, resulting in a dataset with
$782{,}678{,}201$ tweets and $12{,}723{,}995$ distinct accounts. Of
these, there are $3{,}151$ known IO accounts that posted in English or
French (see~{\color{blue}\textit{SI Appendix}},
{\color{blue}Fig.~\customxrref{fig:origin-pie-chart}}).  The great majority of
accounts are unrelated to both these known IO accounts and the French
election, but will provide negative examples for IO classifier
training.



\subsection*{Narrative Detection}

Narratives immediately preceding the election are generated
automatically by dividing this broad content into language groups,
restricting the content and time period to election-related posts
within $1$~wk preceding the election's media blackout date of 5~May
2017, and filtering accounts based on interaction with non-French
media sites pushing narratives expected to harm specific
candidates. Topic modeling~\cite{mccallum2002} is applied to the
separate English and French language corpora, and the resulting topics
are inspected by the authors to identify relevant narratives. Two such
narratives are illustrated in Fig.~\ref{fig:wordcloud_enfr} by the
most-frequent word and emoji usage appearing in tweets included within
the topic. From the English corpus (${N=152{,}203}$), $15$~topics are
generated; from the French (${N=1{,}070{,}158}$), $30$~topics. These
automatically generated topics correspond closely to allegations
claimed to be spread by WikiLeaks, candidate Marine Le~Pen, and
others~\cite{dearden2017, marantz2017}.  The role of bots in spreading
these allegations using the \twitter{\#MacronLeaks} hashtag narrative
is studied by Ferrara~\cite{ferrara2017}.  Two topics, one in English
and one in French, pertaining to unsubstantiated financial allegations
(Fig.~\ref{fig:wordcloud_enfr}) will be used to identify accounts
involved in spreading these narratives, independent of whether the
accounts were used for narrative detection.

\subsection*{IO Account Classification}

Twitter published IO truth data containing $50{,}230$ known IO~account
identities and their multimedia content~\cite{gadde2018}.  IO accounts
represent a tiny fraction of all Twitter accounts, e.g., Twitter's
$50{,}230$ known IO~accounts are only $0.02$\% of its $330$~million
active monthly users, and bots are estimated to comprise $9$--$15$\%
of all Twitter accounts~\cite{varol2017}.  This dataset is used along
with heuristic rules to train a semisupervised
classifier~\cite{ratner2017} using the approach described in
\textit{\ref{sec:methodology}}, and the rulesets are detailed in the
{\color{blue}\textit{SI Appendix}},
{\color{blue}section~\customxrref{app:ioclassifier}}. The classifier
is trained and tested using content from these sources: known
IO~accounts that have tweeted on any topic at least once in either
English or French, $20$~known non-IO mainstream media accounts, and
Snorkel-labeled accounts randomly drawn from our dataset, such~that
$67\%$ are topically relevant and $33\%$ are topically neutral. There
are $3{,}151$ known IO~accounts from Twitter's dataset in our dataset
that have tweeted in English or French. To account for a possible
upper bound of up to $15$\% bots (IO-related or not), we randomly
select $15{,}000$ presumptively false examples with $5{,}000$ each of
three false classes: topically relevant accounts that tweeted at least
once in English, at least once in French, and topically neutral,
randomly chosen from the general dataset (see~{\color{blue}\textit{SI Appendix}},
{\color{blue}Fig.~\customxrref{fig:origin-pie-chart}}).

\subsubsection*{Precision-recall performance}

Precision-recall (P\hbox{-}R) performance of the classifier is computed via cross
validation using the same dataset with a ${90:10}$~split, averaged
over $20$~rounds (Fig.~\ref{fig:classPRCompare}).  Because the number
of known cases is limited, weak supervision from the heuristic Snorkel
labeling functions is used to identify true examples before cross
validation~\cite{ratner2017}. All training is performed with
Snorkel-labeled data, and cross validation is performed both using
both Snorkel-labeled data (solid curves in
Fig.~\ref{fig:classPRCompare}) and omitting Snorkel positives (dashed
curves).  Sensitivity to language-specific training data is also
computed by restricting topically relevant, false examples to specific
languages.  All classifiers exhibit comparably strong performance, and
small differences in relative performance is consistent with the
authors’ expectations. All classifiers detect IO accounts with $96$\%
precision and $79$\% recall at a nominal operating point, $96$\%
area-under-the-PR-curve (AUPRC), and~8\% equal-error rate (EER). The
English-only and French-only classifiers perform slightly better
(nominally 0--3\%) than the combined language model, consistent with
the expectation that models with greater specificity outperform less
specific models.  This strong classifier performance will be combined
with additional inferences---narrative networks, community structure,
and impact estimation---to identify potentially influential
IO~accounts involved in spreading particular narratives.

With this dataset, the original feature space has dimension
$1{,}896{,}163$: $17$ behavior and profile features, $61$ languages,
and $1{,}896{,}085$ $1$- and $2$\hbox{-}grams. {\color{blue}\textit{SI Appendix}},
{\color{blue}section~\customxrref{app:ioclassifier}} lists the behavior, profile
features, and language features.  Grid search over feature
dimensionality is used to identify the best feature set for
dimensionality reduction: $10$ behavioral and profile features, $30$
language features, and $500$ $1$- and $2$\hbox{-}grams.  The most
important features used by the classifier are illustrated by the
relative sizes of feature names appearing in the word~cloud of
Fig.~\ref{fig:feature_imp}. Note that the most important features for
IO~account classification are independent of topic, and pertain
instead to account behavioral characteristics and frequency of
languages other than English or French. This topic independence
suggests the potential applicability of the classifier to other IO
narratives. Furthermore, the diversity of behavioral features suggests
robustness against future changes of any single behavior.

\subsubsection*{Classifier performance comparisons}
\customlabel{sec:classperfcomp}{Classifier Performance Comparisons}

Several online bot classifiers are used to report upon and study
influence campaigns~\cite{mcewan2020, vosoughi2018, ferrara2017},
notably Botometer (formerly BotOrNot)~\cite{davis2016} and
Bot~Sentinel~\cite{botsentinel2020}. Indeed, Rauchfleisch
and~Kaiser~\cite{rauchfleisch2020} assert based on several influential
papers that analyze online political influence that ``Botometer is the
de-facto standard of bot detection in academia'' (p.~2). In spite of
the differences between general, automated bot activity and the
combination of troll and bot accounts used for IO campaigns, comparing
the classifier performance between these different classifiers is
important because it is widespread practice to use such bot
classifiers for insight into IO campaigns.  Therefore, we compare the
P\hbox{-}R performance of our IO classifier to both Botometer and
Bot~Sentinel.  This comparison is complicated by three factors:
1)~neither Botometer nor Bot~Sentinel have published classifier
performance against known IO~accounts; 2)~neither project has posted
open source code; and 3)~known IO~accounts are immediately suspended,
which prevents post~hoc analysis with these online tools.

\begin{table}[b]
\setlength{\tabcolsep}{0.15em}
\newcommand\TopStrut{\rule{0pt}{2.6ex}}       
\newcommand\BotStrut{\rule[-1.2ex]{0pt}{0pt}} 
\newcolumntype{S}{>{\small}r}
\newcolumntype{T}{>{\small\let\twittersize\footnotesize}l}
\caption{Comparison of impact statistics between accounts on the English network:
  tweets (T), retweets (RT), followers (F), first tweet time on
  28~April, PageRank centrality (PR), and causal impact* (CI).}
\label{tab:impact_stats}
\begin{tabular*}{\linewidth}{T@{\extracolsep{\fill}}SSSSSS}
\hline
\multicolumn{1}{l}{Screen name} & \multicolumn{1}{c}{\ T} &
  \multicolumn{1}{c}{RT} &
  \multicolumn{1}{c}{\quad F} &
  \multicolumn{1}{c}{1st time} &
  \multicolumn{1}{c}{\quad PR} &
  \multicolumn{1}{r}{\bf CI*\TopStrut\BotStrut}\\
\hline
\Rowcolor{twitterblue}\twitter{@RT\underscore America} & 39 & 8 & 386k & 12:00 & 2706 &\bf 1.55\TopStrut \\
\twitter{@JackPosobiec} & 28 &123 & 23k & 01:54 & 4690 &\bf 1.43 \\
\Rowcolor{twitterblue}\twitter{@User1}\dag & 8 & 0 & 1.4k & 22:53 & 44 &\bf 0.14 \\
\twitter{@User2}\dag & 12 & 15 & 19k & 12:27 & 151 &\bf 0.41 \\
\Rowcolor{twitterblue}\twitter{@Pamela\underscore Moore13} & 10 & 31 & 56k & 18:46 & 97 &\bf 1.65 \\
\twitter{@TEN\underscore GOP} & 12 & 42 & 112k & 22:15 & 191 &\bf 1.38\BotStrut \\
\hline
\setarstrut{\footnotesize}%
\multicolumn{7}{>{\footnotesize}r}{*Estimate of the causal estimand in Equation~[\ref{eq:individual_impact_causal_estimand}]\TopStrut} \\
\multicolumn{7}{>{\footnotesize}r}{\dag Anonymized screen names of currently active accounts} \\
\restorearstrut
\end{tabular*}
\end{table}

\begin{figure*}[b]
\centering
\includegraphics[width=0.96\linewidth]{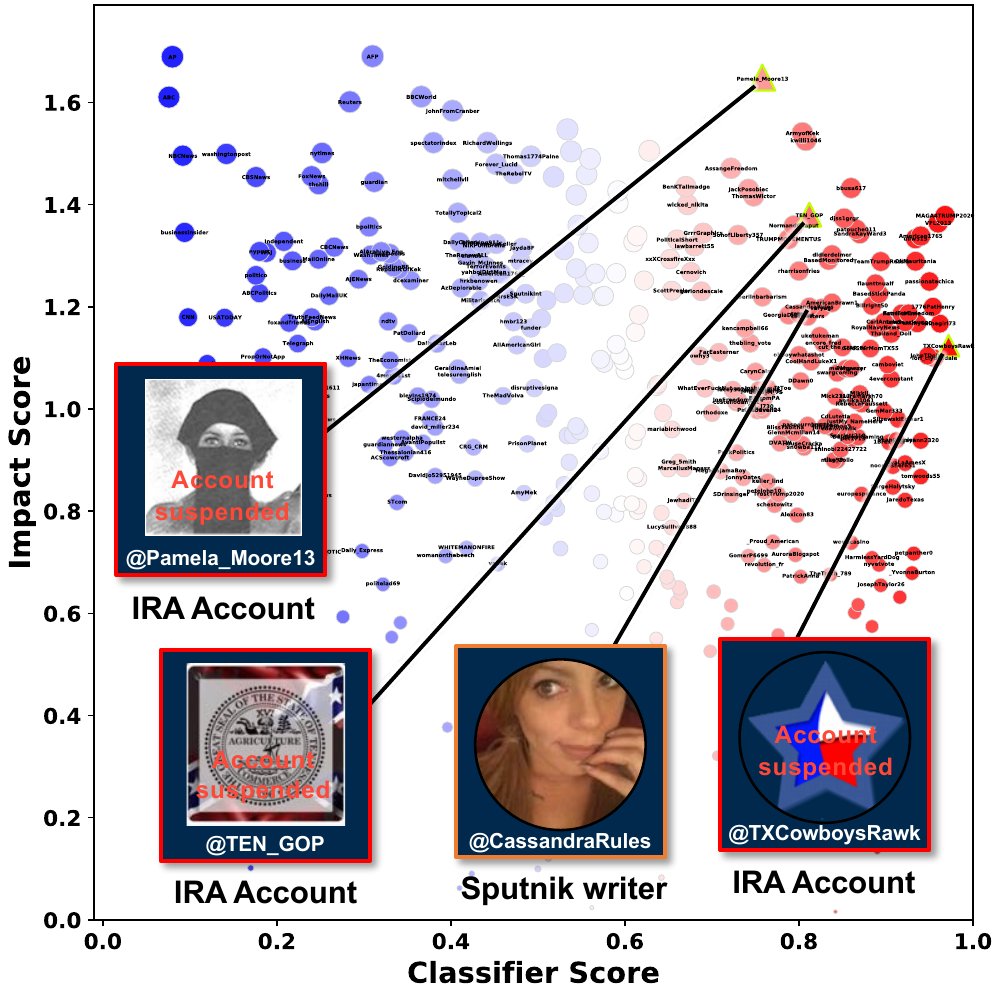}
\caption{Impact versus classifier score, English narrative network
(Fig.~\ref{fig:wordcloud_enfr}). Known IO~accounts run by the Internet
Research Agency (IRA)~\cite{USHPSCI2017a, USHPSCI2017b, gadde2018,
kessler2018, birnbaum2019} are highlighted. Image credits:
Twitter\slash Pamela\underscore Moore13, Twitter\slash TEN\underscore
GOP, Twitter\slash CassandraRules, Twitter\slash
TXCowboysRawk.}\label{fig:en_scatterplot}
\end{figure*}

Therefore, a proxy for known IO~accounts must be used for performance
comparisons. We use the observation that there exists strong
correlation between likely IO~accounts in our narrative network and
membership in specific, distinguishable communities independently
computed using an MCMC-based blockmodel~\cite{peixoto2014b}. Community
membership of accounts in the French language narrative network
(Fig.~\ref{fig:wordcloud_enfr}) are illustrated in
Fig.~\ref{fig:fr_communities}. Five distinct communities are detected,
three of which are identified to have promoted Macron allegation
narratives. The other two narratives promote pro-Macron and
pro-abstention narratives. Accounts in this narrative network are
classified on a $0$--$1$ scale of their similarity to known
IO~accounts, shown in Fig.~\ref{fig:fr_classifier}. Comparing
Figs.\ \ref{fig:fr_communities} and~\ref{fig:fr_classifier} shows that
the great majority of accounts in the `Macron allegation' communities
are classified as highly similar to known IO~accounts, and conversely,
the great majority of accounts in the pro-Macron and pro-abstention
communities are classified as highly dissimilar to known
IO~accounts. This visual comparison is quantified by the account
histogram illustrated in {\color{blue}\textit{SI Appendix}}, {\color{blue}Fig.~\customxrref{fig:fr_hist}}.

Using membership in these Macron allegation communities as a proxy
for known IO~accounts, P\hbox{-}R performance is computed for our IO
classifier, Botometer and Bot~Sentinel (Fig.~\ref{fig:fr_PRcomp}).
Note that Botometer's performance in Fig.~\ref{fig:fr_PRcomp} at a
nominal $50$\% recall is $56$\% precision, which is very close to the
$50$\% Botometer precision performance shown by Rauchfleisch
and~Kaiser using a distinctly different dataset and truthing
methodology~\cite[Fig.~4, ``all'']{rauchfleisch2020}.  Given this
narrative network and truth proxy, both Botometer and Bot~Sentinel
perform nominally at random chance of~$63$\% precision, the fraction
of presumptive IO~accounts. Our IO classifier has precision
performance of $82$--$85$\% over recalls of range $20$--$80$\%, which
exceeds random chance performance by $19$--$22$\%.  These results are
also qualitatively consistent with known issues of false positives and
false negatives in bot detectors~\cite{rauchfleisch2020}, though some
performance differences are also likely caused by the intended design
of Botometer or Bot~Sentinel, which is to detect general bot activity,
rather than the specific IO behavior on~which our classifier is
trained.

\subsection*{Network Discovery}
\customlabel{sec:results_network_discovery}{Network Discovery}

Tweets that match topics of Fig.~\ref{fig:wordcloud_enfr} are
extracted from the collected Twitter data.  French-language tweets
made in the week leading up to the blackout period, 28~April through
5~May 2017, are checked for similarity to the French language
topic. To ensure the inclusion of tweets on
the \twitter{\#MacronLeaks} data dump~\cite{ferrara2017, marantz2017,
smith2018b}, which occurred on the eve of the French media blackout,
English-language tweets from 29~April through 7~May 2017 are compared
to the English topic. In total, the French topic network consists of
$6{,}927$ accounts and the English topic network consists of $1{,}897$
accounts. For visual clarity, network figures are generated on the
most active accounts, $459$ in the French and $752$ in the English
networks.

\subsection*{Impact Estimation}
\customlabel{sec:impactestimation}{Impact Estimation}

Estimation on the causal impact of each account in propagating 
the narrative is performed by computing the estimand in
Eq.~[\ref{eq:individual_impact_causal_estimand}], considering each
account as the source. Unlike existing propagation methods on network 
topology~\cite{smith2014}, causal inference accounts also for 
the observed counts from each account to capture how each
source contributes to the subsequent tweets made by other accounts.
Results demonstrate this method's advantage over traditional impact
statistics based on activity count and network topology alone.

Impact estimation and IO classification on the English narrative
network (Fig.~\ref{fig:wordcloud_enfr}) are demonstrated in
Fig.~\ref{fig:en_network}.  Graph vertices are Twitter accounts sized
by the causal impact score (i.e.\ posterior mean of the causal
estimand) and colored by the IO classifier using the same scale as
Fig.~\ref{fig:fr_classifier}. Redness indicates account behavior and
content like known IO~accounts, whereas blueness indicates the
opposite.  This graph layout reveals two major communities involved in
narrative propagation of unsubstantiated financial allegations during
the French election.  The large community at the top~left comprises
many accounts whose behavior and content is consistent with known
IO~accounts.  The relatively smaller community at the lower~right
includes many mainstream media accounts that publish reports on this
narrative. Within this mainstream journalism-focused community, the
media accounts AP, ABC, RT, and~Reuters are among the most impactful,
consistent with expectation. The most remarkable result, however, is
that known IO~accounts are among the most impactful among the large
community of IO-like accounts. The existence of these IO~accounts was
known previously~\cite{USHPSCI2017a, USHPSCI2017b, gadde2018}, but not
their impact in spreading specific IO narratives.  Also note the
impactful IO accounts (i.e.\ the large red vertices) in the upper
community that appear to target many benign accounts (i.e.\ the white
and blue vertices).

A comparison between the causal impact and traditional impact
statistics is provided in Table~\ref{tab:impact_stats} on several
representative and\slash or noteworthy accounts highlighted in
Fig.~\ref{fig:en_network}. The prominent \twitter{@RT\underscore
America}, a major Russian media outlet, and \twitter{@JackPosobiec}, a
widely reported~\cite{marantz2017} account in spreading this
narrative, corroborate our estimate of their very high causal impact
scores.  This is also consistent with their early participation in
this narrative, high tweet counts, high number of followers, and large
PageRank centralities.  Conversely, \twitter{@User1}
and \twitter{@User2} have low impact statistics, and also receive low
causal impact scores as relatively non-impactful accounts. It is often
possible to interpret why accounts were
impactful. E.g., \twitter{@JackPosobiec} was one of the earliest
participants and has been reported as a key source in pushing the
related \twitter{\#MacronLeaks} narrative~\cite{ferrara2017,
marantz2017} (see~{\color{blue}\textit{SI Appendix}},
{\color{blue}Fig.~\customxrref{fig:influencemacronleaks}}). In that
same narrative, another impactful account \twitter{@UserB} serves as
the initial bridge from the English sub-network into the predominantly
French-speaking sub-network (see~{\color{blue}\textit{SI Appendix}},
{\color{blue}section~\customxrref{sec:cihashtagmacronleaks})}.

Known IO~account~\cite{USHPSCI2017a, USHPSCI2017b, gadde2018,
smith2018b, birnbaum2019} \twitter{@Pamela\underscore Moore13}'s
involvement in this narrative illustrates the relative strength of the
causal impact estimates in identifying relevant
IO~accounts.  \twitter{@Pamela\underscore Moore13} stands out as one
of the most prominent accounts spreading this narrative. Yet `her'
other impact statistics (T, RT, F, PR) are not distinctive, and
comparable in value to the not-impactful account \twitter{@User2}.
Additionally, known IO accounts \twitter{@TEN\underscore GOP}
and~\twitter{@TXCowboysRawk}~\cite{USHPSCI2017a, USHPSCI2017b,
kessler2018, gadde2018}, and Sputik
writer \twitter{@CassandraRules}~\cite{sputnik2020} all stand out for
their relatively high causal impact and IO~account classifier scores
(Fig.~\ref{fig:en_scatterplot}). Causal impact estimation is shown to
find high-impact accounts that do not stand out using traditional
impact statistics. This estimation is accomplished by considering how
the narrative propagates over the influence network, and its utility
is demonstrated using data from known IO~accounts on known IO
narratives. Additional impact estimation results are provided in the
{\color{blue}\textit{SI Appendix}},
{\color{blue}section~\customxrref{sec:cihashtagmacronleaks}}.

\subsection*{Influential IO Account Detection}

The outcome of the automated framework proposed in this manuscript is
the identification of influential IO~accounts in spreading IO
narratives. This is accomplished by combining IO classifier scores
with IO impact scores for a specific narrative
(Fig.~\ref{fig:en_scatterplot}). Accounts whose behavior and content
appear like known IO~accounts and whose impact in spreading an IO
narrative is relatively high are of potential interest. Such accounts
appear in the upper-right side of the scatterplot illustrated in
Fig.~\ref{fig:en_scatterplot}. Partial validation of this approach is
provided by the known IO~accounts discussed above. Many other accounts
in the upper-right side of Fig.~\ref{fig:en_scatterplot} have since
been suspended by Twitter, and some at the time of writing are
actively spreading conspiracy theories about the 2020 coronavirus
pandemic~\cite{donati2020}. These currently-active accounts
participate in IO-aligned narratives across multiple geo-political
regions and topics, and no~matter their authenticity, their content is
used hundreds of times by known IO~accounts~\cite{gadde2018}
(see~{\color{blue}\textit{SI Appendix}},
{\color{blue}section~\customxrref{sec:inflio20172020}}).  Also note
that this approach identifies both managed IO accounts
[e.g., \twitter{@Pamela\underscore Moore13}, \twitter{@TEN\underscore
GOP} and~\twitter{@TXCowboysRawk}~\cite{USHPSCI2017a, USHPSCI2017b,
kessler2018, gadde2018}] as well as accounts of real individuals
[\twitter{@JackPosobiec}
and~\twitter{@CassandraRules}~\cite{marantz2017,sputnik2020}] involved
in the spread of IO narratives.  As an effective tool for situational
awareness, the framework in this manuscript can alert social media
platform providers and the public of influential IO accounts and
networks, and the content they spread.

\section*{Discussion}

We present a framework to automate detection of disinformation
narratives, networks, and influential actors. The framework integrates
NLP, machine learning, graph analytics, and network causal
inference to quantify the impact of individual actors in spreading the
IO narrative.  Application of this framework to several authentic
influence operation campaigns run during the 2017 French elections
provides alerts to likely IO~accounts that are influential in
spreading IO narratives. Our results are corroborated by independent
press reports, US Congressional reports, and Twitter's election
integrity dataset. The detection of IO narratives and high-impact
accounts is demonstrated on a dataset comprising $29$~million Twitter
posts and $1$~million accounts collected in $30$ days leading up to the
2017 French elections. We also measure and compare the classification
performance of a semisupervised classifier for IO~accounts
involved in spreading specific IO narratives. At a representative
operating point, our classifier performs with $96$\%~precision,
79\%~recall, $96$\%~AUPRC, and~8\%~EER. Our classifier precision is
shown to outperform two online Bot detectors by $20$\% (nominally) at
this operating point, conditioned on a network-community-based truth
model.  A causal network inference approach is used to quantify
the impact of accounts spreading specific narratives. This method
accounts for the influence network topology and the observed volume
from each account, and removes the effects of social confounders
(e.g., community membership, popularity). We demonstrate the
approach's advantage over traditional impact statistics based on
activity count (e.g., tweet and retweet counts) and network topology
(e.g., network centralities) alone in discovering high-impact
IO~accounts that are independently corroborated.

\showdataavailability

\showacknow


\section*{References}


\bibliography{io_detection_pnas}

\begin{thebibliography}{10}

\bibitem{king2017}
G. King, J. Pan, M.~E. Roberts (2017) How the {C}hinese government fabricates
  social media posts for strategic distraction, not engaged argument.
\newblock {\em American Political Science Review} 111(3):484--501.
\newblock \doicite:10.1017/S0003055417000144:.

\bibitem{stella2018}
M. Stella, E. Ferrara, M.~D. {D}omenico (2018) Bots increase exposure to
  negative and inflammatory content in online social systems.
\newblock {\em Proc. Natl. Acad. Sci. U.S.A.} 115(49):12435--12440.
\newblock \doicite:10.1073/pnas.1803470115:.

\bibitem{vosoughi2018}
S. Vosoughi, D. Roy, S. Aral (2018) The spread of true and false news online.
\newblock {\em Science} 359(6380):1146--1151.
\newblock \doicite:10.1126/science.aap9559:.

\bibitem{starbird2019}
K. Starbird (2019) Disinformation’s spread: bots, trolls and all of us.
\newblock {\em Nature} 571:449.
\newblock \doicite:10.1038/d41586-019-02235-x:.

\bibitem{rid2020}
T. Rid (2020) {\em Active Measures: The Secret History of Disinformation and
  Political Warfare}.
\newblock (Farrar, Straus and Giroux, New~York NY).

\bibitem{schmidt2020}
M.~S. Schmidt, N. Perlroth (2020) {U.~S.} charges {R}ussian intelligence
  officers in major cyberattacks.
\newblock {\em The New~York Times}.
\newblock
  \webcite[https://www.nytimes.com/2020/10/19/us/politics/russian-intelligence-cyberattacks.html].
  \accesseddaymonthyear{19}{October}{2020}.

\bibitem{mao1937}
M. Tse-tung (1937) {\em On Guerilla Warfare}, FMFRP 12-18.
\newblock (U.~S. Marine Corps).
\newblock
  \webcite[https://www.marines.mil/Portals/1/Publications/FMFRP
  \accesseddaymonthyear{1}{March}{2020}.

\bibitem{putin2014}
V. Putin (2014) The military doctrine of the {R}ussian {F}ederation.
\newblock \webcite[https://rusemb.org.uk/press/2029].
  \accesseddaymonthyear{1}{January}{2018}.

\bibitem{gadde2018}
V. Gadde, Y. Roth (2018) Enabling further research of information operations on
  {T}witter.
\newblock
  \webcite[https://blog.twitter.com/en_us/topics/company/2018/enabling-further-research-of-information-operations-on-twitter.html].
  \accesseddaymonthyear{1}{January}{2020}.

\bibitem{smith2018b}
S.~T. Smith, E.~K. Kao, D.~C. Shah, O. Simek, D.~B. Rubin (2018) Influence
  estimation on social media networks using causal inference in {\em Proc. 2018
  IEEE Statistical Signal Processing Workshop (SSP)}.
\newblock pp. 28--32.
\newblock \doicite:10.1109/SSP.2018.8450823:.

\bibitem{birnbaum2019}
E. Birnbaum (2019) Mueller identified 'dozens' of {US} rallies organized by
  {R}ussian troll farm.
\newblock {\em The Hill}.
\newblock
  \webcite[https://thehill.com/policy/technology/439532-mueller-identified-dozens-of-us-rallies-organized-by-russian-troll-farm].
  \accesseddaymonthyear{18}{May}{2019}.

\bibitem{USHPSCI2017a}
{\relax {US} {H}ouse {P}ermanent {S}elect {C}ommittee on {I}ntelligence} (2017)
  {HPSCI} minority exhibits during open hearing, memorandum.
\newblock
  \webcite[https://democrats-intelligence.house.gov/uploadedfiles/hpsci_minority_exhibits_memo_11.1.17.pdf].
  \accesseddaymonthyear{1}{January}{2018}.

\bibitem{USHPSCI2017b}
{\relax {US} {H}ouse {P}ermanent {S}elect {C}ommittee on {I}ntelligence} (2017)
  Exhibit of the user account handles that {T}witter has identified as being
  tied to {R}ussia's ``{I}nternet {R}esearch {A}gency.''.
\newblock
  \webcite[https://democrats-intelligence.house.gov/uploadedfiles/exhibit_b.pdf].
  \accesseddaymonthyear{1}{January}{2018}.

\bibitem{marantz2017}
A. Marantz (2017) The far-right {A}merican nationalist who tweeted
  \#{M}acron{L}eaks.
\newblock {\em The New~Yorker}.
\newblock
  \webcite[https://www.newyorker.com/news/news-desk/the-far-right-american-nationalist-who-tweeted-macronleaks].
  \accesseddaymonthyear{1}{January}{2018}.

\bibitem{kessler2018}
A. Kessler (2018) Who is {@TEN\_GOP} from the {R}ussia indictment? {H}ere's
  what we found reading 2,000 of its tweets.
\newblock {\em CNN}.
\newblock
  \webcite[https://www.cnn.com/2018/02/16/politics/who-is-ten-gop/index.html].
  \accesseddaymonthyear{1}{March}{2020}.

\bibitem{reimers2019}
N. Reimers, I. Gurevych (2019) Sentence-{BERT}: {S}entence embeddings using
  siamese {BERT}-networks in {\em Proc. 2019 Conf. Empirical Methods in Natural
  Language Processing and the 9th Intl. Joint Conf. Natural Language Processing
  (EMNLP-IJCNLP)}.
\newblock pp. 3973--3983.
\newblock \doicite:10.18653/v1/D19-1410:.

\bibitem{RT2017}
{{RT} en{~}{F}ran{\c c}ais} (2017) \guillemotleft {S}ans moi le 7
  mai\guillemotright, l'abstentionnisme gagne {T}witter.
\newblock {\em RT en~Fran\c cais}.
\newblock
  \webcite[https://francais.rt.com/france/37496-sans-moi-7-mai-abstentionnisme-gagne-twitter].
  \accesseddaymonthyear{24}{April}{2017}.

\bibitem{borger2017}
J. Borger (2017) {US} official says {F}rance warned about {R}ussian hacking
  before {M}acron leak.
\newblock {\em The Guardian}.
\newblock
  \webcite[https://www.theguardian.com/technology/2017/may/09/us-russians-hacking-france-election-macron-leak].
  \accesseddaymonthyear{1}{January}{2018}.

\bibitem{mccallum2002}
A.~K. McCallum (2002) Mallet: {A} machine learning for language toolkit.
\newblock \webcite[http://mallet.cs.umass.edu].
  \accesseddaymonthyear{1}{January}{2018}.

\bibitem{pedregosa2011}
F. Pedregosa, et~al. (2011) Scikit-learn: {M}achine learning in {P}ython.
\newblock {\em J.~Machine Learning Research} 12:2825--2830.
\newblock \arxivcite:1201.0490:.

\bibitem{ratner2017}
A. Ratner, et~al. (2017) Snorkel: {R}apid training data creation with weak
  supervision in {\em Proc. VLDB Endowment}.
\newblock Vol.{}~11, pp. 269--282.
\newblock \doicite:10.14778/3157794.3157797:.

\bibitem{peixoto2014b}
T.~P. Peixoto (2014) Efficient {M}onte {C}arlo and greedy heuristic for the
  inference of stochastic block models.
\newblock {\em Phys. Rev. E} 89:012804.
\newblock \doicite:10.1103/PhysRevE.89.012804:.

\bibitem{kao2019}
E.~K. Kao, S.~T. Smith, E.~M. Airoldi (2019) Hybrid mixed-membership blockmodel
  for inference on realistic network interactions.
\newblock {\em IEEE Trans. Network Science and Engineering} 6(3):336--350.
\newblock \doicite:10.1109/TNSE.2018.2823324:.

\bibitem{kao2017}
E.~K. Kao (2017) Causal inference under network interference: {A} framework for
  experiments on social networks.
\newblock {\em Harvard~University}.
\newblock \arxivcite:1708.08522:.

\bibitem{imbensrubin2015}
G.~W. Imbens, D.~B. Rubin (2015) {\em Causal Inference for Statistics, Social,
  and Biomedical Sciences}.
\newblock (Cambridge University Press).
\newblock \doicite:10.1017/CBO9781139025751:.

\bibitem{dearden2017}
L. Dearden (2017) Emmanuel {M}acron launches legal complaint over offshore
  account allegations spread by {M}arine {L}e {P}en.
\newblock {\em The Independent}.
\newblock
  \webcite[https://www.independent.co.uk/news/world/europe/french-presidential-election-latest-emmanuel-macron-legal-complaint-marine-le-pen-offshore-account-a7717461.html].
  \accesseddaymonthyear{1}{April}{2020}.

\bibitem{ferrara2017}
E. Ferrara (2017) Disinformation and social bot operations in the run up to the
  2017 {F}rench presidential election.
\newblock {\em First Monday} 22(8--7).
\newblock
  \webcite[https://firstmonday.org/ojs/index.php/fm/article/download/8005/6516].
  \accesseddaymonthyear{1}{July}{2020}.

\bibitem{varol2017}
O. Varol, E. Ferrara, C.~A. Davis, F. Menczer, A. Flammini (2017) Online
  human-bot interactions: {D}etection, estimation, and characterization in {\em
  Proc. 11th Intl. AAAI Conf. Web and Social Media (ICWSM 2017)}.
\newblock pp. 280--289.
\newblock
  \webcite[https://www.aaai.org/ocs/index.php/ICWSM/ICWSM17/paper/viewPaper/15587].
  \accesseddaymonthyear{1}{January}{2018}.

\bibitem{mcewan2020}
B. McEwan (2020) How social media misinformation wins---even if you don't
  believe it.
\newblock {\em The Week}.
\newblock
  \webcite[https://theweek.com/articles/890910/how-social-media-misinformation-wins--even-dont-believe].
  \accesseddaymonthyear{1}{March}{2020}.

\bibitem{davis2016}
C.~A. Davis, O. Varol, E. Ferrara, A. Flammini, F. Menczer (2016)
  {B}ot{O}r{N}ot: {A} system to evaluate social bots in {\em Proc. 25th Intl.
  Conf. Companion on World Wide Web}.
\newblock pp. 273--274.
\newblock \doicite:10.1145/2872518.2889302:.

\bibitem{botsentinel2020}
{{B}ot {S}entinel} (2020) Platform developed to detect and track political
  bots, trollbots, and untrustworthy accounts.
\newblock \webcite[https://botsentinel.com].
  \accesseddaymonthyear{1}{March}{2020}.

\bibitem{rauchfleisch2020}
A. Rauchfleisch, J. Kaiser (2020) The false positive problem of automatic bot
  detection in social science research.
\newblock {\em PLOS ONE} 15(10):1--20.
\newblock \doicite:10.1371/journal.pone.0241045:.

\bibitem{smith2014}
S.~T. Smith, E.~K. Kao, K.~D. Senne, G. Bernstein, S. Philips (2014) Bayesian
  discovery of threat networks.
\newblock {\em IEEE Trans. Signal Proc.} 62(20):5324--5338.
\newblock \doicite:10.1109/TSP.2014.2336613:.

\bibitem{sputnik2020}
C. Fairbanks (2020) {C}assandra {F}airbanks. {S}putnik {N}ews.
\newblock \webcite[https://sputniknews.com/authors/cassandra_fairbanks].
  \accesseddaymonthyear{1}{March}{2020}.

\bibitem{donati2020}
J. Donati (2020) {U}.{S}. adversaries are accelerating, coordinating
  coronavirus disinformation, report says.
\newblock {\em The Wall~Street Journal}.
\newblock
  \webcite[https://www.wsj.com/articles/u-s-adversaries-are-accelerating-coordinating-coronavirus-disinformation-report-says-11587514724].
  \accesseddaymonthyear{21}{April}{2020}.

\end{thebibliography}


\begin{thebibliography}{10}

\bibitem{confessore2018}
N. Confessore, G.~J. Dance, R. Harris, M. Hansen (2018) The follower factory.
\newblock {\em The New~York Times}.
\newblock
  \webcite[https://www.nytimes.com/interactive/2018/01/27/technology/social-media-bots.html].
  \accesseddaymonthyear{27}{January}{2018}.

\bibitem{fan2014}
L. Fan, W. Wu, X. Zhai, {et~al.} (2014) Maximizing rumor containment in social
  networks with constrained time.
\newblock {\em Soc. Netw. Anal. Min.} 4:214.
\newblock \doicite:10.1007/s13278-014-0214-4:.

\bibitem{mueller2019}
R.~S. {\relax Mueller, {III}} (2019) {\em Report On The Investigation Into
  {R}ussian Interference In The 2016 Presidential Election}.
\newblock (U.~S. Department of Justice) Vol.{}~1.
\newblock
  \webcite[https://www.justsecurity.org/wp-content/uploads/2019/04/Muelller-Report-Redacted-Vol-I-Released-04.18.2019-Word-Searchable.-Reduced-Size.pdf].
  \accesseddaymonthyear{1}{March}{2020}.

\bibitem{shah2011}
D. Shah, T. Zaman (2011) Rumors in a network: {W}ho's the culprit?
\newblock {\em IEEE Trans. Information Theory} 57(8):5163--5181.
\newblock \doicite:10.1109/TIT.2011.2158885:.

\bibitem{stewart2018}
L.~G. Stewart, A. Arif, K. Starbird (2018) Examining trolls and polarization
  with a retweet network in {\em Proc. ACM WSDM; MIS2: Misinformation and
  Misbehavior Mining on the Web}.
\newblock \webcite[http://snap.stanford.edu/mis2/files/MIS2_paper_21.pdf].
  \accesseddaymonthyear{1}{March}{2020} \doicite:10.475/123_4:.

\bibitem{tambuscio2015}
M. Tambuscio, G. Ruffo, A. Flammini, F. Menczer (2015) Fact-checking effect on
  viral hoaxes: {A} model of misinformation spread in social networks in {\em
  Proc. ACM WWW}.
\newblock \doicite:10.1145/2740908.2742572:.

\bibitem{zhang2015}
H. Zhang, H. Zhang, X. Li, M.~T. Thai (2015) Limiting the spread of
  misinformation while effectively raising awareness in social networks in {\em
  Computational Social Networks, CSoNet 2015}, Lecture Notes in Computer
  Science, eds.{} M. Thai, N. Nguyen, H. Shen.
\newblock (Springer) Vol.{} 9197.
\newblock \doicite:10.1007/978-3-319-21786-4_4:.

\bibitem{chekinov2013}
{\relax Col.~S.~G.}. Chekinov, {\relax Lt.~Gen. S.~A.}. Bogdanov (2013) On the
  nature and content of new generation warfare.
\newblock {\em Military Thought} 10:13--24.
\newblock
  \webcite[http://www.eastviewpress.com/Files/MT_FROM
  \accesseddaymonthyear{1}{March}{2017}.

\bibitem{pugaciauskas2015}
V. Puga{\v c}iauskas (2015) In the post-{S}oviet propaganda sphere.
\newblock {\em J. Baltic Security} 1(1):127--133.
\newblock \webcite[http://www.baltdefcol.org/files/files/JOBS/JOBS.01.1.pdf].
  \accesseddaymonthyear{1}{March}{2017}.

\bibitem{nato2015}
{NATO StratCom} (2015) Analysis of {R}ussia's information campaign against
  {U}kraine, (NATO Strategic Communications Centre of Excellence), Technical
  report.
\newblock \webcite[https://www.stratcomcoe.org/download/file/fid/1886].
  \accesseddaymonthyear{1}{March}{2017}.

\bibitem{giles2016}
K. Giles (2016) Handbook of {R}ussian information warfare, (NATO Defense
  College), Technical Report~9.
\newblock \webcite[http://www.ndc.nato.int/download/downloads.php?icode=506].
  \accesseddaymonthyear{1}{March}{2017}.

\bibitem{kao2020}
J. Kao, M.~S. Li (2020) How {C}hina built a {T}witter propaganda machine then
  let it loose on coronavirus.
\newblock {\em ProPublica}.
\newblock
  \webcite[https://www.propublica.org/article/how-china-built-a-twitter-propaganda-machine-then-let-it-loose-on-coronavirus].
  \accesseddaymonthyear{1}{April}{2020}.

\bibitem{veebel2016}
V. Veebel (2016) Estonia confronts propaganda: {R}ussia manipulates media in
  pursuit of psychological warfare.
\newblock {\em Concordiam per J. European Security and Defense Issues}
  7:14--19.
\newblock
  \webcite[https://www.marshallcenter.org/mcpublicweb/mcdocs/files/College/F_Publications/perConcordiam/pC_V7_SpecialEdition_en.pdf].
  \accesseddaymonthyear{1}{March}{2017}.

\bibitem{borger2017}
J. Borger (2017) {US} official says {F}rance warned about {R}ussian hacking
  before {M}acron leak.
\newblock {\em The Guardian}.
\newblock
  \webcite[https://www.theguardian.com/technology/2017/may/09/us-russians-hacking-france-election-macron-leak].
  \accesseddaymonthyear{1}{January}{2018}.

\bibitem{nyst2018}
C. Nyst, N. Monaco (2018) State-sponsored trolling: how governments are
  deploying disinformation as part of broader digital harassment campaigns,
  (Institute for the Future), Technical report.
\newblock
  \webcite[http://www.iftf.org/fileadmin/user_upload/images/DigIntel/IFTF_State_sponsored_trolling_report.pdf].
  \accesseddaymonthyear{1}{March}{2019}.

\bibitem{birnbaum2019}
E. Birnbaum (2019) Mueller identified 'dozens' of {US} rallies organized by
  {R}ussian troll farm.
\newblock {\em The Hill}.
\newblock
  \webcite[https://thehill.com/policy/technology/439532-mueller-identified-dozens-of-us-rallies-organized-by-russian-troll-farm].
  \accesseddaymonthyear{18}{May}{2019}.

\bibitem{kargar2019}
S. Kargar, A. Rauchfleisch (2019) State-aligned trolling in {I}ran and the
  double-edged affordances of {I}nstagram.
\newblock {\em new media \& society} 21(7):1506--1527.
\newblock \doicite:10.1177/1461444818825133:.

\bibitem{rid2020}
T. Rid (2020) {\em Active Measures: The Secret History of Disinformation and
  Political Warfare}.
\newblock (Farrar, Straus and Giroux, New~York NY).

\bibitem{dearden2017}
L. Dearden (2017) Emmanuel {M}acron launches legal complaint over offshore
  account allegations spread by {M}arine {L}e {P}en.
\newblock {\em The Independent}.
\newblock
  \webcite[https://www.independent.co.uk/news/world/europe/french-presidential-election-latest-emmanuel-macron-legal-complaint-marine-le-pen-offshore-account-a7717461.html].
  \accesseddaymonthyear{1}{April}{2020}.

\bibitem{mcauley2017}
J. McAuley, I. Stanley-Becker (2017) Macron campaign says its emails have been
  subjected to `massive, coordinated' hacking.
\newblock {\em The Washington Post}.
\newblock
  \webcite[https://www.washingtonpost.com/world/macron-campaign-says-its-emails-have-been-subjected-to-massive-coordinated-hacking/2017/05/06/368c0460-31e1-11e7-a335-fa0ae1940305_story.html].
  \accesseddaymonthyear{1}{March}{2020}.

\bibitem{schmidt2020}
M.~S. Schmidt, N. Perlroth (2020) {U.~S.} charges {R}ussian intelligence
  officers in major cyberattacks.
\newblock {\em The New~York Times}.
\newblock
  \webcite[https://www.nytimes.com/2020/10/19/us/politics/russian-intelligence-cyberattacks.html].
  \accesseddaymonthyear{19}{October}{2020}.

\bibitem{akbarpour2018}
M. Akbarpour, M.~O. Jackson (2018) Diffusion in networks and the virtue of
  burstiness.
\newblock {\em Proc. Natl. Acad. Sci. U.S.A.} 115(30):E6996--E7004.
\newblock \doicite:10.1073/pnas.1722089115:.

\bibitem{pennycook2019}
G. Pennycook, D.~G. Rand (2019) Fighting misinformation on social media using
  crowdsourced judgments of news source quality.
\newblock {\em Proc. Natl. Acad. Sci. U.S.A.} 116(7):2521--2526.
\newblock \doicite:10.1073/pnas.1806781116:.

\bibitem{contractor2014}
N.~S. Contractor, L.~A. De{C}hurch (2014) Integrating social networks and human
  social motives to achieve social influence at scale.
\newblock {\em Proc. Natl. Acad. Sci. U.S.A.} 111 (Supplement
  4)(49):13650--13657.
\newblock \doicite:10.1073/pnas.1401211111:.

\bibitem{mccallum2002}
A.~K. McCallum (2002) Mallet: {A} machine learning for language toolkit.
\newblock \webcite[http://mallet.cs.umass.edu].
  \accesseddaymonthyear{1}{January}{2018}.

\bibitem{ratner2017}
A. Ratner, et~al. (2017) Snorkel: {R}apid training data creation with weak
  supervision in {\em Proc. VLDB Endowment}.
\newblock Vol.{}~11, pp. 269--282.
\newblock \doicite:10.14778/3157794.3157797:.

\bibitem{costa-roberts2018}
D. Costa-Roberts (2018) How to spot a {R}ussian bot.
\newblock {\em Mother~Jones}.
\newblock
  \webcite[https://www.motherjones.com/media/2018/08/how-to-identify-russian-bots-twitter].
  \accesseddaymonthyear{1}{March}{2020}.

\bibitem{im2019}
J. Im, et~al. (2019) Still out there: {M}odeling and identifying {R}ussian
  troll accounts on {T}witter.
\newblock \arxivcite:1901.11162:.

\bibitem{zannettou2019}
S. Zannettou, et~al. (2019) Disinformation warfare: {U}nderstanding
  state-sponsored trolls on {T}witter and their influence on the web in {\em
  Proc. 2019 World Wide Web Conf.}
\newblock pp. 218--226.
\newblock \doicite:10.1145/3308560.3316495:.

\bibitem{luceri2020}
L. Luceri, S. Giordano, E. Ferrara (2020) Don't feed the troll: {D}etecting
  troll behavior via inverse reinforcement learning in {\em Proc. 2020 Intl.
  Conf. Web and Social Media (ICWSM)}.
\newblock \arxivcite:2001.10570:.

\bibitem{varol2017}
O. Varol, E. Ferrara, C.~A. Davis, F. Menczer, A. Flammini (2017) Online
  human-bot interactions: {D}etection, estimation, and characterization in {\em
  Proc. 11th Intl. AAAI Conf. Web and Social Media (ICWSM 2017)}.
\newblock pp. 280--289.
\newblock
  \webcite[https://www.aaai.org/ocs/index.php/ICWSM/ICWSM17/paper/viewPaper/15587].
  \accesseddaymonthyear{1}{January}{2018}.

\bibitem{liaw2002}
A. Liaw, M. Wiener (2002) Classification and regression by {R}andom{F}orest.
\newblock {\em R news} 2(3):18--22.

\bibitem{chen2016}
T. Chen, C. Guestrin (2016) Xgboost: A scalable tree boosting system in {\em
  Proc. ACM SIGKDD Intl. Conf. Knowledge Discovery and Data Mining (KDD)}.
\newblock pp. 785--794.
\newblock \doicite:10.1145/2939672.2939785:.

\bibitem{pedregosa2011}
F. Pedregosa, et~al. (2011) Scikit-learn: {M}achine learning in {P}ython.
\newblock {\em J.~Machine Learning Research} 12:2825--2830.
\newblock \arxivcite:1201.0490:.

\bibitem{geurts2006}
P. Geurts, D. Ernst, L. Wehenkel (2006) Extremely randomized trees.
\newblock {\em Machine Learning} 63(1):3--42.
\newblock \doicite:10.1007/s10994-006-6226-1:.

\bibitem{marantz2017}
A. Marantz (2017) The far-right {A}merican nationalist who tweeted
  \#{M}acron{L}eaks.
\newblock {\em The New~Yorker}.
\newblock
  \webcite[https://www.newyorker.com/news/news-desk/the-far-right-american-nationalist-who-tweeted-macronleaks].
  \accesseddaymonthyear{1}{January}{2018}.

\bibitem{USHPSCI2017a}
{\relax {US} {H}ouse {P}ermanent {S}elect {C}ommittee on {I}ntelligence} (2017)
  {HPSCI} minority exhibits during open hearing, memorandum.
\newblock
  \webcite[https://democrats-intelligence.house.gov/uploadedfiles/hpsci_minority_exhibits_memo_11.1.17.pdf].
  \accesseddaymonthyear{1}{January}{2018}.

\bibitem{USHPSCI2017b}
{\relax {US} {H}ouse {P}ermanent {S}elect {C}ommittee on {I}ntelligence} (2017)
  Exhibit of the user account handles that {T}witter has identified as being
  tied to {R}ussia's ``{I}nternet {R}esearch {A}gency.''.
\newblock
  \webcite[https://democrats-intelligence.house.gov/uploadedfiles/exhibit_b.pdf].
  \accesseddaymonthyear{1}{January}{2018}.

\bibitem{gadde2018}
V. Gadde, Y. Roth (2018) Enabling further research of information operations on
  {T}witter.
\newblock
  \webcite[https://blog.twitter.com/en_us/topics/company/2018/enabling-further-research-of-information-operations-on-twitter.html].
  \accesseddaymonthyear{1}{January}{2020}.

\bibitem{smith2018b}
S.~T. Smith, E.~K. Kao, D.~C. Shah, O. Simek, D.~B. Rubin (2018) Influence
  estimation on social media networks using causal inference in {\em Proc. 2018
  IEEE Statistical Signal Processing Workshop (SSP)}.
\newblock pp. 28--32.
\newblock \doicite:10.1109/SSP.2018.8450823:.

\bibitem{kao2017}
E.~K. Kao (2017) Causal inference under network interference: {A} framework for
  experiments on social networks.
\newblock {\em Harvard~University}.
\newblock \arxivcite:1708.08522:.

\bibitem{basse2016}
G. Basse, A. Feller (2016) Analyzing multilevel experiments in the presence of
  peer effects.
\newblock \arxivcite:1608.06805:.

\bibitem{kim2015}
D.~A. Kim, et~al. (2015) Social network targeting to maximise population
  behaviour change: a cluster randomised controlled trial.
\newblock {\em The Lancet} 386(9989):145--153.
\newblock \doicite:10.1016/S0140-6736(15)60095-2:.

\bibitem{christakis2010}
S.~C. Mednick, N.~A. Christakis, J.~H. Fowler (2010) The spread of sleep
  behavior influences drug use in adolescent social networks.
\newblock {\em PLoS One} 5(3):e9775.
\newblock \doicite:10.1371/journal.pone.0009775:.

\bibitem{sobel2006}
M.~E. Sobel (2006) What do randomized studies of housing mobility demonstrate?
\newblock {\em J. American Statistical Association} 101(476):1398--1407.
\newblock \doicite:10.1198/016214506000000636:.

\bibitem{gui2015}
H. Gui, Y. Xu, A. Bhasin, J. Han (2015) Network {A/B} testing: {F}rom sampling
  to estimation in {\em Proc. 24th Intl. Conf. World Wide Web}.
\newblock (International World Wide Web Conferences Steering Committee), pp.
  399--409.
\newblock \doicite:10.1145/2736277.2741081:.

\bibitem{bond2012}
R.~M. Bond, et~al. (2012) A 61-million-person experiment in social influence
  and political mobilization.
\newblock {\em Nature} 489(7415):295--298.
\newblock \doicite:10.1038/nature11421:.

\bibitem{bakshy2012}
E. Bakshy, D. Eckles, R. Yan, I. Rosenn (2012) Social influence in social
  advertising: Evidence from field experiments in {\em Proc. 13th ACM Conf.
  Electronic Commerce}.
\newblock pp. 146--161.
\newblock \doicite:10.1145/2229012.2229027:.

\bibitem{parker2011}
B.~M. Parker (2011) Design of network experiments.
\newblock
  \webcite[http://www.newton.ac.uk/programmes/DAE/seminars/090111301.pdf].
  \accesseddaymonthyear{1}{January}{2013}.

\bibitem{coronges2012}
K. Coronges, et~al. (2012) The influences of social networks on phishing
  vulnerability in {\em Proc. 45th Intl. Conf. System Science (HICSS)}.
\newblock (IEEE), pp. 2366--2373.
\newblock \doicite:10.1109/HICSS.2012.657:.

\bibitem{banerjee2013}
A. Banerjee, A.~G. Chandrasekhar, E. Duflo, M.~O. Jackson (2013) The diffusion
  of microfinance.
\newblock {\em Science} 341(6144):1236498.
\newblock \doicite:10.1126/science.1236498:.

\bibitem{acemoglu2010}
D. Acemoglu, A. Ozdaglar, A. ParandehGheibi (2010) Spread of (mis)information
  in social networks.
\newblock {\em Games and Economic Behavior} 70(2):194--227.
\newblock \doicite:10.1016/j.geb.2010.01.005:.

\bibitem{manski1993}
C.~F. Manski (1993) Identification of endogenous social effects: {T}he
  reflection problem.
\newblock {\em Rev. Economic Studies} 60(3):531--542.
\newblock \doicite:10.2307/2298123:.

\bibitem{david1996}
O. David, R.~A. Kempton (1996) Designs for interference.
\newblock {\em Biometrics} 52(2):597--606.
\newblock \doicite:10.2307/2532898:.

\bibitem{azais1993}
J.~M. Azais, R.~A. Bailey, H. Monod (1993) A catalogue of efficient
  neighbour-designs with border plots.
\newblock {\em Biometrics} 49(4):1252--1261.
\newblock \doicite:10.2307/2532269:.

\bibitem{rubin1974}
D.~B. Rubin (1974) Estimating causal effects of treatments in randomized and
  nonrandomized studies.
\newblock {\em J. Educational Psychology} 66(5):688--701.
\newblock \doicite:10.1037/h0037350:.

\bibitem{imbensrubin2015}
G.~W. Imbens, D.~B. Rubin (2015) {\em Causal Inference for Statistics, Social,
  and Biomedical Sciences}.
\newblock (Cambridge University Press).
\newblock \doicite:10.1017/CBO9781139025751:.

\bibitem{bowers2013}
J. Bowers, M.~M. Fredrickson, C. Panagopoulos (2013) Reasoning about
  interference between units: A general framework.
\newblock {\em Political Analysis} 21(1):97--124.
\newblock \doicite:10.1093/pan/mps038:.

\bibitem{athey2015}
S. Athey, D. Eckles, G.~W. Imbens (2018) Exact {$p$}-values for network
  interference.
\newblock {\em J. Am. Stat. Assoc.} 113(521):230--240.
\newblock \doicite:10.1080/01621459.2016.1241178:.

\bibitem{aronow2017}
P.~M. Aronow, C. Samii (2017) Estimating average causal effects under general
  interference, with application to a social network experiment.
\newblock {\em Ann. Appl. Stat.} 11(4):1912--1947.
\newblock \doicite:10.1214/16-AOAS1005:.

\bibitem{ugander2013}
J. Ugander, B. Karrer, L. Backstrom, J. Kleinberg (2013) Graph cluster
  randomization: {N}etwork exposure to multiple universes in {\em ACM SIGKDD
  Intl. Conf. Knowledge Discovery and Data Mining (KDD)}.
\newblock pp. 329--337.
\newblock \doicite:10.1145/2487575.2487695:.

\bibitem{sussman2017}
D.~L. Sussman, E.~M. Airoldi (2017) Elements of estimation theory for causal
  effects in the presence of network interference.
\newblock \arxivcite:1702.03578:.

\bibitem{li2020}
S. Li, S. Wager (2020) Random graph asymptotics for treatment effect estimation
  under network interference.
\newblock \arxivcite:2007.13302:.

\bibitem{vanderweele2014}
T.~J. VanderWeele, E.~J. Tchetgen, M.~E. Halloran (2014) Interference and
  sensitivity analysis.
\newblock {\em Statist. Sci.} 29(4):687--706.
\newblock \doicite:10.1214/14-STS479:.

\bibitem{ogburn2014}
E.~L. Ogburn, T.~J. {V}ander{W}eele (2014) Causal diagrams for interference.
\newblock {\em Statist. Sci.} 29(4):559--578.
\newblock \doicite:10.1214/14-STS501:.

\bibitem{shalizi2011}
C.~R. Shalizi, A.~C. Thomas (2011) Homophily and contagion are generically
  confounded in observational social network studies.
\newblock {\em Sociological Methods \& Research} 40(2):211--239.
\newblock \doicite:10.1177/0049124111404820:.

\bibitem{aral2009}
S. Aral, L. Muchnik, A. Sundararajan (2009) Distinguishing influence-based
  contagion from homophily-driven diffusion in dynamic networks.
\newblock {\em Proc. Natl. Acad. Sci. U.S.A.} 106(51):21544--21549.
\newblock \doicite:10.1073/pnas.0908800106:.

\bibitem{omalley2014}
A.~J. O'Malley, F. Elwert, J.~N. Rosenquist, A.~M. Zaslavsky, N.~A. Christakis
  (2014) Estimating peer effects in longitudinal dyadic data using instrumental
  variables.
\newblock {\em Biometrics} 70(3):506--515.
\newblock \doicite:10.1111/biom.12172:.

\bibitem{eckles2016}
D. Eckles, R.~F. Kizilcec, E. Bakshy (2016) Estimating peer effects in networks
  with peer encouragement designs.
\newblock {\em Proc. Natl. Acad. Sci. U.S.A.} 113(27):7316--7322.
\newblock \doicite:10.1073/pnas.1511201113:.

\bibitem{rubin1980}
D.~B. Rubin (1980) Randomization analysis of experimental data: {T}he {F}isher
  randomization test comment.
\newblock {\em J. Am. Stat. Assoc.} pp. 587--589.
\newblock \doicite:10.2307/2287653:.

\bibitem{toulis2013}
P. Toulis, E. Kao (2013) Estimation of causal peer influence effects in {\em
  Proc. 30th Intl. Conf. Machine Learning}.
\newblock pp. 1489--1497.

\bibitem{rubin1991}
D.~B. Rubin (1991) Practical implications of modes of statistical inference for
  causal effects and the critical role of the assignment mechanism.
\newblock {\em Biometrics} 47(4):1213--1234.
\newblock \doicite:10.2307/2532381:.

\bibitem{rubin1978}
D.~B. Rubin (1978) Bayesian inference for causal effects: {T}he role of
  randomization.
\newblock {\em Ann. Stat.} pp. 34--58.
\newblock \webcite[https://www.jstor.org/stable/2958688].

\bibitem{hoff2002}
P.~D. Hoff, A.~E. Raftery, M.~S. Handcock (2002) Latent space approaches to
  social network analysis.
\newblock {\em J. Am. Stat. Assoc.} 97(460):1090--1098.
\newblock \doicite:10.1198/016214502388618906:.

\bibitem{wang1987}
Y.~J. Wang, G.~Y. Wong (1987) Stochastic blockmodels for directed graphs.
\newblock {\em J. American Statistical Association} 82(397):8--19.
\newblock \doicite:10.1080/01621459.1987.10478385:.

\bibitem{airoldi2008}
E.~M. Airoldi, D.~M. Blei, S.~E. Fienberg, E.~P. Xing (2008) Mixed membership
  stochastic blockmodels.
\newblock {\em J. Machine Learning Research} 9(1981--2014):3.

\bibitem{kao2019}
E.~K. Kao, S.~T. Smith, E.~M. Airoldi (2019) Hybrid mixed-membership blockmodel
  for inference on realistic network interactions.
\newblock {\em IEEE Trans. Network Science and Engineering} 6(3):336--350.
\newblock \doicite:10.1109/TNSE.2018.2823324:.

\bibitem{aiello2001}
W. Aiello, F. Chung, L. Lu (2001) A random graph model for power law graphs.
\newblock {\em Experimental Mathematics} 10(1):53--66.
\newblock \doicite:10.1080/10586458.2001.10504428:.

\bibitem{chung2002}
F. Chung, L. Lu (2002) The average distances in random graphs with given
  expected degrees.
\newblock {\em Proc. Natl. Acad. Sci. U.S.A.} 99(25):15879--15882.
\newblock \doicite:10.1073/pnas.252631999:.

\bibitem{lovasz2006}
L. Lov{\'a}sz, B. Szegedy (2006) Limits of dense graph sequences.
\newblock {\em J. Combinatorial Theory, Series B} 96(6):933--957.
\newblock \doicite:10.1016/j.jctb.2006.05.002:.

\bibitem{frangakis1983}
C.~E. Frangakis, D.~B. Rubin (1999) Addressing complications of
  intention-to-treat analysis in the combined presence of all-or-none
  treatment-noncompliance and subsequent missing outcomes.
\newblock {\em Biometrika} 86(2):365--379.
\newblock \doicite:10.1093/biomet/86.2.365:.

\bibitem{gelman2008}
A. Gelman, A. Jakulin, M.~G. Pittau, Y. Su (2008) A weakly informative default
  prior distribution for logistic and other regression models.
\newblock {\em The Annals of Applied Statistics} 2(4):1360--1383.

\bibitem{gelman1996}
A. Gelman, X. Meng, H. Stern (1996) Posterior predictive assessment of model
  fitness via realized discrepancies.
\newblock {\em Statistica Sinica} pp. 733--760.

\bibitem{rubin2004}
D.~B. Rubin (2004) {\em Multiple imputation for nonresponse in surveys}, Wiley
  Classics.
\newblock (John Wiley \& Sons, Hoboken NJ) Vol.{}~81.

\bibitem{forastiere2020}
L. Forastiere, E.~M. Airoldi, F. Mealli (2020) Identification and estimation of
  treatment and interference effects in observational studies on networks.
\newblock {\em Journal of the American Statistical Association} pp. 1--18.

\bibitem{han2020}
S. Han, D.~B. Rubin (2020) Contrast specific propensity scores.
\newblock \arxivcite:2007.01253:.

\bibitem{ferrara2017}
E. Ferrara (2017) Disinformation and social bot operations in the run up to the
  2017 {F}rench presidential election.
\newblock {\em First Monday} 22(8--7).
\newblock
  \webcite[https://firstmonday.org/ojs/index.php/fm/article/download/8005/6516].
  \accesseddaymonthyear{1}{July}{2020}.

\bibitem{langville2005}
A.~N. Langville, C.~D. Meyer (2005) A survey of eigenvector methods for web
  information retrieval.
\newblock {\em SIAM review} 47(1):135--161.
\newblock \doicite:10.1137/S0036144503424786:.

\bibitem{peixoto2014a}
T.~P. Peixoto (2014) The graph-tool {P}ython library.
\newblock \webcite[http://graph-tool.skewed.de].
  \accesseddaymonthyear{1}{March}{2017}.

\bibitem{smith2014}
S.~T. Smith, E.~K. Kao, K.~D. Senne, G. Bernstein, S. Philips (2014) Bayesian
  discovery of threat networks.
\newblock {\em IEEE Trans. Signal Proc.} 62(20):5324--5338.
\newblock \doicite:10.1109/TSP.2014.2336613:.

\bibitem{donati2020}
J. Donati (2020) {U}.{S}. adversaries are accelerating, coordinating
  coronavirus disinformation, report says.
\newblock {\em The Wall~Street Journal}.
\newblock
  \webcite[https://www.wsj.com/articles/u-s-adversaries-are-accelerating-coordinating-coronavirus-disinformation-report-says-11587514724].
  \accesseddaymonthyear{21}{April}{2020}.

\bibitem{donovan2020}
J. Donovan (2020) Here’s how social media can combat the coronavirus
  `infodemic'.
\newblock {\em MIT Technology Review}.
\newblock
  \webcite[https://www.technologyreview.com/2020/03/17/905279/facebook-twitter-social-media-infodemic-misinformation/].
  \accesseddaymonthyear{1}{April}{2020}.

\bibitem{linvill2020}
D. Linvill, P. Warren (2020) {T}he {R}ussia {T}weets.
\newblock \webcite[https://russiatweets.com].
  \accesseddaymonthyear{1}{April}{2020}.

\end{thebibliography}

\end{document}



\ifpnastypesetdefaults\else\ifpnastypesetarxiv
\makeatletter
\@input{io_detection_pnas_final_aux.tex}%
\makeatother
\fi\fi

\maketitle

\SItext


\subsection{Additional Background}

Exploitation of social media and digital communications by world
powers to achieve their political objectives~\cite{confessore2018,
fan2014, mueller2019, shah2011, stewart2018, tambuscio2015, zhang2015,
chekinov2013, pugaciauskas2015, nato2015, giles2016, kao2020,
veebel2016, borger2017, nyst2018, birnbaum2019, kargar2019, rid2020}
at unprecedented scales, speeds, and reach presents a rising threat,
especially to democratic societies~\cite{confessore2018, mueller2019,
stewart2018, tambuscio2015, dearden2017, mcauley2017, schmidt2020}.
Situational awareness of influence campaigns and a better
understanding of the mechanism behind social
influence~\cite{akbarpour2018, shah2011} provide necessary
capabilities for potential responses~\cite{pennycook2019, zhang2015,
fan2014}.  In other applications, social influence can even be
harnessed to promote knowledge and best practices in public policy
settings for social good~\cite{contractor2014}.

\subsection{Narrative Detection Using Topic Modeling}

Tweets were collected that are potentially relevant to a
representative scenario in which actual IO~accounts were expected to
be active: the 2017 French election. A large dataset containing both
English and French language corpora was created from this collection.
From these corpora, $15$ English language topics and $30$ French
topics are generated automatically using a topic modeling
algorithm~\cite{mccallum2002}.  The set of generated English topics
includes several relevant to the French election as well as topics on
U.S. politics and other world events. A selection of the English
topics is shown in Table~\ref{tab:en_topics}. The first topic relates
to unsubstantiated financial allegations; this is the topic used for
network discovery in the {\em\href{\mainpaperDOI}{\color{blue}main
  paper}}.  The generated French topics are
predominantly focused on the French election, including a French
language version of unsubstantiated financial allegations. A subset of
these topics is shown in Table~\ref{tab:fr_topics}. Note that in both
tables and the sequel, the notation `:emoji\underscore symbol:' specifies an
emoji symbol.

\subsection{IO Account Classifier and Feature Engineering}
\label{app:ioclassifier}

\begin{figure}
\centering
\halign to\hsize{&\hbox to\textwidth{\hss#\hss}\cr
\kern-2pc\includegraphics[height=.4\textheight]{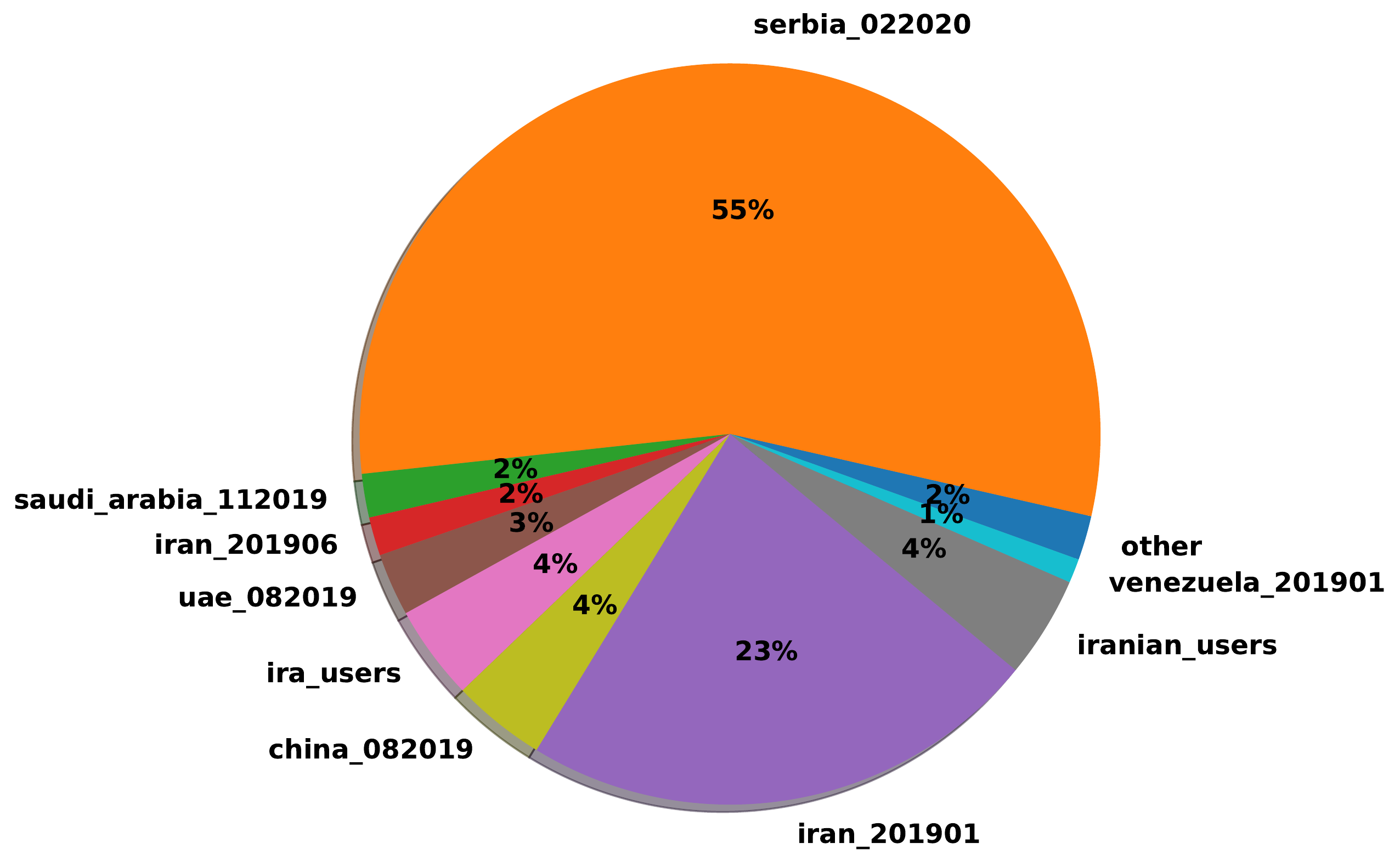}\cr
(A)\cr
\noalign{\vskip\abovecaptionskip}%
\includegraphics[height=.4\textheight]{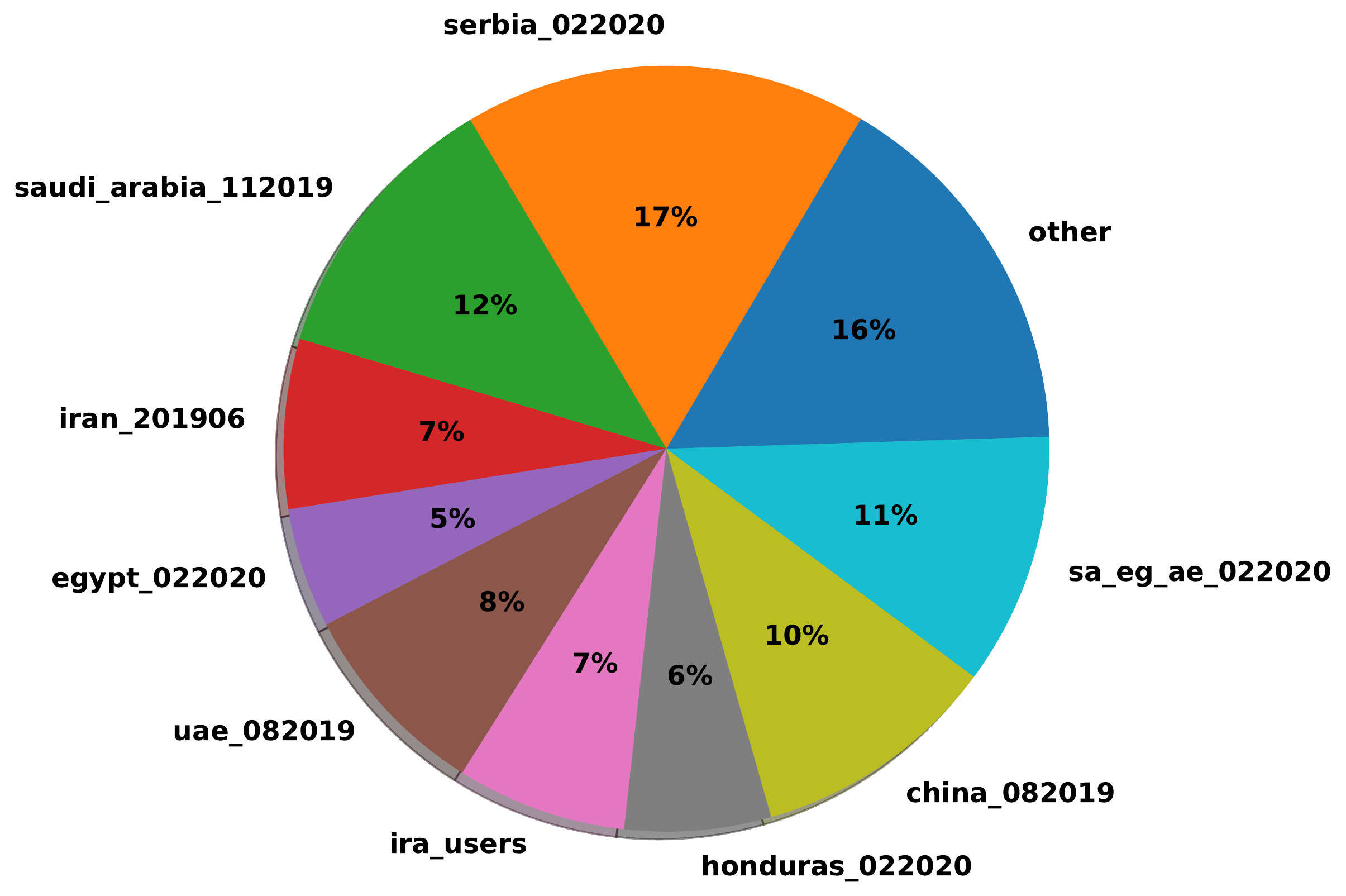}\cr
(B)\cr}
\caption{(A)~Fraction of our $3{,}151$ accounts by~origin from Twitter's
datasets~\cite{gadde2018} used for classifier training and
testing. (B)~Account fraction by~origin of all released Twitter
datasets by March~2020.}\label{fig:origin-pie-chart}
\end{figure}

The data used to train and test the IO classifier includes $3{,}151$
known IO~accounts released by Twitter and $15{,}000$ randomly selected
accounts from three subsets of the targeted collection dataset:
accounts that tweeted on the French election in English, accounts that
tweeted on the French election in French, and topic- and
language-neutral accounts.  Each subset contributes $5{,}000$ accounts
to the training dataset.  Multiple feature categories are used:
account behavior features, language features across all content, and
features derived from the content itself.  The origin of these
accounts from Twitter's dataset is illustrated in
Fig.~\ref{fig:origin-pie-chart}.

\subsubsection{Heuristics for Semisupervised Learning}
\label{app:snorkel}

Because the quantity of known IO~accounts is relatively small, a
semisupervised training strategy based on Snorkel~\cite{ratner2017}
is employed. In this approach, training data from known truth are
augmented with weakly labeled training data provided by heuristic
``labeling functions'' (Table~\ref{tab:snorkel_labels}), and the
learning model learns and incorporates labeling function inaccuracies.
Our Snorkel labeling functions are based on reported characteristics
of IO~account behavior, content, and metadata~\cite{costa-roberts2018,
im2019, zannettou2019, luceri2020}.  These Snorkel heuristics were
then refined by applying them to a small validation set. One hundred
accounts were randomly selected out of the $5{,}000$ that tweeted on
the French election in English. An attempt was made to label each
account as either a real person or an IO~account by examining their
Twitter profile and current tweets. However, many accounts were either
suspended or difficult to determine with confidence to which category
they belong.  Of these $100$ randomly chosen accounts, $24$ were
labeled as IO~accounts, $31$ were labeled non-IO, and the remaining
were discarded from the validation set.

Though $9$--$15\%$ of all Twitter accounts are estimated to be
bots~\cite{varol2017}, it is possible that the charged topics analyzed
in the {\em\href{\mainpaperDOI}{\color{blue}main paper}}
contain a higher proportion of bots and possible IO
accounts. To account for this possibility in classifier training, all
accounts that have a $70\%$ or higher likelihood of being IO-like,
according to Snorkel, are labeled as IO~accounts, and accounts below
the $70\%$ threshold are labeled as non-IO~accounts. This results in
roughly $30\%$ of the training accounts on the French election and
$15\%$ of the topic-neutral accounts being labeled IO~accounts. The
proportions of each training data subset that falls above the
threshold is given for three different points in
Table~\ref{tab:snorkel_fracs}.

\subsubsection{Classifier Design Comparisons}
\label{app:class_comp}

\begin{figure}
\centering
\includegraphics[width=1\linewidth]{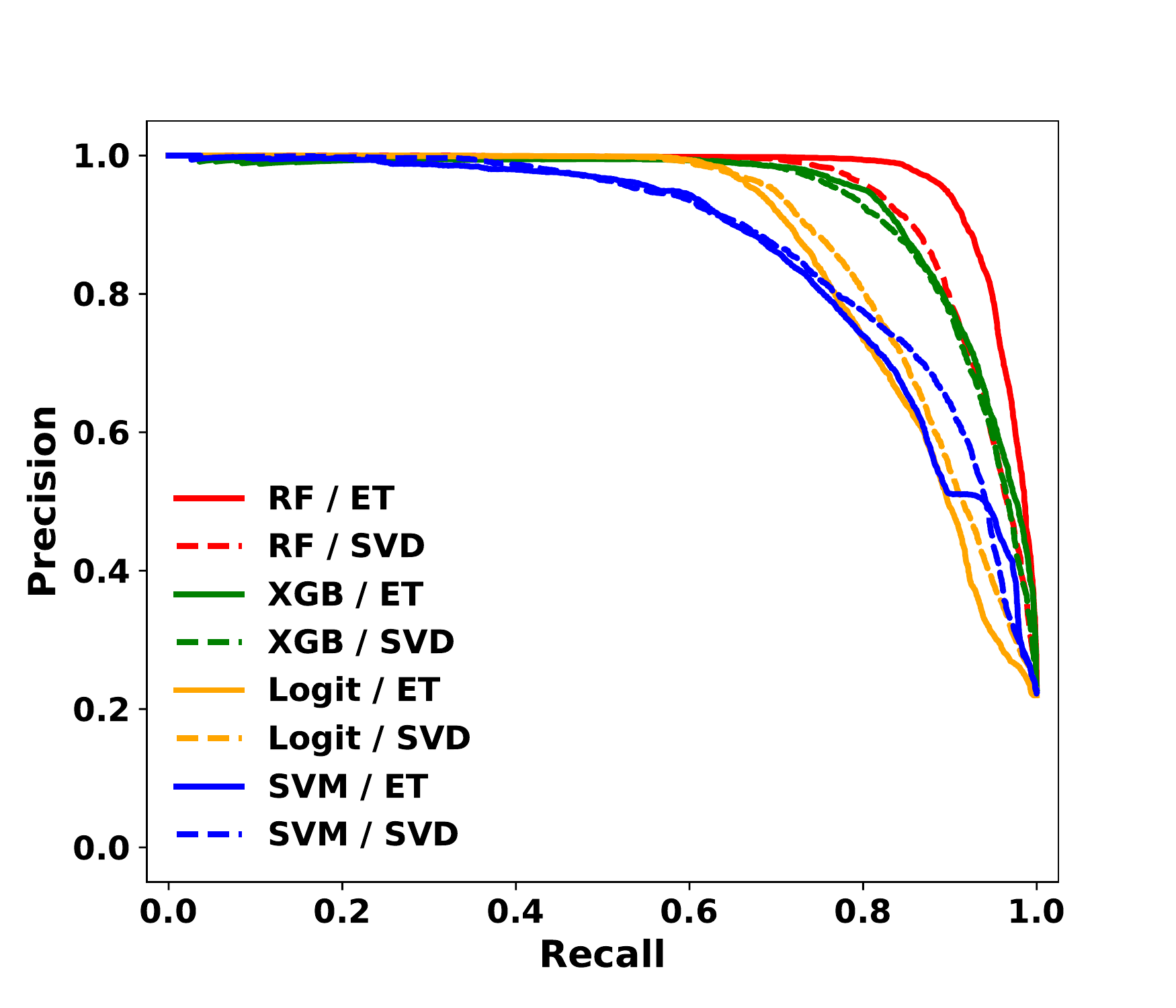}
\caption{Classifier P\hbox{-}R performance across many classifier algorithms. Validation set selected from known IO~accounts and negatively labeled Snorkel data.}\label{fig:PRcomp_all_drop}
\end{figure}

\begin{figure}
\centering
\includegraphics[width=1\linewidth]{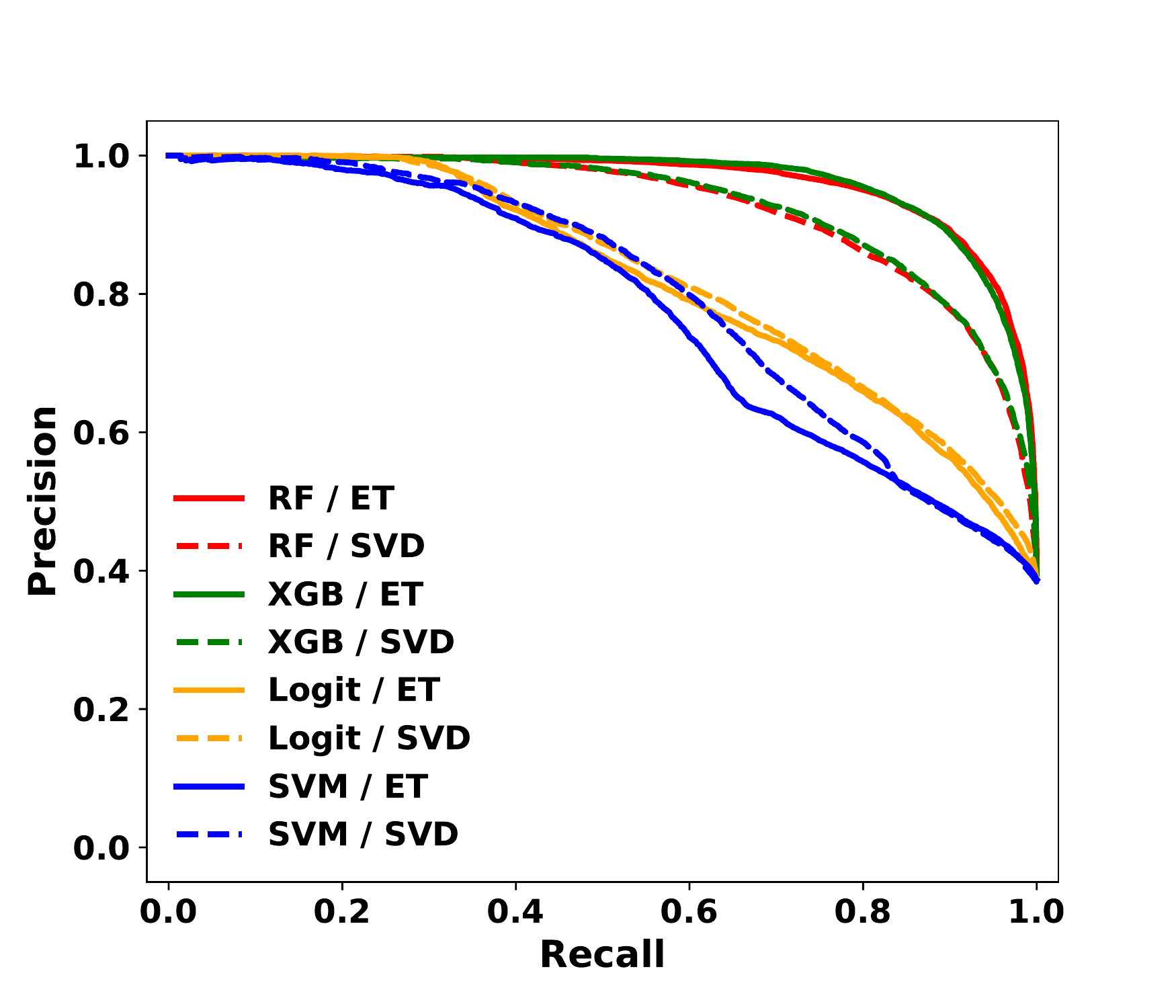}
\caption{Classifier P\hbox{-}R performance across many classifier algorithms. Validation set selected from all data.}\label{fig:PRcomp_all_keep}
\end{figure}

\begin{figure}
\centering
\includegraphics[width=1\linewidth]{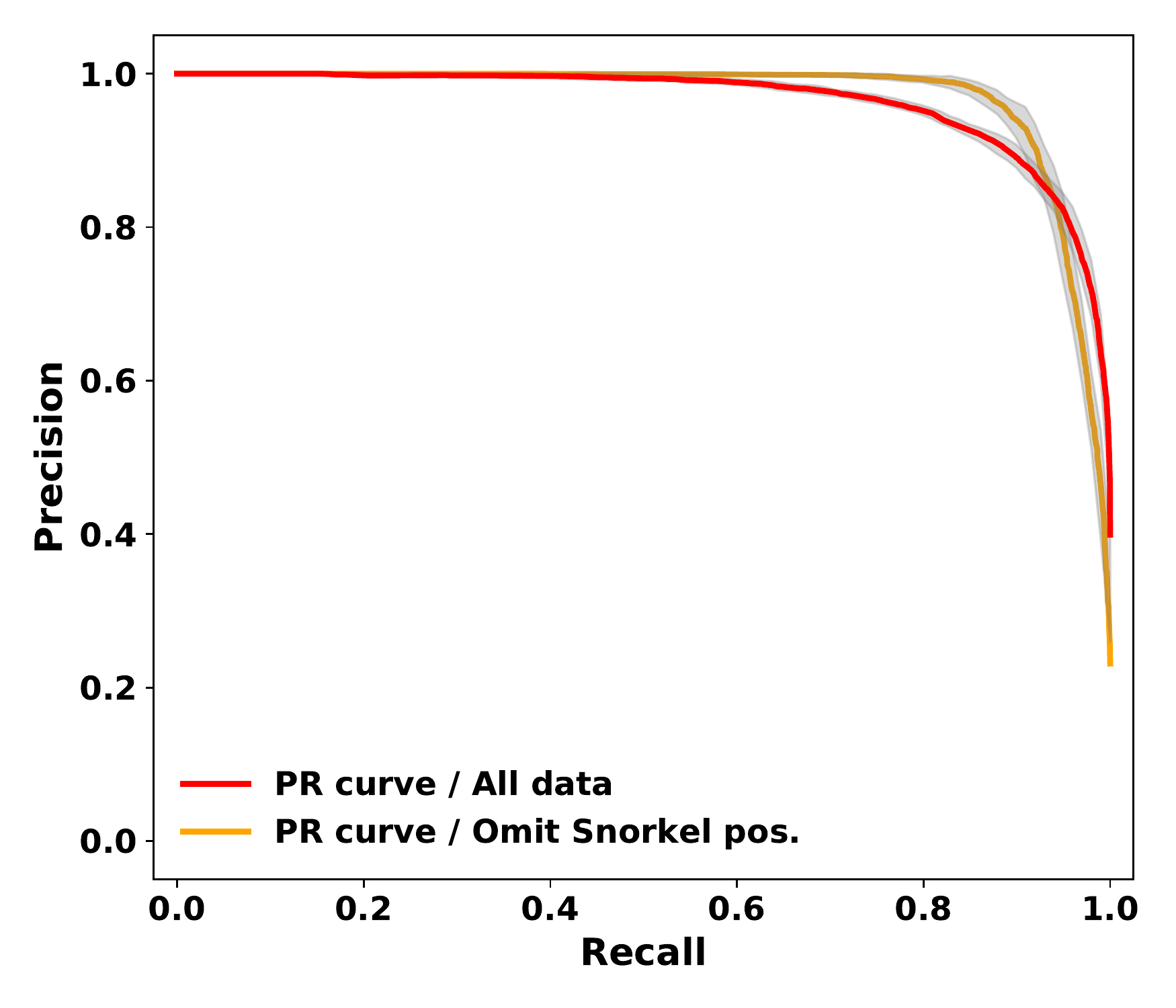}
\caption{RF~/~ET classifier precision--recall (P\hbox{-}R)
  performance with cross-validation error over twenty \mathsffamily ${90:10}$
splits; all data
(\thinspace{\color{pythonred}\legendbar{1.5pt}}\thinspace), and
Snorkel positives omitted
(\thinspace{\color{pythonorange}\legendbar{1.5pt}}\thinspace). Maximum
standard deviation (gray
region, \thinspace{\color{light-gray}\legendbar{2.0pt}}\thinspace) is
$0.016$ and $0.032$, respectively.}\label{fig:PR_error}
\end{figure}

\begin{figure}
\centering
\includegraphics[width=1\linewidth]{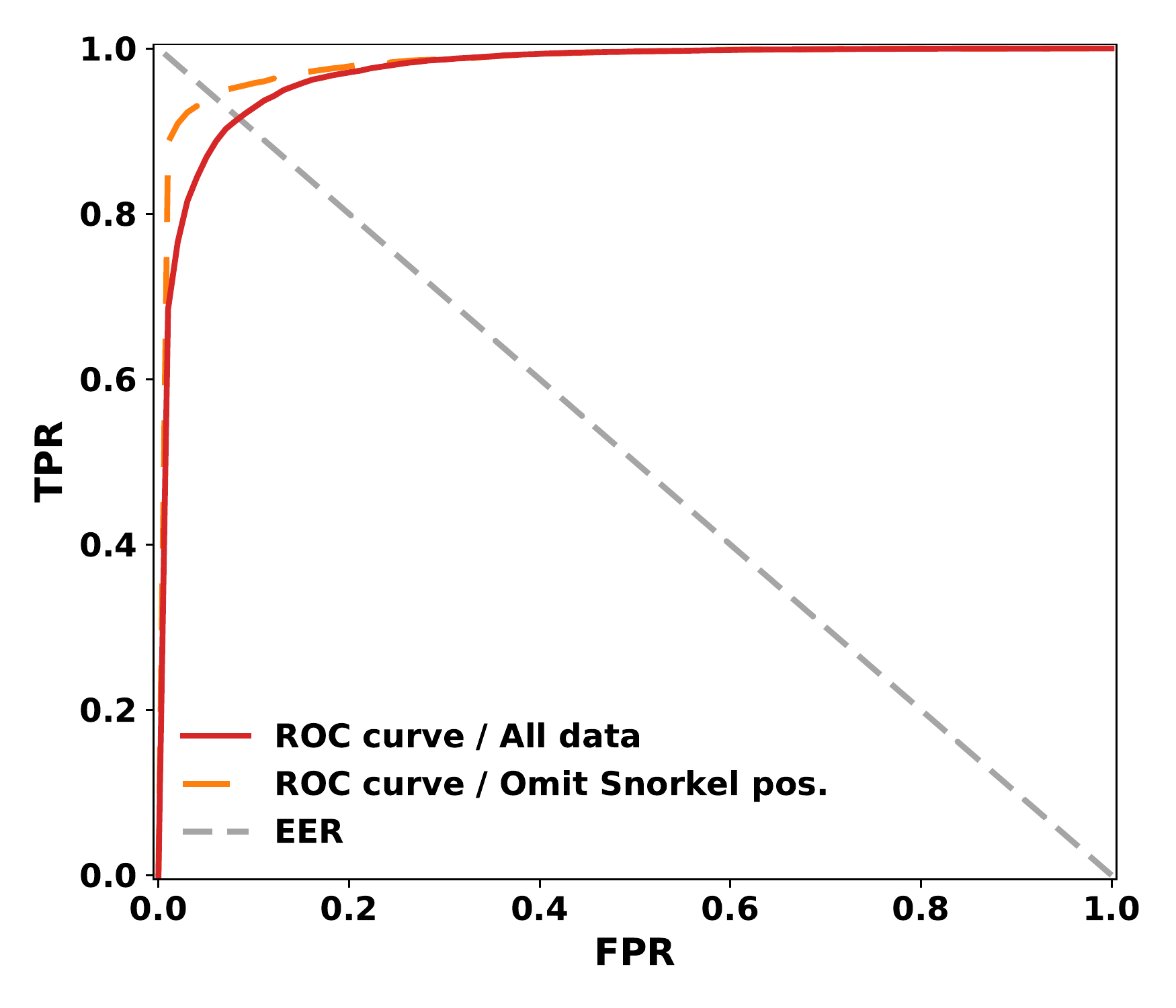}
\caption{RF~/~ET classifier receiver operating characteristic (ROC)
  performance: true positive rate (TPR, a.k.a.\ recall) versus false
  positive rate (FPR); all data
  (\thinspace{\color{pythonred}\legendbar{1.5pt}}\thinspace), and
  Snorkel positives omitted
  (\thinspace{\color{pythonorange}\legendbardashed{1.5pt}}\thinspace). Classifier
  EER is determined by the point at which \mathsffamily
  $1-\hbox{TPR}=\hbox{FPR}$ (dashed gray curve, \thinspace{\color{light-gray}\legendbardashed{1.5pt}}\thinspace).}\label{fig:EER}
\end{figure}

\begin{figure}
\centering
\includegraphics[width=0.8\linewidth]{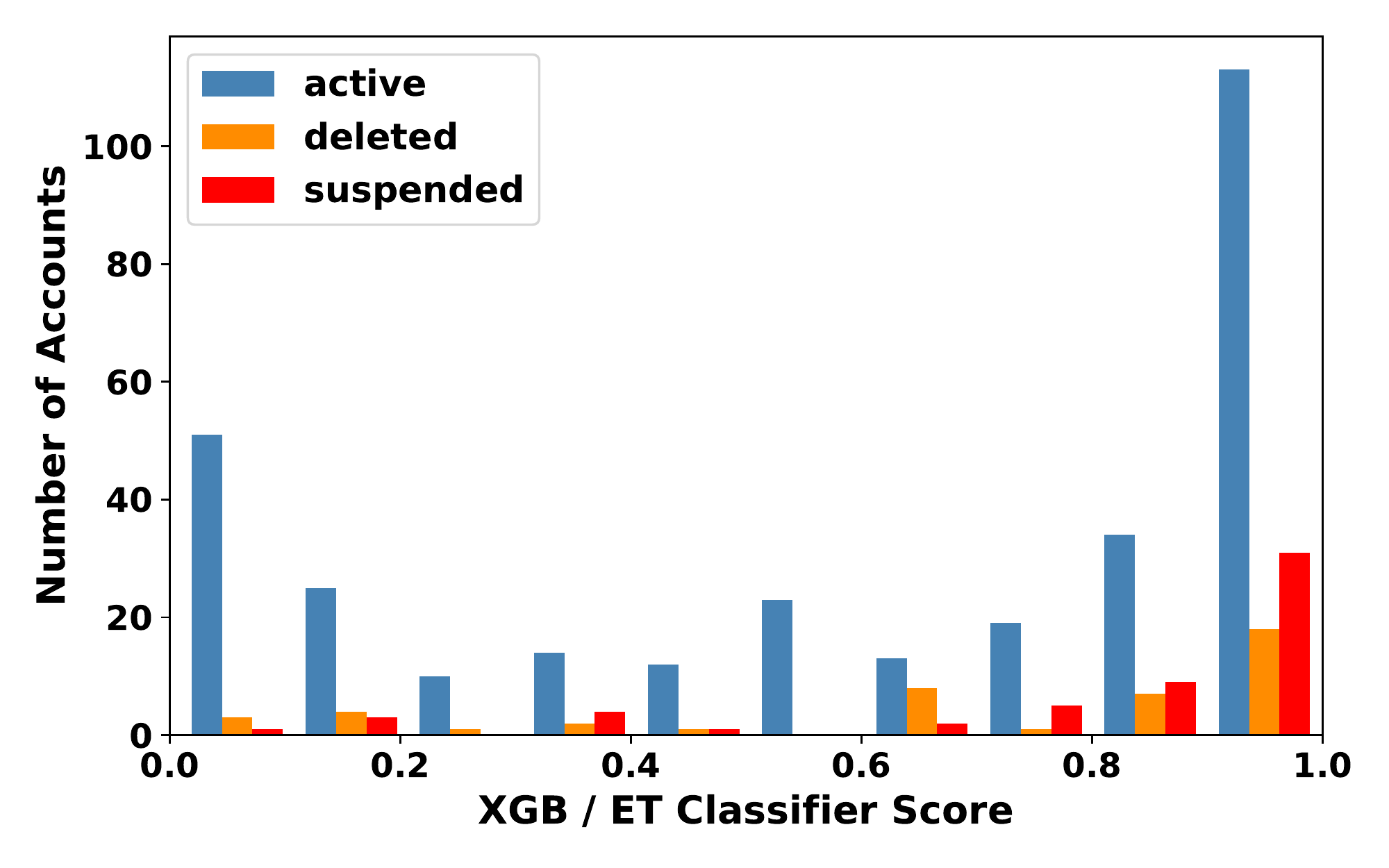}
\caption{Classifier score histograms (xgBoost\slash Extra-Trees) for
active, suspended, and deleted accounts in the French narrative
network.}\label{fig:fr_hist_xgb}
\end{figure}

Classifier design is conducted by comparing the relative performance
of four different classifier models, Random Forest~\cite{liaw2002},
Logistic Regression, xgBoost~\cite{chen2016},
and~SVM~\cite{pedregosa2011}, and two dimensionality reduction
approaches, Extra-Trees~\cite{geurts2006} and~SVD. The performance of
the all classifier design combinations over a grid of classifier
parameters is evaluated via averaged cross validation over twenty
${90:10}$ splits. Cross validation is performed both over $10\%$ of
all training data (Fig.~\ref{fig:PRcomp_all_keep}) and after
discarding all accounts that were labeled as IO-like by Snorkel
heuristics (Fig.~\ref{fig:PRcomp_all_drop}). Of the eight combinations
and for both methods of cross validation, the best performing method
is composed of dimensionality reduction using the Extra-Trees method
followed by a Random Forest classifier
(Figs.\ \ref{fig:PRcomp_all_drop},~\ref{fig:PRcomp_all_keep}). We
compare the P\hbox{-}R and ROC curves for Random Forest, Extra-Trees
across the two methods of cross validation in Figs.\ \ref{fig:PR_error}
and~\ref{fig:EER}, respectively. The standard deviation from
cross-validation is shown in Fig.~\ref{fig:PR_error}, in which it is
seen that the maximum standard deviation for when all data is used
is~$1.6\%$, and~$3.2\%$ when Snorkel positives are omitted.
Although xgBoost with Extra-Trees performs as well as the Random
Forest, Extra-Trees classifier when the validation set includes $10\%$
of all training data, it performs significantly worse when the
Snorkel-labeled IO~accounts are discarded from the validation
set. Further, examination of the performance of xgBoost, Extra-Trees
on accounts in the narrative networks shows that the classifier scores
tend to be skewed towards extremes, rather than being distributed more
evenly across the interval $[0,1]$ (Fig.~\ref{fig:fr_hist_xgb}).

\begin{figure}
\centering
\includegraphics[width=0.8\linewidth]{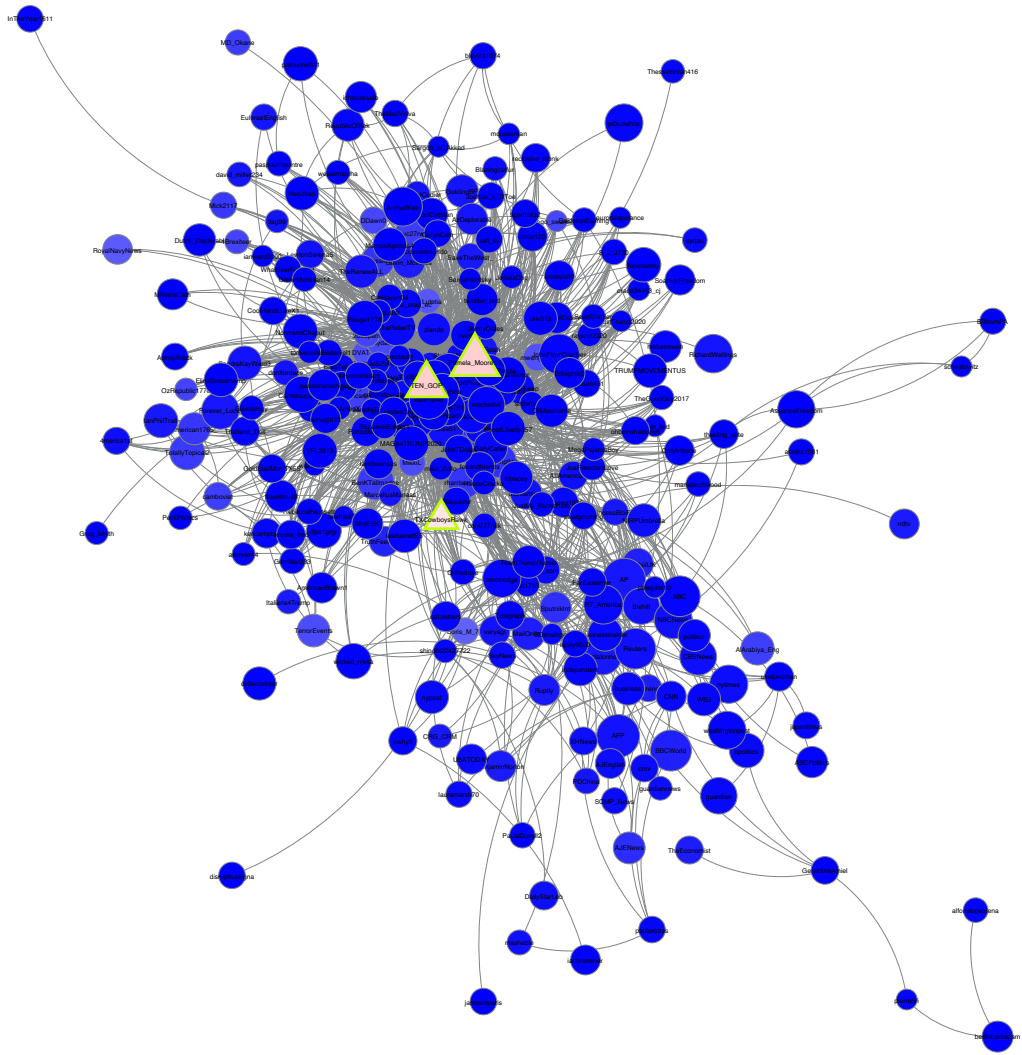}
\caption{Classification on English narrative network ({\color{blue}{\em main paper}},
{\color{blue}Fig.~\protect\customxrref{fig:en_network}}) without Snorkel labeling
functions; RF~/~ET classifier. Classifier overfitting without Snorkel
is apparent because classifier scores for all accounts not within
Twitter's known IO dataset are all clustered near zero. See also
Table~\ref{tab:no_snorkel_classifier_top_features} for the feature
importances without Snorkel.}\label{fig:en_network_no_snorkel}
\end{figure}

To assess the necessity of using Snorkel heuristics in our
semisupervised learning approach, we compute classifier performance
without Snorkel labeling functions. As expected in this
learning problem with limited truth data, we observe that when the
classifier is trained on a deterministic combination of $3{,}151$
positive examples and $15{,}000$ negative examples, there is strong
evidence of classifier overfitting whereby the classifier simply
learns the boundary between these two classes. Consequentially, the
classical, supervised classifier learns previously observed IO
behavior and is unable to recognize new IO accounts. This overfitting
is observed both in Fig.~\ref{fig:en_network_no_snorkel} and
Table~\ref{tab:no_snorkel_classifier_top_features}, in which the
English narrative network ({\color{blue}{\em main paper}},
{\color{blue}Fig.~\customxrref{fig:en_network}}) is labeled as entirely non-IO-like
with the exception of the three known IO accounts within this network.
Furthermore, in contrast to the balanced principal classifier features
obtained using a semisupervised approach
(Tables \ref{tab:num_feats}--\ref{tab:words_feats}), the features of
the strictly supervised classifier are dominated by largest component
of the training data (Fig.~\ref{fig:origin-pie-chart}). For example,
the most important supervised classifier feature is the Serbian
language, and $8$ out of the top-$10$ content features are related to
Serbia, all corresponding to the fact that $55\%$ of our training data
is comprised of Twitter's Serbian IO dataset.  We conclude from both
classifier design principles and these results that a
semisupervised approach is necessary to avoid overfitting in the IO
classifier problem.

\subsubsection{Sensitivity Analysis and Performance Statistics}

\begin{figure}
\centering
\includegraphics[width=0.8\linewidth]{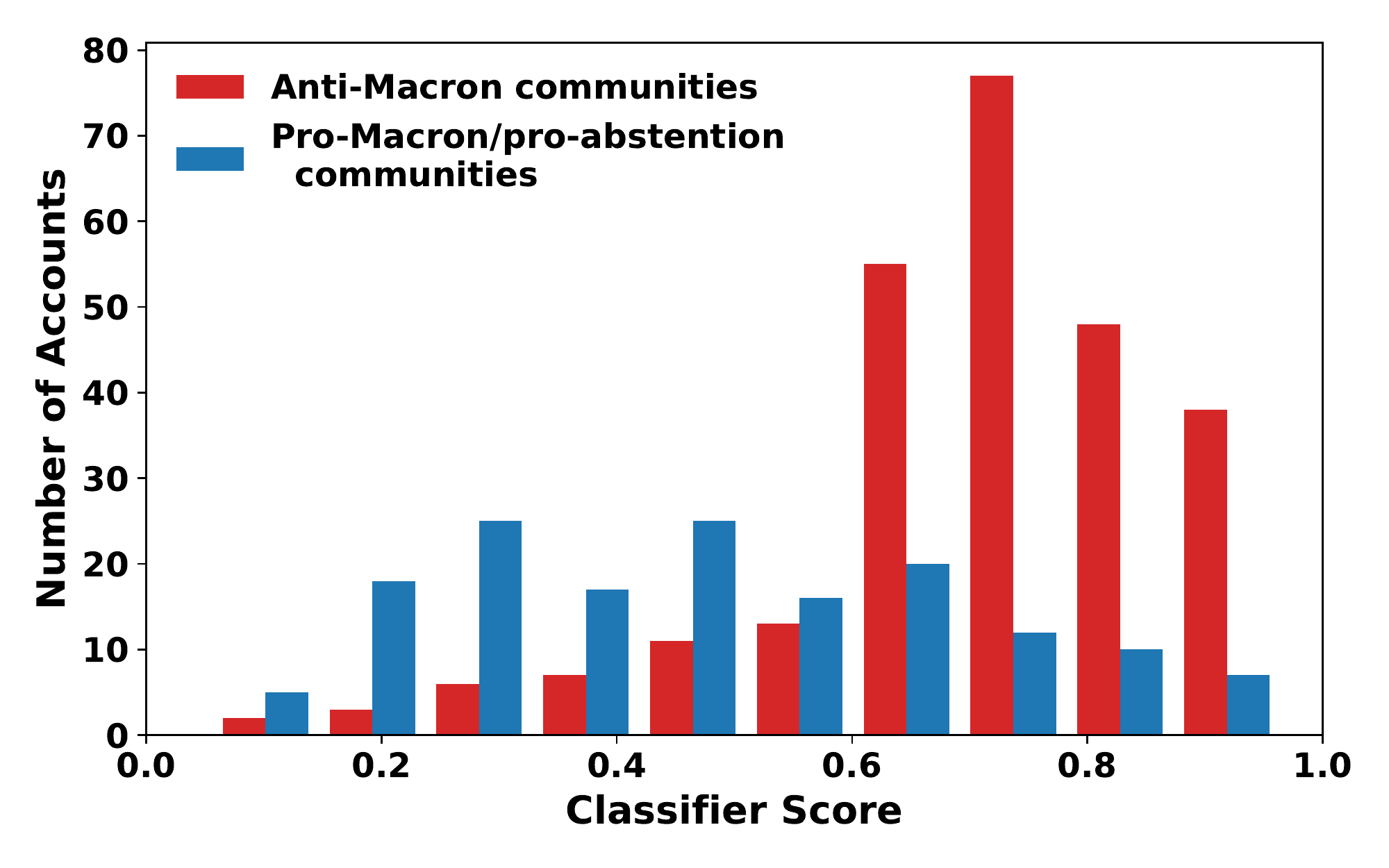}
\caption{Relative classifier score frequencies between communities
({\color{blue}{\em main paper}}, {\color{blue}Fig.~\protect\customxrref{fig:fr_classifier}}).}\label{fig:fr_hist}
\end{figure}

Snorkel is used to provide additional, semisupervised training data
by labeling IO positives. A Snorkel score is computed for each
account, and in the {\em\href{\mainpaperDOI}{\color{blue}main paper}},
all accounts that exceed a score of
$70\%$ are treated as IO truth within classifier training. This $70\%$
threshold is determined by both a sensitivity analysis that compares
classifier performance across a range of metrics: precision--recall
(P\hbox{-}R) at a fixed classifier score threshold ($0.6$ based on
Fig.~\ref{fig:fr_hist}), area-under-the-PR-curve (AUPRC), the
equal-error rate (EER, Table~\ref{tab:snorkel_sensitivity}), and the
fraction of training accounts that are labeled positively by Snorkel
(Table~\ref{tab:snorkel_fracs}). Performance is comparable across the
three bracketed thresholds considered---$50\%$, $70\%$,
and~$90\%$---though the $50\%$ threshold has the worst performance for
most of the statistics. Retaining all training data within
cross-validation sets (contra omitting Snorkel positives) is most
consistent with the objective of measuring classifier performance on
actual accounts within a narrative network. The relative performance
comparisons of Table~\ref{tab:snorkel_sensitivity} (using all data)
and Table~\ref{tab:snorkel_fracs} show that a $70\%$ threshold has the
best AUPRC, best recall, with a small ($0.5$--$1.5\%$) trade-off in
EER and precision, and also the fraction of IO~accounts at this
threshold is consistent with published estimates of bot activity as
discussed in~\ref{app:snorkel}.

\subsubsection{Feature Engineering}

The IO classifier is trained on three categories of features: account
behavior, languages used in tweets, and all $1$- and $2$\hbox{-}grams
that appear more than $15$~times across all account tweets in the
training set.  Initially, the feature set is composed of
$17$~behavioral features, $60$~language features, and $1.8$~million
$1$- and $2$\hbox{-}grams. To limit the per~account content used to
generate these content features, a maximum of $10{,}000$ randomly
selected tweets are chosen from each account. The total number of
tweets used in the English and~French classifier is $40{,}155{,}545$.

The standard machine-learning dimensionality reduction step is used to
improve classifier performance in problems like this one that have a
very large feature space relative to the number of training
samples. Grid search optimization is used to determine the
best-performing dimensionality reduction approach, and the relative
importance of each feature. In the dataset used in the
{\em\href{\mainpaperDOI}{\color{blue}main paper}}, the
Extra-Trees algorithm is used to reduce the feature space to
$10$~behavioral features (Table~\ref{tab:num_feats}), $30$~languages
(Table~\ref{tab:langs_feats}), and $500$ $1$- and $2$\hbox{-}grams
(Table~\ref{tab:words_feats}).  Additionally, the behavioral,
language, and $1$-and $2$\hbox{-}gram feature spaces are each reduced
independently of each other to ensure adequate representation by each
feature category.

\subsubsection{Classifier Scores versus Account Status}

\begin{figure}
\centering
\includegraphics[width=0.8\linewidth]{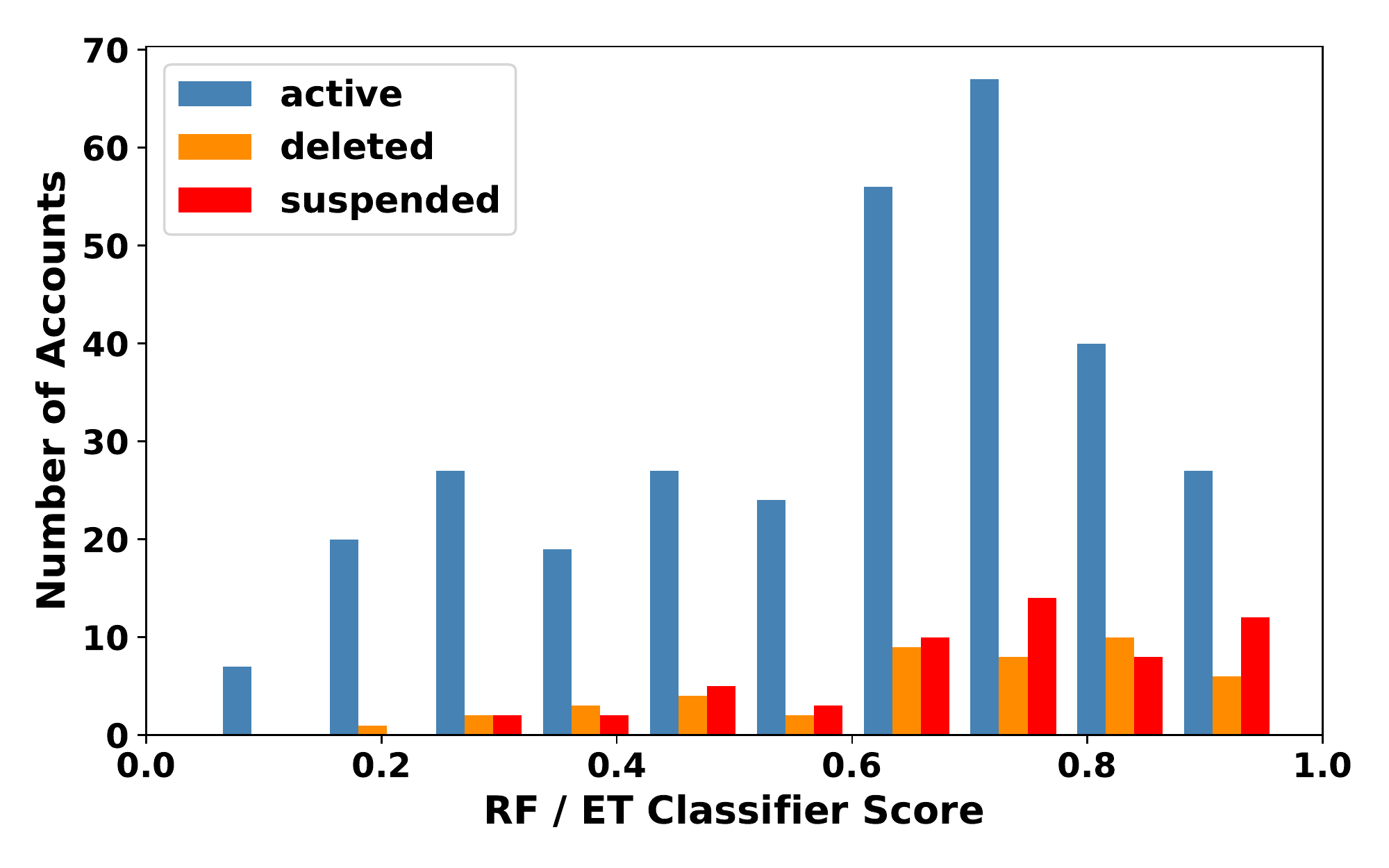}
\caption{Classifier score histograms (Random Forest\slash Extra-Trees)
for active, suspended, and deleted accounts in the French narrative
network.}\label{fig:fr_hist_rf}
\end{figure}

As noted in section~\ref{app:class_comp}, the Random Forest with Extra
Trees combination is the best performing classifier model. Its
optimality is further justified upon examination of the distribution
of the classifier scores across accounts in the French narrative
network (Fig.~\ref{fig:fr_hist_rf}). The accounts are divided by
current (March~2020) status as reported by the Twitter API: active,
suspended, and deleted. The suspended and deleted accounts are skewed
toward higher classifier scores, showing a correlation between
accounts detected by the classifier and behavior that results in
suspension from Twitter.

Although both the Random Forest\slash Extra-Trees and xgBoost\slash
Extra-Trees classifiers have near-identical performance in
Fig.~\ref{fig:PRcomp_all_keep}, the latter's performance on the French
narrative network is not nearly as promising. In
Fig.~\ref{fig:fr_hist_xgb}, the classifier scores over the French
narrative network tend towards extremes for all three account status
categories, in stark contrast to the more realistic distribution seen
in Fig.~\ref{fig:fr_hist_rf}.

\subsubsection{Validation of Community-Based Proxy Truth}
\label{sec:validationcommunity}

As discussed in the {\em\href{\mainpaperDOI}{\color{blue}main paper}},
{\color{blue}section~\customxrref{sec:classperfcomp}}, membership in
the ``Macron allegations'' community of the French narrative network
({\color{blue}{\em main paper}},
{\color{blue}Fig.~\customxrref{fig:fr_classifier}}) is used as a proxy
for known IO~accounts where independent truth is unavailable.  To
establish the narratives used by accounts in the Macron allegations
community, topic modeling is performed on tweets from accounts in the
community over the week preceding the media blackout (28~April to
5~May 2017). A selection of the generated topics is given in
Table~\ref{tab:fr_topics_antiMacron}. Three representative tweets from
each topic are shown in Table~\ref{tab:macron_allegation_tweets},
which illustrate the stance of accounts within this community. Topic
modeling is performed on the tweets in the pro-Macron and
pro-Abstention communities over the same time period. A selection from
those topics is given in Table~\ref{tab:fr_topics_proMacron} and
Table~\ref{tab:fr_topics_abstention}, respectively.

To validate this proxy assumption and quantify its accuracy in the
absence of the underlying truth, we hypothesize that: 1)~the Macron
allegations narrative is used by actual IO~accounts in the 2017
French election; and 2)~the distribution of known IO~accounts is
higher in IO narrative networks. The first hypothesis is confirmed by
numerous independent news reports~\cite{borger2017, dearden2017,
marantz2017, mcauley2017} and direct observation~\cite{USHPSCI2017a,
USHPSCI2017b, gadde2018, smith2018b, birnbaum2019, mueller2019,
schmidt2020}. Though we do not have the ability to independently
establish the validity of the second hypothesis, we can show that the
validity of our results are consistent with this hypothesis.

The distribution of our classifier scores is computed across both the
``Macron allegations'' and ``pro-Macron''\slash ``pro-abstention''
communities in the French narrative network
(Fig.~\ref{fig:fr_hist}). There is a distinct disparity between these
histograms. Classifier scores of accounts in the Macron allegations
community are relatively small at lower (non-IO-like) scores, and rise
sharply to very high relative frequencies above a classifier score
of~$0.6$---the expected range of IO-like scores. In contrast,
classifier scores across the other communities are more evenly
distributed and slightly skewed towards lower (non-IO-like) scores,
also expected for narrative network communities dominated by
pro-Macron activity.

\subsection{Network Potential Outcome Framework for Causal Inference}
\label{SI_causalImpact}

Interference takes place in causal inference when the treatment
applied to one unit affects the outcomes of other units due to their
interactions and influence. An example of rising importance is
treating individuals on a social network.  This section introduces the
mathematical framework of causal inference under network
interference~\cite{kao2017}, used in the
{\em\href{\mainpaperDOI}{\color{blue}main paper}},
{\color{blue}section \customxrref{sec:methodology_impact_estimation}}.
Network potential outcomes are the fundamental quantity used to
capture various types of causal effects such as network impact in
Eq.~[\customxrref{eq:individual_impact_causal_estimand}] of the main
paper. Realistically, many network potential outcomes will be
unobserved and a complete randomization over the different treatment
exposures is infeasible, making the estimation of the causal estimands
challenging. Bayesian imputation of the missing network potential
outcomes provides a natural conceptual solution. Theory for this
imputation is developed based on the critical assumptions of
unconfounded treatment assignment and an unconfounded influence
network, leading to the key ignorability condition for the treatment
exposure mechanism. Driven by this theory, a rigorous design and
analysis procedure is proposed for causal inference under network
interference.

\subsubsection{Introduction}

Interference, in the context of causal inference, refers to the
situation when the outcome of a ``unit'' is affected not only by its own
treatment, but also by the treatment of other units. Interference on a
network of influence, known as network interference, is of rising
importance. Common examples are experiments and observational studies
on a social network where treatment effects propagate through peer
influence, spread of know'ledge, or social benefits, etc. This
phenomenon is also known as spillover effects and social
contagions. The application areas are numerous, e.g., public health,
education, and policy~\cite{basse2016, kim2015, christakis2010,
sobel2006}; social media and marketing~\cite{gui2015, bond2012,
bakshy2012, parker2011}; network security~\cite{coronges2012,
shah2011}; and economics~\cite{banerjee2013, acemoglu2010,
manski1993}.

Traditionally, interference has been viewed as a nuisance in causal
experiments, and earlier works propose experimental designs that
render the interference effects ignorable on restricted, simple block
structures~\cite{david1996, azais1993}. More recent work detects and
estimates interference effects on networks, many of them building on
Rubin's potential outcome framework~\cite{rubin1974, imbensrubin2015}. To detect network
interference effects, Bowers et~al.~\cite{bowers2013} propose
hypothesis testings using potential outcome models with specific
primary and peer effects, and Athey et~al.~\cite{athey2015} propose
exact $p$-value tests by constructing artificial experiments on the
original experimental units such that the null hypothesis is sharp. To
estimate network interference effects, Aronow and
Samii~\cite{aronow2017} propose inverse probability weighting when the
probability of the specific exposure condition can be computed,
Ugander et~al.~\cite{ugander2013} develop a cluster randomization
approach that leads to a closed-form solution on the probabilities of
specific neighborhood exposures, and Sussman and
Airoldi~\cite{sussman2017} propose exclusion restrictions on the
potential outcomes and derive design conditions that lead to unbiased
estimators. Li and Wager demonstrate practicality of non-parametric 
estimators of average primary and peer effects through random graph
asymptotic analysis~\cite{li2020}. 

Social confounders present another source of challenge for
causal inference under network interference. Early work by
Manski~\cite{manski1993} demonstrates unidentifiability of the peer
effects in the presence of social confounders under linear outcome
models. Recent works in causal inference show that confounding social
covariates lead to unidentifiability and biased estimates of causal
effects~\cite{vanderweele2014}, especially on social
networks~\cite{ogburn2014, shalizi2011}, and how longitudinal
studies~\cite{aral2009, omalley2014} and design of experiment for
specific peer effects~\cite{eckles2016} provide a way forward.

\subsubsection{Definitions}

The key quantities are the potential outcomes of each unit in the
study under different treatment conditions. They serve as the basic
building blocks for causal inference in the potential outcome
framework. Most existing works assume an absence of interference and a
simple binary treatment assignment, which is the Stable Unit Treatment
Value Assumption (SUTVA)~\cite{rubin1980}. Under SUTVA, the potential
outcomes for each unit $i$ are a function of its own treatment and are
denoted as $Y_i(Z_i)$, where $Z_i$ is a binary indicator for whether
the treatment is assigned to unit $i$. The potential outcomes of all
$N$ units in an experiment, $\Y$, can be partitioned into two vectors
of $N$ components: $\bm{Y}(0)$ for all outcomes under control and
$\bm{Y}(1)$ for all outcomes under treatment. $\Y$ can also be
partitioned according to whether it is observed. In an experiment, a
unit is either under control or under treatment. Therefore, half of
all potential outcomes are observed, denoted as $\bm{Y}_{\rm obs}$. The
other half of the potential outcomes are unobserved, denoted as
$\bm{Y}_{\rm mis}$. As a result, causal inference is fundamentally a
missing data problem, with a rich body of work on rigorous design and
analysis for estimating and imputing the missing
outcomes~\cite{imbensrubin2015}.

Under network interference, the potential outcome definitions need to
be generalized to encompass the different treatment exposure
conditions via the network, starting with the units:
\begin{definition}[Finite Population on a Network of Influence]
\label{def:networkPopulation}
The study takes place on a finite population of $N$ units where their
influence on each other is represented as a ${N\times N}$ influence
matrix~$\A$. Each element of the influence matrix,
${A_{ij}\in\mathbb{R}}$, represents the strength of influence unit~$i$
has on unit~$j$. One may visualize this population network as a graph,
$\Graph=(\VSet, \ESet)$, of which the node set~$\VSet$ consists of the
units of the study (${|\VSet|=N}$) and the edge set~$\ESet$ represents
the none-zero entries of the influence matrix~$\A$.
\end{definition}
The outcomes of each unit not only depend on its own treatment but
also on exposure to treatments on other units propagated through the
influence network.

\begin{definition}[Network Potential Outcomes] 
\label{def:networkNetworkPO}
Under network interference, the outcomes of a unit~$i$, change
according to its exposure to the treatment on the finite
population~$\Z$, through the network of influence~$\A$. The network
potential outcomes of~$i$ are denoted as~$Y_i(\Z, \A)$.
\end{definition}

\noindent Sometimes, it is clearer to denote the treatment on certain 
units separately. For example, one may want to denote the treatment 
on unit $i$ itself. In such cases, the following notational convention is 
adopted:  $Y_i(\Z, \A) \equiv Y_i(Z_i, \Z_{-i}, \A)$, where $\Z_{-i}$ is the 
treatment vector excluding the $i$th element. 

\begin{definition}[Network Potential Outcome Sets]
\label{def:networkPOSet}
The entire set of network potential outcomes for unit~$i$ is~$\YSet_i
= \{\,Y_i(\Z=\z, \A=\bm{a})\,\}$ for~all ${\z\in\ZSetClosed}$,
${\bm{a}\in\ASetClosed}$, where $\ZSetClosed$ is the set of all
possible assignments on the closed neighborhood of~$i$ and
$\ASetClosed$ is the set of all possible influence networks on the
closed neighborhood of~$i$. The set of all network potential outcomes
on the finite population with~$N$ units is~$\YSet
= \{\,\YSet_1, \YSet_2, \ldots, \YSet_N\,\}$. $\YSet$ can also be
partitioned based on whether it is observed. In an experiment, for
each unit, only one of the numerous possible neighborhood treatment
in~$\ZSetClosed$ is realized. The set of observed network potential
outcomes, typically of size~$N$, is denoted as~$\YobsSet$. Most of the
network potential outcomes will be unobserved. The set of unobserved
outcomes, its size depending on the size of each unit's closed
neighborhood, is denoted as~$\YmisSet$.
\end{definition}

Lastly, covariates $\X$ on the network potential outcome units play a
vital role in principled design and analysis for estimating and
imputing the unobserved outcomes. For $k$-dimensional covariates on
the units, the matrix $\X$ has ${N\times k}$ elements. Under the scope
here, the treatment vector $\Z$ is limited to binary indicators for
treatment versus control, but can be easily generalized for
multi-level treatments.

\subsubsection {Causal Estimands Using Network Potential Outcomes}
\label{sec:CEUnderNetworkInteference}

The network potential outcomes serve as the building blocks for
defining appropriate causal estimands to answer various causal
questions under network interference. In addition to the causal
estimand in {\color{blue}{\em main paper}},
{\color{blue}Eq.~[\customxrref{eq:individual_impact_causal_estimand}]},
this section gives more example causal estimands, each focusing on
quantifying the effect of a particular kind of exposure to
treatment. Many more causal estimands may be defined using the network
potential outcomes, but these demonstrate the flexibility of the
network potential outcomes in expressing causal quantities under
network interference.\\

\noindent\textbf{Primary Causal Effect Estimands:}
If the primary treatment causal effect is the quantity of interest
(i.e., want to separate it from the peer influence effects), an
appropriate set of conditional causal estimands for each unit $i$ are:
\begin{equation}
\label{eq:primaryEffectConditionalCausalEstimand}
	\xi_i(\z) \equiv Y_i(Z_i = 1, \Zopen  = \z , \A) -  Y_i(Z_i = 0,  \Zopen = \z ,\A)
\end{equation}
where $\z \in \ZSetOpen$, is a member of the set of all possible
assignments on $\Zopen$. This set of conditional estimands capture the
causal effect of the treatment on unit $i$ conditioning on a
particular treatment assignment $\z$ on the other units. This estimand
focuses on the causal effect of receiving the treatment itself, by
fixing the exposure through the influence network (i.e., the peer
effect). In the absence of any exposure to treatment on peers,
$\xi_i(\bm{0})$, the classical (i.e., under SUTVA) unit level causal
effect, $Y_i(1)-Y_i(0)$, is recovered. If one wants to estimate the
average primary treatment causal effect on unit $i$ under all possible
neighborhood treatment, the causal estimand becomes:
\begin{equation}
	\xi_i^\ave \equiv \frac{1}{2^{N-1}} \sum_{\z \in \ZSetOpen} {\xi_i(\z)}
\end{equation}
For a population of size $N$, the average primary treatment causal effect on the population is simply:
\begin{equation}
\label{eq:populationAveragePrimaryEffect}
	\xi^\ave \equiv \frac{1}{N} \sum_{i=1:N}{\xi_i^\ave}
\end{equation}

\noindent\textbf{$k$ Treated Neighbor Causal Effect Estimands:}
If the peer influence effect is the focus of interest, a natural
quantity to consider is the causal effect of having $k$ of the
neighbors of unit $i$ treated. Assuming unit $i$ has at least $k$
neighbors, an appropriate set of conditional causal estimands are:
\begin{equation}
\label{eq:kNeighborIndividualCausalEstimand}
	\delta_{i, k} (z) \equiv {|\N_{-i}|  \choose  k}^{-1} 
	\sum_{\z \in \ZkSet} {Y_i(Z_i=z,\Zopen = \z, \A) - Y_i(Z_i = z, \Zopen = \bm{0}, \A)}
\end{equation}
where $z \in \{0, 1\}$ are the possible assignments on unit $i$,
$\N_{-i}$ the open neighborhood of $i$, and $\ZkSet$ the set of all
treatment assignments where exactly $k$ of $i$'s neighbors are treated
(i.e., $\sum{\Z_{\N_{-i}}} = k$). This set of conditional estimands
capture the average causal effect on unit $i$ for having $k$ of its
neighbors treated, while conditioning the treatment assignment $z$ on
$i$ itself. Similar to the previous example, the $k$ treated neighbor
causal effect averaged over unit $i$'s own treatment can be expressed
in the following estimand:
\begin{equation}
	\delta_{i, k}^\ave \equiv \frac{1}{2} \sum_{z=\{0,1\}} {\delta_{i, k} (z)}
\end{equation}
The population here is a bit more nuanced, because not all units have
at least $k$ neighbors and therefore can not possibly receive such
peer treatment. Therefore, the population average effect should only
be averaged over the units that have at least $k$ neighbors. Defining
$\VSet_{\geq k}$ to be the set of units with at~least $k$
neighbors, the average $k$ treated neighbor causal effect on the
population is:
\begin{equation}
	\delta_ k^\ave \equiv \frac{1}{|\VSet_{\geq k}|} \sum_{i \in \VSet_{\geq k} }{\delta_{i, k}^\ave}
\end{equation}
The idea of capturing causal peer effects based on the number of
neighbors being treated has been proposed by other work. Ugander
et~al.~\cite{ugander2013} define a similar condition called the
"absolute $k$-neighborhood exposure" where unit $i$ meets this
neighborhood treatment condition if $i$ is treated and at~least
$k$ of $i$'s neighbors are treated. Adopting Ugander et~al.'s peer
treatment condition gives the following estimand for unit $i$:
\begin{equation}
	\tilde{\delta}_{i, k} \equiv \left[ \sum_{l=k}^{|\N_{-i}|} {|\N_{-i}| \choose  l}\right]^{-1} \;
	\sum_{l=k}^{|\N_{-i}|} \; \sum_{\z \in \ZlSet} {Y_i(Z_i=1,\Zopen = \z, \A) - Y_i(Z_i = 1, \Zopen = \bm{0}, \A)}
\end{equation}
One may want to know the causal effect of having a certain fraction of
the neighbors treated (e.g., 30\% of the neighbors treated), instead of
the absolute number of neighbors. Ugander et~al.~\cite{ugander2013}
define a version of such treatment condition called the "fractional
q-neighborhood exposure". A causal estimand for fractional
neighborhood treatment can be defined by simply mapping the fractional
criterion to an absolute number for each unit $i$. For example, a
$q\%$ neighborhood treatment for unit $i$ could map to a $k$ neighbor
treatment with $k=\lceil\frac{q}{100} |\N_{-i}|)\rceil$. In any case,
the individual $k$ treated neighbor causal estimand in equation
(\ref{eq:kNeighborIndividualCausalEstimand}) serves as the basic
building block for these types of neighborhood treatment causal
estimands.\\

\noindent\textbf{Influence Network Manipulation Causal Estimands:}
Sometimes, one may be able to manipulate the influence network to
achieve the desired outcome. Causal effects of network manipulation
can be expressed using network potential outcomes as building
blocks. This may seem counterintuitive at first because the influence
network itself is not a "treatment". However, under network
interference, the influence network leads to exposure to treatment on
peers, so manipulating the influence network alters the "social
treatment". The influence network and the treatment assignment
together can be viewed as an assignment of "social
treatment". Consider the causal estimand below on the average total
effect of manipulating the influence network from $\A$ to $\A'$, given
a particular treatment assignment $\z$,:
\begin{equation}
	\zeta_{\A}(\bm{z}) \equiv \frac{1}{N} \sum_{i \in 1:N} {Y_i(\Z=\z, \bm{A}' ) -  Y_i(\Z=\z, \bm{A} )}
\end{equation}
This estimand may quantify the effect of weakening the disinformation
network through account suspension and warning. This estimand
highlights the flexibility and expressiveness of the network potential
outcomes as the basic building block for causal inference under
network interference.

\subsubsection{Bayesian Imputation of Missing Outcomes and Unconfoundedness Assumptions}
\label{sec:imputation_and_unconfoundedness_assumption}

Under network interference, most of the potential outcomes in the
causal estimands will be unobserved. Furthermore, a complete
randomized assignment of the different exposures to neighborhood
treatment is typically infeasible, as the structure of the influence
network impacts each unit's chance to receive a certain
exposure \cite{toulis2013}. These make the estimation of causal
estimands challenging under network interference. A natural solution
is to perform Bayesian imputation of missing potential outcomes. Also
known as the Bayesian predictive inference for causal effects, this
method has been well established for causal inference in the absence
of interference (i.e., SUTVA holds) \cite{rubin1991, rubin1978}.

As the potential outcomes serve as the building blocks for each causal
estimand, computing the posterior distribution of the unobserved
potential outcomes also gives the posterior distribution of any causal
estimands. In the regular case under SUTVA, the posterior distribution
of interest is $P(\bm{Y}_{\rm mis} | \bm{X},\bm{Z},\bm{Y}_{\rm obs})$. Inference of
the unobserved potential outcomes from the observed potential
outcomes, the unit covariates $\X$, and the treatment assignment
vector $\Z$, is typically done with a potential outcome
model. Modeling and inference of this posterior distribution is
greatly simplified when the treatment assignment mechanism can be
ignored (i.e., $\Z$ can be dropped from the posterior). Rubin shows
how the unconfounded treatment assignment assumption leads to this
ignorability \cite{rubin1991, rubin1978}.

In the case under network interference, the posterior distribution is
on the expanded sets of network potential outcomes and includes the
influence network: $P(\YmisSet
| \bm{X}, \bm{Z}, \bm{A}, \YobsSet)$. Similar to the regular case
under SUTVA, this posterior can also be greatly simplified when the
neighborhood treatment mechanism can be ignored (i.e., both $\Z$ and
$\A$ can be dropped from the posterior). This section shows how the
assumptions of unconfounded treatment assignment and unconfounded
influence network lead to this more extended ignorability. Under
network interference, although Bayesian imputation of the missing network
potential outcomes offers a practical solution, the procedure needs to
respect the key unconfoundedness assumptions in order to avoid incorrect
causal estimates.

\begin{assumption}[Unconfounded Treatment Assignment Assumption Under Network Interference]
\label{as:unconfoundedTreatmentAssignmentUnderNetworkInterference}
Conditional on the relevant unit covariates $\X$ and the influence
network $\A$, the treatment assignment $\Z$ is independent from the
potential outcomes $\YSet$:
\begin{equation}
\label{eq:networkInterferenceUnconfoundedTreatment}
	 P(\Z | \bm{X}, \bm{A}, \YSet) =  P(\Z | \bm{X}, \bm{A}).
\end{equation} 
\end{assumption}
This assumption can be met in real-world experiments through complete
randomization of~$\Z$ or including possible confounders in the
conditional unit covariates~$\X$. Sometimes, experiments on a social
network target units with certain network characteristics for
treatment, in order to achieve a desirable overall outcome. For
example, a researcher may target the most influential units (e.g.,
high degree nodes) in order to maximize peer influence
effects. Assumption~\ref{as:unconfoundedTreatmentAssignmentUnderNetworkInterference}
holds under such treatment assignment strategies because the
treatment~$\bm{Z}$ only depends on the influence network,~$\bm{A}$,
and relevant unit covariates $\X$.  This paper estimates the impact of
each unit as a potential source (i.e., being treated), so the
treatment assignment does not depend on the potential outcomes,
satisfying
Assumption~\ref{as:unconfoundedTreatmentAssignmentUnderNetworkInterference}.

\begin{assumption}[Unconfounded Influence Network Assumption Under Network Interference]
\label{as:unconfoundedInfluenceNetworkUnderNetworkInterference}
Conditional on the relevant unit covariates $\X$, the influence
network $\A$ is independent from the potential outcomes $\YSet$:
\begin{equation}
\label{eq:networkInterferenceUnconfoundedInfluenceNetwork}
	 P(\A | \bm{X},\YSet) =  P(\A | \bm{X}).
\end{equation}
\end{assumption}

\noindent This assumption is met if the formation of the influence
network has no correlation with the potential outcomes. However, this
is often not true in real-world
experiments~\cite{shalizi2011}. Intuitively, correlation between the
potential outcomes and the influence network may arise from certain
characteristics of a unit that are correlated with both its outcomes
as well as its relationships with other units on the network. Activity
level and group memberships are examples of such characteristics. For
example, an account's node degree on the retweet network may be
positively correlated with its potential outcomes (i.e., tweet count
on the IO narrative) because hubs in the influence network may tweet
more in general. Similarly, an account's membership to an IO community
may be positively correlated with its potential outcomes. Therefore,
meeting Assumption~\ref{as:unconfoundedInfluenceNetworkUnderNetworkInterference}
will likely require including such confounding characteristics in the
conditional unit covariates $\X$. A method to achieve this will be
formally proposed in
Theorem~\ref{thm:unconfoundedInfluenceNetworkByConditioning}, but
first, we introduce the ignorability condition.

\begin{theorem}[Ignorable Treatment Exposure Mechanism Under Network Interference]
\label{thm:ignorabilityUnderNetworkInterference}
If the unconfounded treatment assignment assumption and the
unconfounded influence network assumption are both met, the treatment
exposure mechanism is ignorable and does not enter the posterior
distribution of the missing potential outcomes:
\begin{equation}
	 P(\YmisSet | \bm{X}, \bm{Z}, \bm{A},\YobsSet) = P(\YmisSet | \bm{X},\YobsSet).
\end{equation}
\end{theorem}
\noindent Theorem~\ref{thm:ignorabilityUnderNetworkInterference} is
proved via factorization and application of
Assumptions~\ref{as:unconfoundedTreatmentAssignmentUnderNetworkInterference}
and~\ref{as:unconfoundedInfluenceNetworkUnderNetworkInterference}:
\begin{align}
	 P(\YmisSet | \bm{X}, \bm{Z}, \bm{A},\YobsSet) &= \frac{ P(\Z
					    | \bm{X}, \bm{A}, \YSet) P(\YSet
					    | \bm{X}, \bm{A})}{\int P(\Z
					    | \bm{X}, \bm{A}, \YSet) P(\YSet
					    | \bm{X}, \bm{A})
					    \,d\YmisSet} \nonumber \\
					    &= \frac{ P(\Z | \bm{X}, \bm{A})
					    P(\YSet | \bm{X}, \bm{A})}{P(\Z
					    | \bm{X}, \bm{A}) \int P(\YSet
					    | \bm{X}, \bm{A})
					    \,d\YmisSet} \nonumber \\
					    &= \frac{P(\YSet
					    | \bm{X}, \bm{A})}{\int P(\YSet
					    | \bm{X}, \bm{A})
					    \,d\YmisSet} \nonumber \\
					    &= P(\YmisSet | \bm{X},
					    \bm{A},\YobsSet) \nonumber \\
					    &= \frac{ P(\A
					    | \bm{X}, \YSet) P(\YSet
					    | \bm{X})}{\int P(\A
					    | \bm{X},\YSet) P(\YSet
					    | \bm{X})
					    \,d\YmisSet} \nonumber \\
					    &= \frac{ P(\A | \bm{X})
					    P(\YSet | \bm{X})}{P(\A
					    | \bm{X}) \int P(\YSet
					    | \bm{X})
					    \,d\YmisSet} \nonumber \\
					    &= \frac{P(\YSet
					    | \bm{X})}{\int P(\YSet
					    | \bm{X})
					    \,d\YmisSet} \nonumber \\
					    &= P(\YmisSet
					    | \bm{X},\YobsSet).
\end{align}

Finally,
Assumption~\ref{as:unconfoundedInfluenceNetworkUnderNetworkInterference}
can be satisfied by conditioning on network parameters through
parametric modeling.

\begin{theorem}[Unconfounded Influence Network by Conditioning on Network Parameters]
\label{thm:unconfoundedInfluenceNetworkByConditioning}
The unconfounded influence network assumption in
Assumption~\ref{as:unconfoundedInfluenceNetworkUnderNetworkInterference},
$P(\A |
\bm{X},\YSet) = P(\A | \bm{X})$, is met if:
\begin{enumerate}
\item The distribution of the influence network $\A$ can be
  characterized by a model $H_G$ with nodal parameters $\X_G$ and
  population parameters $\bm{\Theta}_G$: $\A \sim H_G(\X_G,
  \bm{\Theta}_G)$;
\item The potential outcomes $\YSet$ correlate with the influence
  network $\A$ only through a subset of the nodal parameters
  $\tilde{\X}_G \in \X_G$ and population parameters
  $\tilde{\bm{\Theta}}_G \in \bm{\Theta}_G$;
\item The unit covariates $\X$ contain these network parameters
  $\tilde{\X}_G$ and $\tilde{\bm{\Theta}}_G$.
\end{enumerate}
\end{theorem}
Theorem~\ref{thm:unconfoundedInfluenceNetworkByConditioning} leverages
the assumption that the influence network~$\A$ can be characterized by
a model with nodal (i.e., unit-specific) parameters~$\X_G$ (e.g.,
expected degree, community membership, position in the latent space,
etc.)\ and population parameters~$\bm{\Theta}_G$ (e.g.,
inter-community interaction, sparsity, etc.). Most if not all of the
currently well-known network models in the network inference community
can be described in this way, including the latent space
models~\cite{hoff2002}, the latent class models such as the membership
blockmodels~\cite{wang1987, airoldi2008, kao2019}, degree distribution
models~\cite{aiello2001, chung2002}, and the
graphon~\cite{lovasz2006}. Typically, only a subset of the nodal
parameters, $\tilde{\X}_G$, may be correlated with the potential
outcomes, like the unit-specific characteristics such as activity
level and community membership in the IO narrative influence
network. Some of the population parameters, $\tilde{\bm{\Theta}}_G$,
like sparsity, may be correlated with the potential outcomes as
well. Intuitively, conditioning on these parameters breaks the
correlation between the potential outcomes and the influence network,
therefore meeting the unconfounded influence network
assumption. Often, these network model parameters are not readily
observed, but they can be estimated from the social network data
collected in the experiment. This is similar in spirit to work by
Frangakis and Rubin~\cite{frangakis1983} on latent ignorability where
the treatment mechanism is ignorable by conditioning on the latent
compliance covariate.

\subsubsection {Theory to Practice: Design and Analysis Under Network Interference}

Driven by the theoretical framework developed above, the experimental
and analytical procedure for Bayesian imputation of missing potential
outcomes and causal estimands is summarized in the following steps.

\begin{enumerate}
\item Define the population, the treatment, and the network potential
  outcomes. Collect prior information on the underlying influence
  network. This can be from data on interactions between the units
  (e.g., emails, tweets, phone calls, etc.)\ or a survey on the social
  network (e.g., list of relationships). One may only have partial or
  prior information on the influence network, in which case the
  network will need to be imputed during analysis.
\item Propose an appropriate network model, such as the ones mentioned
  in the previous section~\cite{hoff2002, aiello2001, chung2002,
  lovasz2006, wang1987, airoldi2008, kao2019}, and estimate the model
  parameters using the observed influence network or the prior
  distribution on the influence network.
\item Propose an appropriate potential outcome model including all the
  possible confounding covariates, guided by
  Assumptions~\ref{as:unconfoundedTreatmentAssignmentUnderNetworkInterference}
  and~\ref{as:unconfoundedInfluenceNetworkUnderNetworkInterference},
  and Theorem~\ref{thm:unconfoundedInfluenceNetworkByConditioning}, in
  order to meet the ignorable treatment exposure mechanism condition
  specified in
  Theorem~\ref{thm:ignorabilityUnderNetworkInterference}. Some of
  these may be the estimated network parameters from the previous step
  and the possible treatment assignment confounders. Perform Bayesian
  inference to compute the posterior distribution of the potential
  outcome model parameters, jointly with the influence network if it
  is not fully known, as typically is the case. Weakly informative
  priors on the model parameters have shown to improve convergence
  stability while minimizing any bias on the posterior
  distribution~\cite{gelman2008}.  Some model parameters such as
  propagated effects on the network (each $\gamma_k$ in {\em\href{\mainpaperDOI}{\color{blue}main paper}},
  {\color{blue}Eq.~[\customxrref{eq:PO_PoissonGLM}]}) should
  respect social phenomenologies such as decaying exposure effects
  with each additional hop in the propagation. This can be
  accomplished with truncated priors to restrict the feasible
  parameter range. Lastly, adequacy of the potential outcome model in
  describing the observed outcomes can be evaluated via statistical
  tests such as the posterior predictive check~\cite{gelman1996}.
\item Using the potential outcome model and the parameter posterior distribution 
  from the step above, impute the missing network potential outcomes in the 
  causal estimands of interest. This will finally provide estimates on the
  desired causal estimands. Typically, one would want to quantify the uncertainty on
  the estimated causal estimand. This can be accomplished by multiple 
  imputation~\cite{rubin2004} of the missing potential outcomes, by imputing them with
  independent samples of model parameters from their posterior distribution.
\end{enumerate}
This procedure accounts for potential confounders through covariate
adjustment, where accuracy depends on the adequacy of the potential
outcome model. Additional robustness to model mis-specification can be
achieved through balancing of confounding covariates across different
treatment exposure conditions in the causal estimand. This can be done
via treatment design in experiments or matching in observational
studies, as a desirable future expansion on this framework. Propensity score
matching for multiple treatment conditions such as the case here with numerous
treatment exposure conditions through the network is a challenging and relevant
research topic with recent work by Forastiere et~al.~\cite{forastiere2020} and 
Han and Rubin~\cite{han2020}.

\subsection{Impact Estimation on the \#MacronLeaks Narrative}
\label{sec:cihashtagmacronleaks}

\begin{figure}
\normalsize
\centerline{\includegraphics[width=0.7\linewidth]{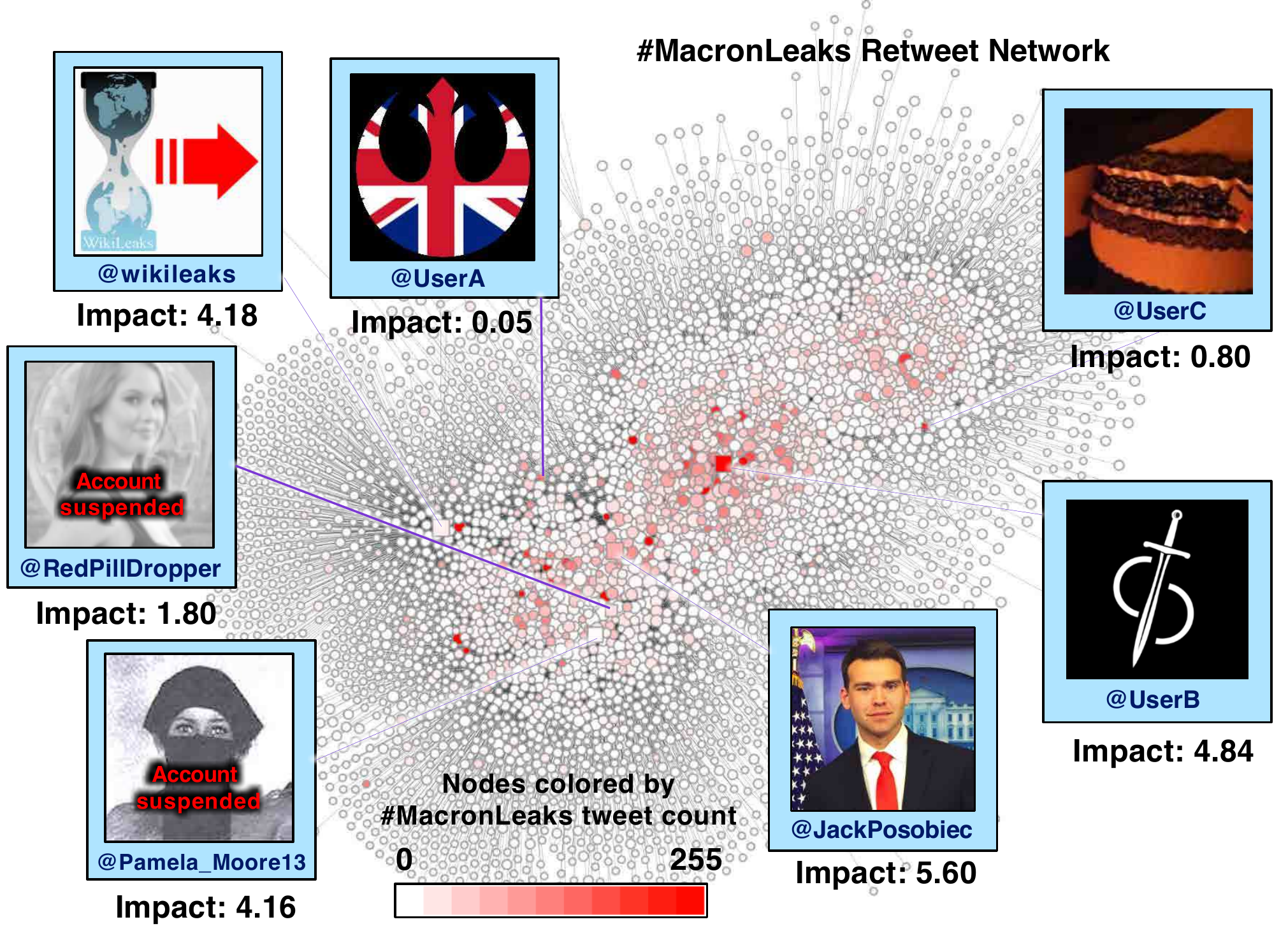}}
\caption{Causal impact of highlighted accounts on the \twitter{\#MacronLeaks} narrative
  network. Vertices are accounts and edges are retweets. Vertex color
  indicates the number of tweets and the vertex size corresponds to
  in-degrees. Causal impact is the average number of additional tweets
  generated by an user's participation ({\color{blue}{\em main paper}},
  {\color{blue}Eq.~[\protect\customxrref{eq:individual_impact_causal_estimand}]}). Image
  credits: Twitter\slash wikileaks, Twitter\slash RedPillDropper,
  Twitter\slash Pamela\underscore Moore13, Twitter\slash
  JackPosobiec.\label{fig:influencemacronleaks}}
\end{figure}

Further support of this framework's efficacy is provided by its
application to a well-known, highly visible, and distinctive IO
narrative network defined by a simple Twitter
hashtag, \twitter{\#MacronLeaks}~\cite{ferrara2017, marantz2017,
smith2018b} (Fig.~\ref{fig:influencemacronleaks}). Vertex color in
Fig.~\ref{fig:influencemacronleaks} indicates the number of times each
account tweets the hashtag \twitter{\#MacronLeaks}; vertex size is the
in-degree, i.e., the number of retweets received by each account.
Compared to the financial allegation narrative detected via topic
modeling in the {\em\href{\mainpaperDOI}{\color{blue}main paper}},
this narrative is on a related but more focused event of the leaking
of candidate Macron's emails. While many IO narratives do not align
nicely with a hashtag and need to be detected via topic modeling,
the \twitter{\#MacronLeaks} narrative is an instance where the hashtag
became a prominent signature for the narrative. The hashtag was widely
used in many tweets and retweets, as reflected by the higher magnitude
impact statistics in Table~\ref{tab:macronleaks_impact_estimation},
compared to Table~\customxrref{tab:impact_stats} in the main
paper. Among the accounts with high estimated impact, there exists
independent confirmation of the prominence of the two accounts
\twitter{@wikileaks} and \twitter{@JackPosobiec} in pushing the
\twitter{\#MacronLeaks} narrative~\cite{ferrara2017, marantz2017}. The relatively
high causal impact of these accounts shown in
Table~\ref{tab:macronleaks_impact_estimation} is consistent with
independently reported articles about this narrative.  Further, a new
finding is the high-impact spreading of this IO narrative by the known
IO~account \twitter{@Pamela\underscore Moore13}~\cite{USHPSCI2017a, USHPSCI2017b,
gadde2018}.

Comparison between impact statistics in
Table~\ref{tab:macronleaks_impact_estimation} highlights the advantage
of causal impact estimation in quantifying impact over activity- and
topologically based statistics such as retweet (RT) volume and
PageRank centrality~\cite{langville2005,peixoto2014a}. The limitation
of activity count statistics as indicators for impact is seen
in \twitter{@UserA} and \twitter{@UserC}, both having high tweet and
retweet counts but little impact due to their positions on and
connectivities to the network with low centrality. Follower count and
PageRank centrality clearly highlight \twitter{@wikileaks}'s impact on
the network, which is consistent with its high causal impact. However,
accounts such as \twitter{@JackPosobiec}, \twitter{@UserB}, and known
IO~account \twitter{@Pamela\underscore Moore13} have only a medium level of
PageRank centrality, but high causal
impact. \twitter{@JackPosobiec}, being one of the earliest
participant, has been reported as a key source in pushing
the \twitter{\#MacronLeaks}
narrative~\cite{ferrara2017, marantz2017}. \twitter{@UserB} serves as a bridge into
the predominantly French-speaking subgraph (the cluster seen in the
middle of Fig.~\ref{fig:influencemacronleaks}). Causal impact is able
to detect these prominent accounts that do not stand out in other
impact statistics by modeling the narrative propagation on the
network. Unlike existing propagation methods on network 
topology~\cite{smith2014} alone, causal inference accounts also for 
the observed outcomes at each node.

\subsection{Influential IO-like Content 2017 vs.\ 2020}
\label{sec:inflio20172020}

Many of the high-impact accounts with behaviors and content similar to
known IO~accounts (upper-right corner of
{\em\href{\mainpaperDOI}{\color{blue}main paper}}, {\color{blue}
Fig.~\customxrref{fig:en_scatterplot}}) have been suspended by Twitter
since~2017.  However, several remain actively engaged at the time of
writing in narratives also used by IO~accounts
in~2020~\cite{donati2020, donovan2020, kao2020}. For example,
high-impact, IO-like accounts that posted on the 2017 French election
narrative network ({\color{blue}{\em main paper}},
{\color{blue}Figs.\ \customxrref{fig:wordcloud_enfr}
and~\customxrref{fig:fr_classifier}}) are actively posting three years
later on COVID\hbox{-}19 conspiracy theories. Additional validation of
this paper's approach is suggested by examining how the content of
such active accounts has been used by known IO~accounts. Inspecting a
small-but-representative sample of $3$ active accounts appearing in
the upper-right corner of {\em\href{\mainpaperDOI}{\color{blue}main
paper}}, {\color{blue}Fig.~\customxrref{fig:en_scatterplot}} shows
that content from these accounts has been used by known IO~accounts
hundreds of times ($9$ in one case, $472$ in another, and $589$ in
another) for each of these active accounts~\cite{gadde2018}.
Furthermore, we observe that our classifier is robust to whether or
not these accounts are directly retweeted by known IO~accounts.  Only
$5$~tweets from these $3$~representative accounts ($1$, $1$, and~$3$
tweets for~each account) were included in our classifier as ``positive''
examples used by known IO~accounts, whereas $555$~tweets (total for
all~$3$) were retweeted by accounts included as negative examples in
the training data. Because the classifier training data is comprised
of $40$~million tweets, the presence of $5$ tweets from these specific
accounts does not have a large effect on their positive
classification, especially that their content appears $555$~times in
non-IO ``negative'' training data. Finally, note that we do not assert
or imply that these are IO~accounts, but merely observe that their
content and behavior is quantifiably similar to known IO~accounts,
independent of their authenticity.  This observation is consistent
with the fact that content from these specific accounts has also been
used hundreds of times (as retweets) by known
IO~accounts~\cite{gadde2018, linvill2020}.


\begin{table}
\newcommand\TopStrut{\rule{0pt}{2.6ex}}       
\newcommand\BotStrut{\rule[-1.2ex]{0pt}{0pt}} 
\def\minicenteredpp#1{\vtop{\hsize0.15\linewidth \parindent0pt \hangindent0pt \baselineskip\normalbaselineskip \begin{centering}\textsl{#1}\end{centering}}}
  \begin{center}
    \caption{Select English topics.}
    \label{tab:en_topics}
    \begin{tabular}{>{\quad\columncolor{twitterblue}}p{0.1667\linewidth} >{\quad}p{0.1667\linewidth} >{\quad\columncolor{twitterblue}}p{0.1667\linewidth} >{\quad}p{0.1667\linewidth}} 
\hline 
\rowcolor{white}
\multicolumn{1}{c}{\textbf{Topic 1}} & \multicolumn{1}{c}{\textbf{Topic 2}} & \multicolumn{1}{c}{\textbf{Topic 3}} & \multicolumn{1}{c}{\textbf{Topic 4}\TopStrut\BotStrut} \\
\rowcolor{white}
\multicolumn{1}{c}{\minicenteredpp{Anti-Macron, financial allegations\BotStrut}}
 & \multicolumn{1}{c}{\minicenteredpp{May 1st protest in Paris\BotStrut}} & \multicolumn{1}{c}{\minicenteredpp{Brexit\BotStrut}}
 & \multicolumn{1}{c}{\minicenteredpp{Russia and Middle East\BotStrut}}\\
\hline
macron & police & brexit & syria\TopStrut \\
tax & antifa &  news & russia \\
documents & paris & win & sptnkne.ws \\
emmanuel & day & amp & syrian \\
evasion & mayday2017 & twitter.com & isis \\ 
engaging & on.rt.com & britain & russian \\
putin & protesters & brussels  & turkey \\
trump & breaking  & theresa & sputniknews.com \\
huge & video & ukip & military \\
disobedientmedia.com & anti & labour & amp \\
on.rt.com & amp & macron & attack \\
prove & arrested & election & army \\
busted & officers & europe & war \\
news & news & pen & israel \\
cheat & violent & on.rt.com & news \\
rally & violence & ge2017 & putin \\
breaking & texas & telegraph.co.uk & killed \\
phone & portland & juncker & on.rt.com  \\
harrisburg & left & :united\underscore kingdom: & ukraine \\
nato & pen & politics & nato\BotStrut \\
\hline
    \end{tabular}
  \end{center}
\end{table}


\begin{table}
\newcommand\TopStrut{\rule{0pt}{2.6ex}}       
\newcommand\BotStrut{\rule[-1.2ex]{0pt}{0pt}} 
\def\minicenteredpp#1{\vtop{\hsize0.15\linewidth \parindent0pt \hangindent0pt \baselineskip\normalbaselineskip \begin{centering}\textsl{#1}\end{centering}}}
  \begin{center}
    \caption{Select French topics.}
    \label{tab:fr_topics}
    \begin{tabular}{>{\quad\columncolor{twitterblue}}p{0.1667\linewidth} >{\quad}p{0.1667\linewidth} >{\quad\columncolor{twitterblue}}p{0.1667\linewidth} >{\quad}p{0.1667\linewidth}} 
\hline 
\rowcolor{white}
\multicolumn{1}{c}{\textbf{Topic 1}} & \multicolumn{1}{c}{\textbf{Topic 2}} & \multicolumn{1}{c}{\textbf{Topic 3}} & \multicolumn{1}{c}{\textbf{Topics 4}\TopStrut\BotStrut} \\
\rowcolor{white}
\multicolumn{1}{c}{\minicenteredpp{Anti-Macron, financial allegations\BotStrut}}
 & \multicolumn{1}{c}{\minicenteredpp{Final debate of 2017 election\BotStrut}} & \multicolumn{1}{c}{\minicenteredpp{Police violence at May 1st protests\BotStrut}}
 & \multicolumn{1}{c}{\minicenteredpp{Islamophobia during election\BotStrut}}\\
\hline
macron & macron & paris & macron\TopStrut \\
compte & d\'{e}bat & crs & uoif \\
journaliste & 2017led\'{e}bat & policiers & youtube.com \\
fiscale & marine & gauche & watch \\
plainte & pen & mai & islamistes \\
offshore & debat2017 & 1er & voter \\
:red\underscore circle: & soir & bless\'{e}s & soutenu \\
mort & france & 1ermai & jamaismacron \\
documents & plateau &extr\^{e}me & france \\
macrongate & menace & policier & musulmans \\
emmanuel & quitter & macron & islamiste \\
porte & 2017led\'{e}bat & br\^{u}l\'{e} & soutien \\
twitter.com & debat & police & :france: \\
bahamas & brigitte & france & emmanuel \\
\'{e}vasion & mlp & :france: & jamais \\
 preuves & lepen & molotov & :thumbs\underscore down: \\
 france & tours & violences & juifs \\
 pen & faire & th\'{e}o & fr\`{e}res \\
 fraude & bout & ordre & lyon \\
 soci\'{e}t\'{e} & heure & cgt & 2017l\'{e}debat\BotStrut \\
 \hline
    \end{tabular}
  \end{center}
\end{table}

\begin{table}
\caption{IO~account heuristics~\cite{costa-roberts2018, im2019, zannettou2019, luceri2020} as Snorkel labeling functions.}
\label{tab:snorkel_labels}
\hrule
\vskip6pt
\beginpython
# Account has no profile description
@labeling_function()
def profile_length(x):
    return IO if x.profile_length==0 else ABSTAIN

# Account has repeated interactions with external news sites
@labeling_function()
def external_news__interactions(x):
    return IO if x.num_external__news_interactions > 5 else ABSTAIN

# Account follows large number of accounts
@labeling_function()
def num_following(x):
    return IO if x.following_count > 3000 else ABSTAIN

# On average, every tweet includes a link
@labeling_function()
def num_links(x):
    return IO if x.avg_num_links > 1 else ABSTAIN

# Account tweeted in many languages
@labeling_function()
def many_langs(x): 
    return IO if np.count_nonzero(x.num_langs_used) > 10 else ABSTAIN

# Account rarely favorited tweets
@labeling_function()
def few_faves(x):
    return IO if x.num_faves <20 else ABSTAIN

# Account favorited large number of tweets
@labeling_function()
def too_many_faves(x):
    return IO if x.num_faves >30000 else ABSTAIN

#Account tweeted in an undetermined language often
@labeling_function()
def many_und_tweets(x):
    return IO if x.und > 0.05 else ABSTAIN

# Account has follow, following counts indicative of real user
@labeling_function()
def normal_people_ff_ratio(x):
    return REAL if x.follower_count < 500 and 0.75 < x.followers_following_ratio < 4 else ABSTAIN

# Account interacted with external news source once or never
@labeling_function()
def no_external_news__interactions(x):
    return REAL if x.num_news__news_interactions < 2 else ABSTAIN

#Account seldom included links in tweets
@labeling_function()
def few_tweets_w_links(x):
    return REAL if 0.05 < x.ratio_tweets_w_links_all_tweets < 0.15 else ABSTAIN

# Number of likes by account in normal range
@labeling_function()
def normal_num_likes(x):
    return REAL if 500 < x.num_faves < 10000 else ABSTAIN

# Profile description of a normal length
@labeling_function()
def normal_profile_len(x):
    return REAL if x.profile_length > 50 else ABSTAIN

# Many legitimate organizations have large number of followers, don't want to classify them as IO~accounts
@labeling_function()
def org_num_followers(x):
    return REAL if x.follower_count > 60000 else ABSTAIN
$endpython
\vskip6pt
\hrule
\end{table}

\begin{table}
\newcommand\TopStrut{\rule{0pt}{2.6ex}}       
\newcommand\BotStrut{\rule[-1.2ex]{0pt}{0pt}} 
  \begin{center}
    \caption{Proportion of training data subsets labeled as IO~accounts by Snorkel heuristics.}
    \label{tab:snorkel_fracs}
    \begin{tabular}{l *{3}{r<{\qquad}}} 
\hline
 & \multicolumn{1}{c}{50\% threshold} & \multicolumn{1}{c}{70\% threshold} & \multicolumn{1}{c}{90\% threshold\TopStrut\BotStrut} \\
\hline
\rowcolor{twitterblue}French election, English & 38\% & 32\% & 23\%\TopStrut\BotStrut \\
French election, French & 34\% & 28\% & 22\%\TopStrut\BotStrut \\
\rowcolor{twitterblue}Topic and language neutral & 30\% & 15\% & 8\%\TopStrut\BotStrut \\
\hline
    \end{tabular}
  \end{center}
\end{table}

\begin{table}
\newcommand\TopStrut{\rule{0pt}{2.6ex}}       
\newcommand\BotStrut{\rule[-1.2ex]{0pt}{0pt}} 
  \begin{center}
    \caption{Sensitivity comparison, Snorkel thresholds for IO labeling and cross validation set selection.}
    \label{tab:snorkel_sensitivity}
    \begin{tabular}{l *{4}{r<{\hquad}}} 
\hline
& \multicolumn{1}{c}{Precision} & \multicolumn{1}{c}{Recall} & \multicolumn{1}{c}{AUPRC} & \multicolumn{1}{c}{EER\TopStrut\BotStrut} \\
\hline
\rowcolor{twitterblue} 50\% threshold, all data & 94.3\% & 78.7\% & 95.8\% & 10.5\%\TopStrut \\
50\% threshold, omit Snorkel positives & 88.2\% & 91.6\% & 96.9\% & 6.1\% \\
\rowcolor{twitterblue} 70\% threshold, all data & 95.6\% & 78.9\% & 96.4\% & 8.1\% \\
70\% threshold, omit Snorkel positives & 91.6\% & 90.3\% & 96.6\% & 5.6\% \\
\rowcolor{twitterblue} 90\% threshold, all data & 97.1\% & 73.5\% & 96.0\% & 7.6\% \\
90\% threshold, omit Snorkel positives & 95.8\% & 90.5\% & 97.4\% & 5.0\%\BotStrut \\
\hline
    \end{tabular}
  \end{center}
\end{table}

\begin{table}
\newcommand\TopStrut{\rule{0pt}{2.6ex}}       
\newcommand\BotStrut{\rule[-1.2ex]{0pt}{0pt}} 
  \begin{center}
    \caption{Behavioral features in descending importance.}
    \label{tab:num_feats}
    \begin{tabular}{>{\quad\columncolor{twitterblue}}p{0.25\linewidth} >{\quad}p{0.25\linewidth} >{\quad\columncolor{twitterblue}}p{0.25\linewidth} } 
    \hline
num\underscore external\underscore news\underscore interactions & num\underscore tweets\underscore in\underscore time\underscore range & follower\underscore count\TopStrut \\
avg\underscore num\underscore chars & sd\underscore num\underscore tweets\underscore per\underscore day & profile\underscore length \\
avg\underscore num\underscore hashtag\underscore chars & ratio\underscore retweets\underscore w\underscore links\underscore all\underscore tweets & \\
num\underscore faves\BotStrut & avg\underscore num\underscore tweets\underscore per\underscore day & \\
\hline
    \end{tabular}
  \end{center}
\end{table}

\begin{table}
  \begin{center}
    \caption{Language features in descending importance (Twitter
    language codes).}
    \label{tab:langs_feats}
    \begin{tabular}{l b l b l b} 
    \hline
en & de & ht & nl & pl & eu \\
sr & fr & tl & cs & no & ca \\
it & sl & ro & da & lt & fi \\
und & pt & ru & sv & lv & hu \\
in & es & et & tr & cy & hi \\
\hline
    \end{tabular}
  \end{center}
\end{table}

\begin{table}
\newcommand\TopStrut{\rule{0pt}{2.6ex}}       
\newcommand\BotStrut{\rule[-1.2ex]{0pt}{0pt}} 
  \begin{center}
    \caption{Top 150 1- and 2-grams* (out of 500) in descending order of importance.}
    \label{tab:words_feats}
    \begin{tabular}{>{\hquad\columncolor{twitterblue}}p{0.25\linewidth} >{\hquad}p{0.25\linewidth} >{\hquad\columncolor{twitterblue}}p{0.25\linewidth}} 
    \hline
en:serbia & en:serbia\textvisiblespace austria & en:investigation\TopStrut \\
en:macron & en:vu\v{c}i\'{c}& en:nursultan \\
fr:lepen & en:aleksandar\textvisiblespace vu\v{c}i\'{c}& en:friends \\
en:france & en:serbian\textvisiblespace president & en:18 \\
fr:macron & en:campaign & en:jeff\textvisiblespace sessions \\
en:president\textvisiblespace serbia & en:mueller & en:nursultan\textvisiblespace nazarbayev \\
en:lepen & en:country & en:smarttraffic \\
fr:france & en:government & en:jeff \\
en:belgrade & en:germany & en:robert\textvisiblespace mueller \\
fr:pen & en:sessions & en:foreign \\
en:serbian & en:attack & en:euro\textvisiblespace atlantic \\
en:serbia\textvisiblespace amp & en:israel & en:follow \\
en:aleksandar & en:meeting & en:terrorist \\
fr:marine & en:presidential & en:beautiful \\
en:election & en:merkel & en:girl \\
en:forum & en:aleksandar\textvisiblespace vucic & en:palace\textvisiblespace serbia \\
fr:fillon & en:nazarbayev\textvisiblespace 55qsospayr & en:close \\
en:le & fr:merkel & en:congress \\
en:president & en:strike & en:presidentielle2017 \\
fr:emmanuel & en:night & en:probe \\
fr:emmanuel\textvisiblespace macron & en:exercise\textvisiblespace organised & en:organised\textvisiblespace nato \\
en:french & en:55qsospayr & en:countries \\
en::serbia: & en:report & en:firing \\
en:trump & fr:lepen\textvisiblespace macron & en:arrested \\
en:amp & en:team & en:china \\
en:le\textvisiblespace pen & en:american & en:press \\
en:pen & en:french\textvisiblespace election & en:missile \\
en:vote & en:fbi & en:khashoggi \\
en:people & en:life & en:visit \\
en:2 & en:emmanuel & en:presidential\textvisiblespace election \\
en:discussed & en:watch & en:syrian \\
fr:faut & en:emmanuel\textvisiblespace macron & en:nazarbayev \\
en:video & en:leaks & en:congress\textvisiblespace participants \\
en:day & en:war & en:minister \\
fr:macron\textvisiblespace lepen & en:hosting\textvisiblespace largest & en:heart \\
en:vucic & en:special & en:proof \\
en:win & en:nato\textvisiblespace euro & en:citizens \\
en:business\textvisiblespace forum & en:participants\textvisiblespace president & en:prince \\
en:austria & en:photo\textvisiblespace congress & en:share \\
fr:parti & en:atlantic\textvisiblespace di & en:crisis \\
en:russian & en:family & en:family\textvisiblespace photo \\
en:breaking & en:iran & en:president\textvisiblespace nursultan \\
en:syria & en:killed & en:turkey \\
en:obama & en:children & en:business \\
en:news & en:emails & en:participants \\
en:love & en:largest\textvisiblespace disaster & en:happening \\
en:support & en:kosovo & en:disaster\textvisiblespace response \\
en:live & en:robert & en:including \\
en:eu & en:saudi & en:france\textvisiblespace macron \\
en:amp\textvisiblespace hosting\BotStrut & en:attacks & fr:cqfd \\
\hline
\setarstrut{\footnotesize}%
\multicolumn{3}{>{\footnotesize}r}{\mathsffamily *$2$-grams are represented by two
    words separated by the visible space character `\textvisiblespace'.\TopStrut} \\
\restorearstrut
    \end{tabular}
  \end{center}
\end{table}

\begin{table}
\newcommand\TopStrut{\rule{0pt}{2.6ex}}       
\newcommand\BotStrut{\rule[-1.2ex]{0pt}{0pt}} 
  \begin{center}
    \caption{Top 36 classifier features* without Snorkel,
    descending order of importance. Note that classifier overfitting
    without Snorkel labeling functions is apparent from the prevalence
    of Serbian-related features that correspond to the dominance of
    known IO accounts in our training data from Twitter's Serbian
    dataset (Fig.~\ref{fig:origin-pie-chart}).}
    \label{tab:no_snorkel_classifier_top_features}
    \begin{tabular}{>{\hquad\columncolor{twitterblue}}p{0.25\linewidth} >{\hquad}p{0.25\linewidth} >{\hquad\columncolor{twitterblue}}p{0.25\linewidth}} 
    \hline
lang:sr & en:macron & en:business\textvisiblespace forum\TopStrut \\
num\_tweets\_in\_time\_range & en::serbia: & en:forum \\
following\_count & en:france & en:vu\v{c}i\'c \\
num\_faves & en:serbian & en:discussed \\
en:serbia & follower\_count & fr:lepen \\
sd\_num\_tweets\_per\_day & en:serbia\textvisiblespace amp & en:serbian\textvisiblespace president \\
lang:sl & lang:fr & fr:macron \\
lang:und & avg\_num\_chars & en:vucic \\
en:president\textvisiblespace serbia & en:aleksandar\textvisiblespace vu\v{c}i\'c & lang:es \\
ratio\_retweets\_w\_links\_all\_tweets & en:belgrade & en:nursultan \\
ratio\_retweets\_all\_tweets & en:aleksandar & profile\_length \\
lang:en\BotStrut & avg\_num\_tweets\_per\_day & en:election \\
\hline
\setarstrut{\footnotesize}%
\multicolumn{3}{>{\footnotesize}r}{\mathsffamily *$2$-grams are represented by two
    words separated by the visible space character `\textvisiblespace'.\TopStrut} \\
\restorearstrut
    \end{tabular}
  \end{center}
\end{table}

\begin{table}
\newcommand\TopStrut{\rule{0pt}{2.6ex}}       
\newcommand\BotStrut{\rule[-1.2ex]{0pt}{0pt}} 
\def\minicenteredpp#1{\vtop{\hsize0.15\linewidth \parindent0pt \hangindent0pt \baselineskip\normalbaselineskip \begin{centering}\textsl{#1}\end{centering}}}
  \begin{center}
    \caption{Select French topics from the three Macron allegations communities.}
    \label{tab:fr_topics_antiMacron}
    \begin{tabular}{>{\quad\columncolor{twitterblue}}p{0.1667\linewidth} >{\quad}p{0.1667\linewidth} >{\quad\columncolor{twitterblue}}p{0.1667\linewidth} >{\quad}p{0.1667\linewidth}} 
\hline 
\rowcolor{white}
\multicolumn{1}{c}{\textbf{Topic A}} & \multicolumn{1}{c}{\textbf{Topic B}}& \multicolumn{1}{c}{\textbf{Topic C}\TopStrut\BotStrut} \\
\rowcolor{white}
\multicolumn{1}{c}{\minicenteredpp{Macron, police violence at May 1st protests\BotStrut}}
 & \multicolumn{1}{c}{\minicenteredpp{Macron, immigration and Islam\BotStrut}} & \multicolumn{1}{c}{\minicenteredpp{Macron, financial allegations\BotStrut}} \\
\hline
macron & macron  & macron \TopStrut\\
policiers & france  & france \\
gauche & fran\c{c}ais & soutien \\
paris &oradour & marie \\
crs & marine & emmanuel \\ 
france & shoah  & uoif \\
extr\^{e}me & loi & garaud \\
mai & veut & marine \\
1er & 2017led\'{e}bat & pen \\
bless\'{e}s & fronti\`{e}res & voter \\
1ermai & faire &  compte \\
police & emmanuel & porte \\
policier & guerre & candidat \\
br\^{u}l\'{e} & campagne & plainte \\
twitter.com & immigration & fiscale \\
ordre & ans & appelle \\
guillon &  islamiste &  islamistes \\
mort & travail & hollande \\
:france: & \'{e}trangers &  grande \\
forces & vote & 2017led\'{e}bat \BotStrut\\
 \hline
    \end{tabular}
  \end{center}
\end{table}

\begin{table}
\newcommand\TopStrut{\rule{0pt}{2.6ex}}       
\newcommand\BotStrut{\rule[-1.2ex]{0pt}{0pt}} 
\def\minicenteredpp#1{\vtop{\hsize0.15\linewidth \parindent0pt \hangindent0pt \baselineskip\normalbaselineskip \begin{centering}\textsl{#1}\end{centering}}}
  \begin{center}
    \caption{Representative tweets of the Topics in Table~\ref{tab:fr_topics_antiMacron}.}
    \label{tab:macron_allegation_tweets}
    \begin{tabular}{>{\columncolor{twitterblue}\small}p{0.3\linewidth} >{\small}p{0.3\linewidth} >{\columncolor{twitterblue}\small}p{0.3\linewidth}} 
\hline 
\rowcolor{white}
\multicolumn{1}{c}{\textbf{Topic A}} & \multicolumn{1}{c}{\textbf{Topic B}}& \multicolumn{1}{c}{\textbf{Topic C}\TopStrut\BotStrut} \\
\rowcolor{white}
\multicolumn{1}{c}{\minicenteredpp{Macron, police violence at May 1st protests\BotStrut}}
 & \multicolumn{1}{c}{\minicenteredpp{Macron, immigration and Islam\BotStrut}} & \multicolumn{1}{c}{\minicenteredpp{Macron, financial allegations\BotStrut}} \\
\hline
\retweettext{\TopStrut Rottweiller83}{rottweiller83}{535820307}{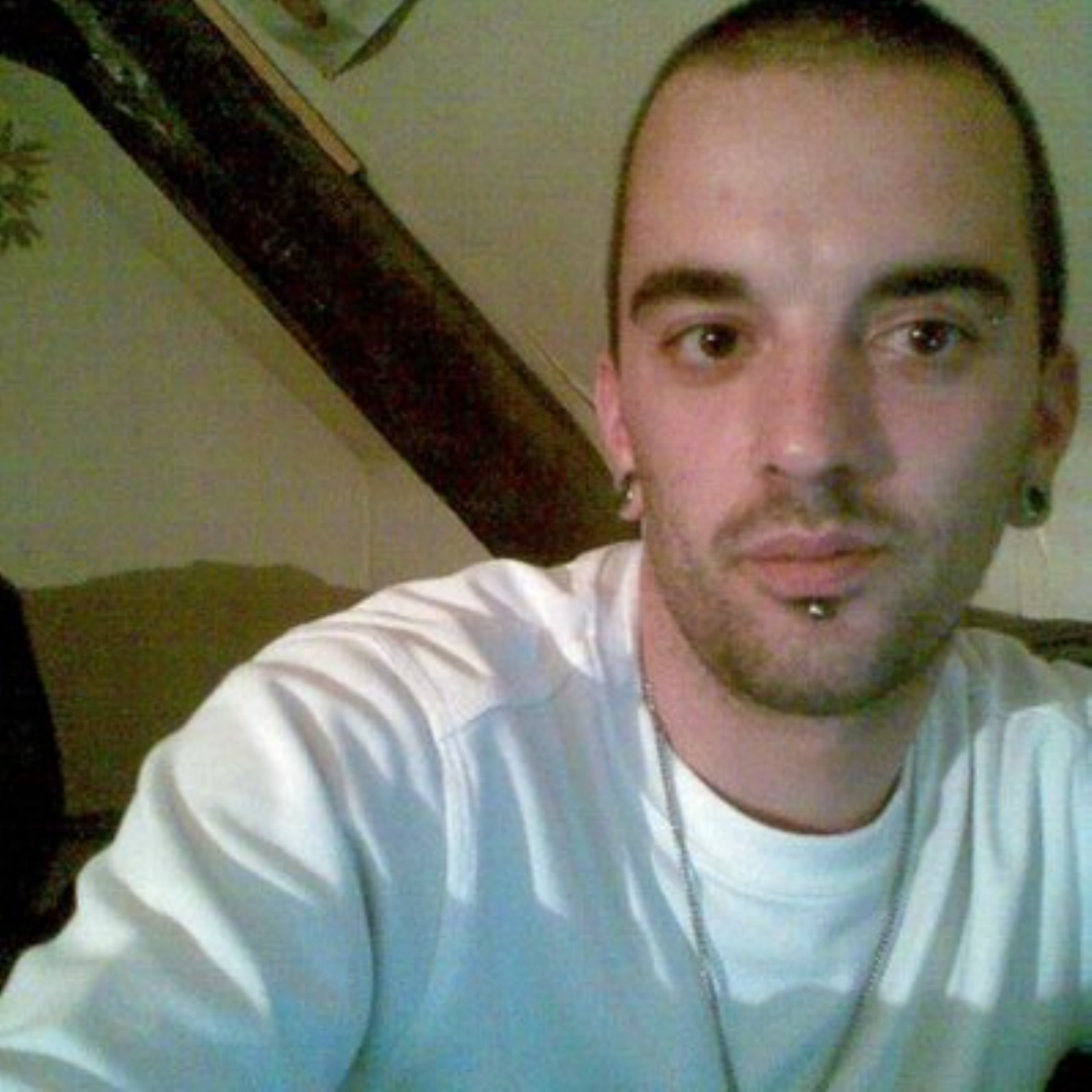}{5:26~AM - 28~April 2017} {Du mat\'{e}riel \twitterhlt{\#ToutSaufMacron} a \'{e}t\'{e} vol\'{e} lors d'une agression ultra violente de 2 jeunes par un gang pro- Macron. Nous allons porter plainte :france:} {}{} 
&
\tweettext{\TopStrut M-J :latin_cross: :star_of_david: :france:}{Cal_369}{799704104004100096}{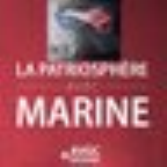}{2:20~PM - 30~April 2017} {@EmmanuelMacron Rothschild : appels de fond pour vous financer. Cach\'{e} au public ! \twitterhlt{\#MacronNon} \par
\strut\par \urlcite[http://www.valeursactuelles.com/politique/coulisses-politiques-quand-rothschild-sponsorise-macron-72133]} {}{} 
&
\retweettext{\TopStrut Alain Thomas}{AlainThomas1}{4856695311}{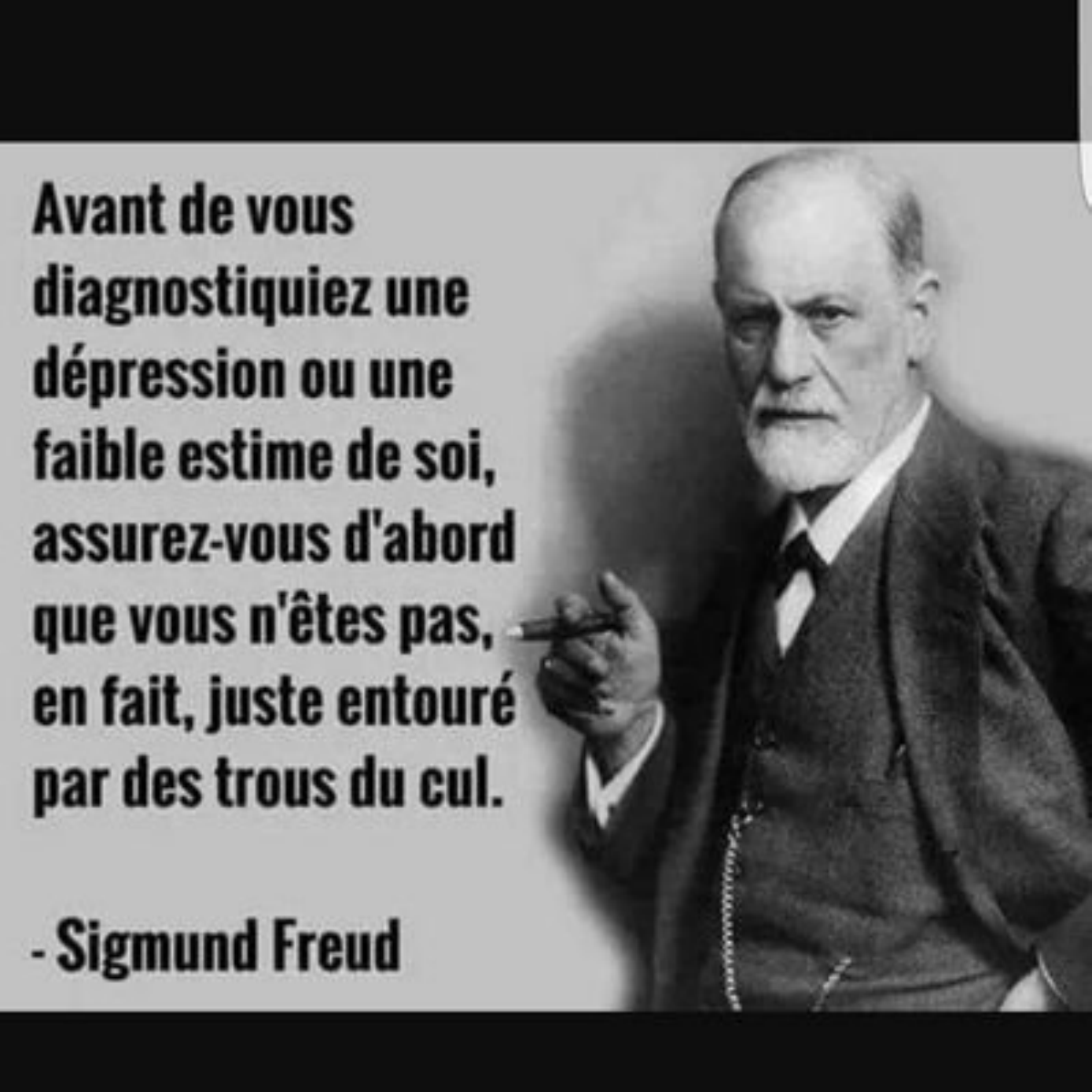}{1:15~PM- 28~April 2017} {Du tr\`{e}s tr\`{e}s lourd sur @Macron , banquier pourri , ``enflure bancaire'' financ\'{e} par Goldman Sachs ! Gravissime ! TWEET ET RETWEET} {}{} 
\\
\retweettext{Julia :france: :pig: :sun: :butterfly:}{mamititi31}{789035544021983232}{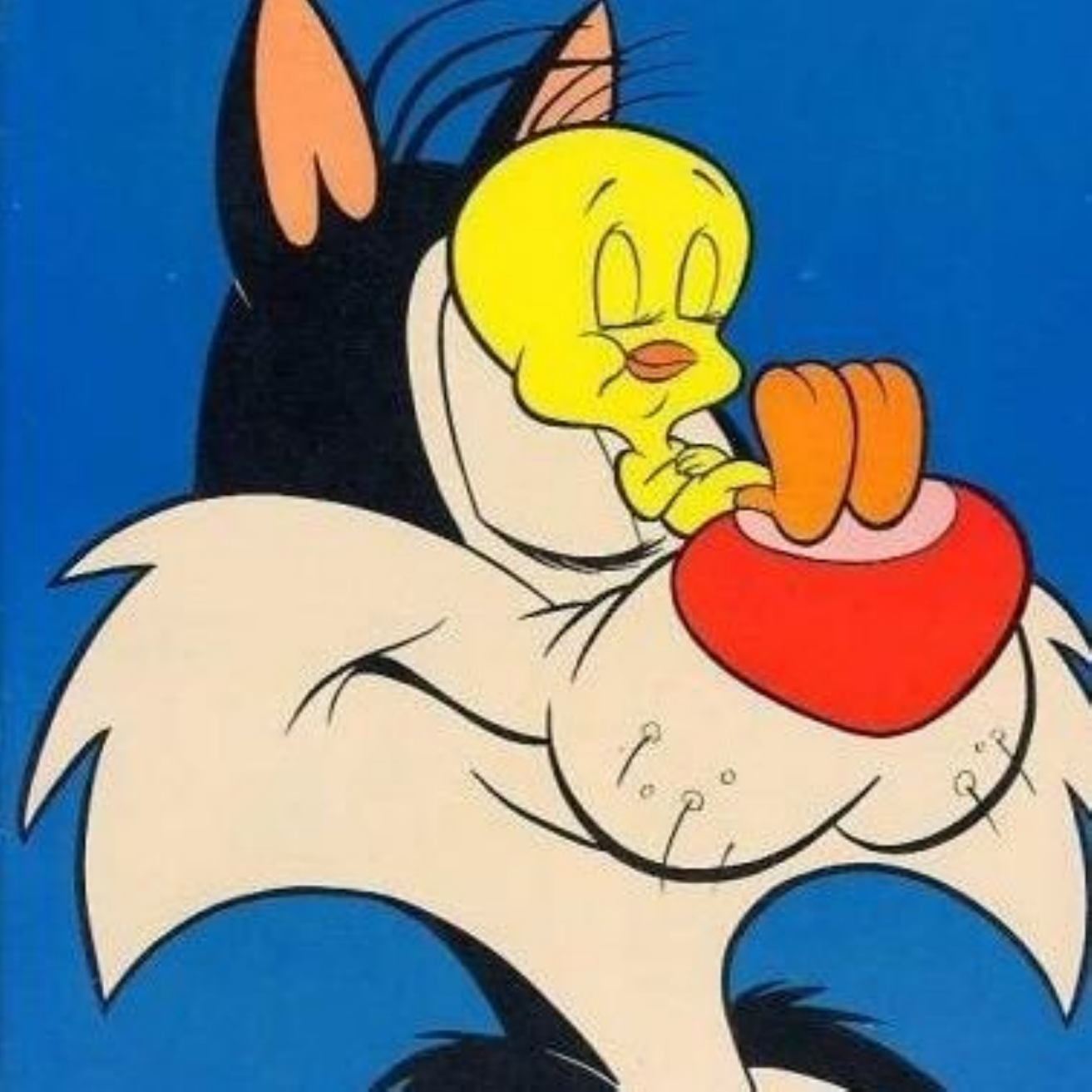}{7:33~PM - 1~May 2017} {Les vrais facistes sont ceux qui,\`{a} l'extr\^{e}me gauche,manifestent et cassent en ce moment,refusant le r\'{e}sultat du 1er tour des pr\'{e}sidentielles} {}{} 
&
\retweettext{l'oranaise}{L_oranaise_}{2983945431}{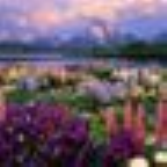}{4:09~PM - 4~May 2017} {voil\`{a} ce qui nous attend avec \twitterhlt{\#Macron} et apr\`{e}s ca voile obligatoire pour les femmes et jeunes filles? :angry\underscore face: :angry\underscore face: :angry\underscore face:  } {}{} 
&
\retweettext{l'oranaise}{Pascal Azoulay}{768475921112104960}{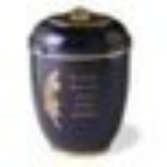}{8:25~AM - 4~May 2017} {\#MacronGates : Voici 1 copie pour un ch\`{e}que d\'{e}pos\'{e} par Macron dans la banque Nevis as an Offshore Asset Prot C'est bien } {}{} 
\\ 
\retweettext{Frexit_2017}{avril_sylvie}{775389437672947717}{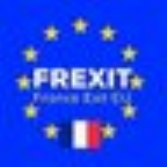}{7:33~PM - 1~May 2017} {La porte-parole de \twitterhlt{\#Macron} appelle \`{a} la violence contre Marine et nos forces de l'ordre. Effarant. \par
\strut\par \centerline{\includegraphics[width=0.4\linewidth]{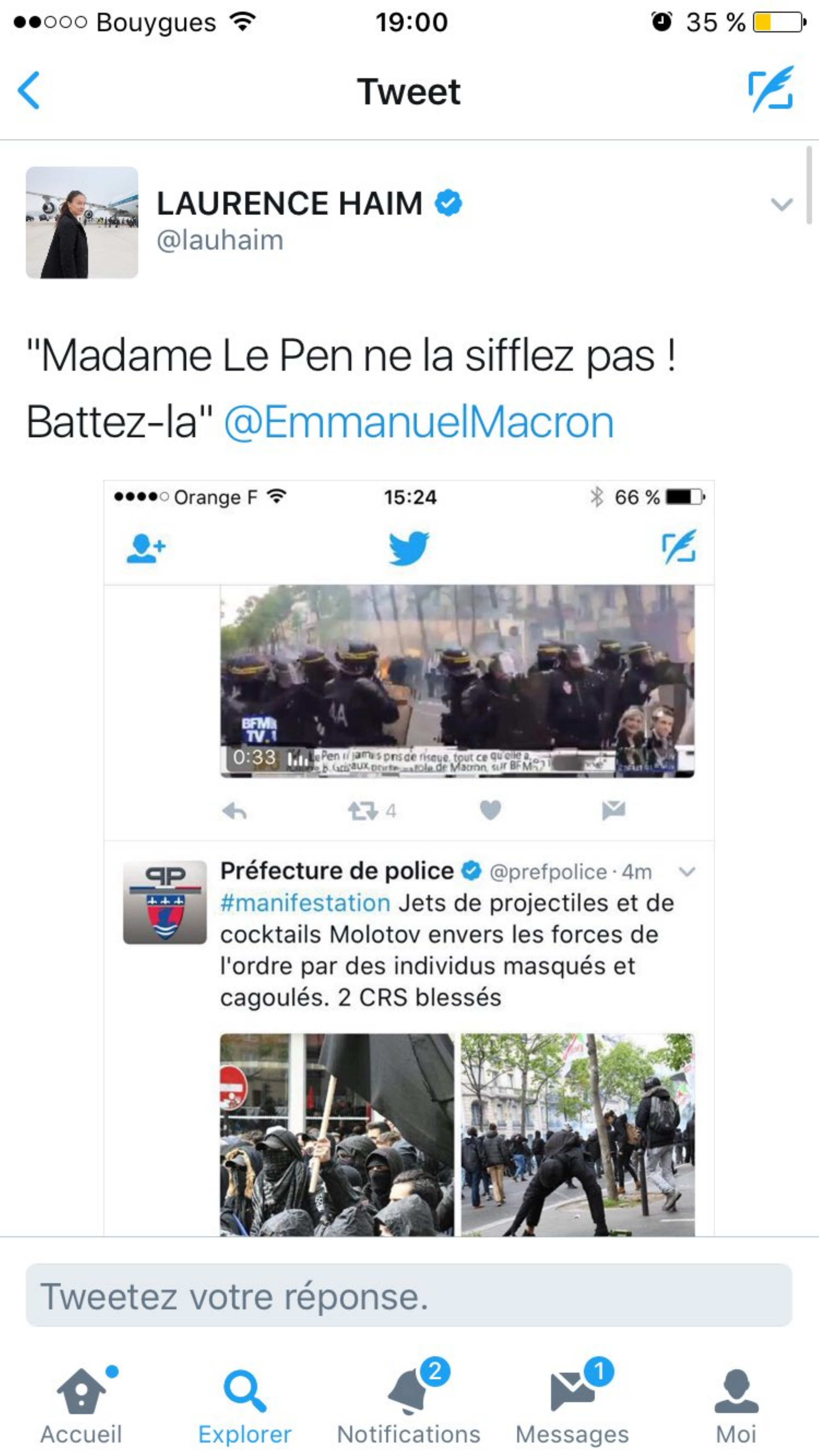}}} {}{}
&
\retweettext{MAXIMUS DECIMUS}{lorquaphilip}{304858372}{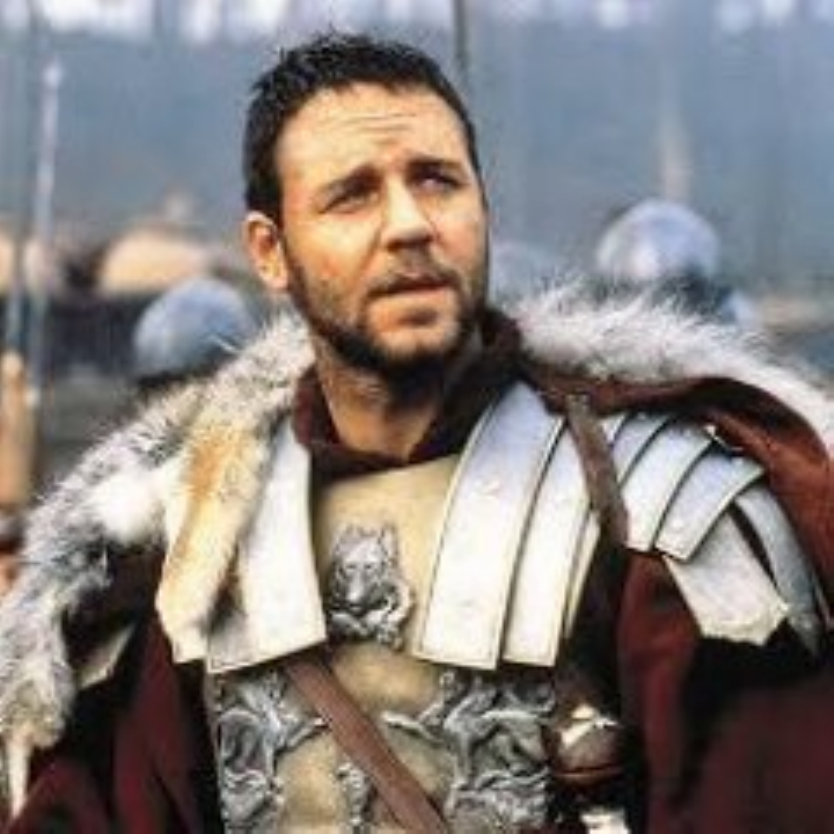}{7:33~PM - 1~May 2017} {Apr\`{e}s avoir ni\'{e} l'existence de la culture fran\c{c}aise, Macron nie d\'{e}sormais l'existence de la France.} {}{} 
&
\tweettext{M-J :latin_cross: :star_of_david: :france:}{Cal_369}{799704104004100096}{Cal_369_profile.pdf}{1:00~AM - 4~May 2017} {@JackPosobiec \twitterhlt{\#2017LeDebat} C'est donc l\`{a} que sont fric non d\'{e}clar\'{e} \`{a} pris la fuite ! \par
\strut\par\urlcite[https://www.les-crises.fr/emmanuel-macron-36-millions-deuros-de-revenus-cumules-patrimoine-negatif/]} {}{} \\ 
 \hline
    \end{tabular}
  \end{center}
\end{table}

\begin{table}
\newcommand\TopStrut{\rule{0pt}{2.6ex}}       
\newcommand\BotStrut{\rule[-1.2ex]{0pt}{0pt}} 
\def\minicenteredpp#1{\vtop{\hsize0.15\linewidth \parindent0pt \hangindent0pt \baselineskip\normalbaselineskip \begin{centering}\textsl{#1}\end{centering}}}
  \begin{center}
    \caption{Select French topics from the pro-Macron/anti-Le Pen/anti-abstention community.}
    \label{tab:fr_topics_proMacron}
    \begin{tabular}{>{\quad\columncolor{twitterblue}}p{0.1667\linewidth} >{\quad}p{0.1667\linewidth} >{\quad\columncolor{twitterblue}}p{0.1667\linewidth} >{\quad}p{0.1667\linewidth}} 
\hline 
\rowcolor{white}
\multicolumn{1}{c}{\textbf{Topic A}} & \multicolumn{1}{c}{\textbf{Topic B}}& \multicolumn{1}{c}{\textbf{Topic C}\TopStrut\BotStrut} \\
\rowcolor{white}
\multicolumn{1}{c}{\minicenteredpp{pro-Macron and anti-Le Pen\BotStrut}}
 & \multicolumn{1}{c}{\minicenteredpp{pro-Macron and anti-abstention \BotStrut}} & \multicolumn{1}{c}{\minicenteredpp{Final election debate: lead-up and event \BotStrut}} \\
\hline
ensemble & macron & pen \TopStrut\\
france & voter & 2017ledebat \\
veux & vote & marine \\
macron & pen & macron \\
projet & tour & lepen \\
europe & appelle & d\'{e}bat \\
r\'{e}publique & m\'{e}lenchon & programme \\
pays & faire & 2017led\'{e}bat \\
national & france & euro \\
fran\c{c}ais & insoumis & mlp \\
mai & blanc & debat2017 \\
jevotemacron & abstention & mme \\
politique & mai & projet \\
macrondirect & marine & soir \\
libert\'{e} & lepen & jevotemacron \\
:france: & faut & jamais \\
porte & emmanuel & france \\
macronpresident & 1er & madame \\
2017led\'{e}bat & twitter.com & faire \\
jt20h & dimanche & twitter.com  \BotStrut\\
 \hline
    \end{tabular}
  \end{center}
\end{table}
%
%
%
%
%

\begin{table}
\newcommand\TopStrut{\rule{0pt}{2.6ex}}       
\newcommand\BotStrut{\rule[-1.2ex]{0pt}{0pt}} 
\def\minicenteredpp#1{\vtop{\hsize0.15\linewidth \parindent0pt \hangindent0pt \baselineskip\normalbaselineskip \begin{centering}\textsl{#1}\end{centering}}}
  \begin{center}
    \caption{Select French topics from the pro-abstention community.}
    \label{tab:fr_topics_abstention}
    \begin{tabular}{>{\quad\columncolor{twitterblue}}p{0.1667\linewidth} >{\quad}p{0.1667\linewidth} >{\quad\columncolor{twitterblue}}p{0.1667\linewidth} >{\quad}p{0.1667\linewidth}} 
\hline 
\rowcolor{white}
\multicolumn{1}{c}{\textbf{Topic A}} & \multicolumn{1}{c}{\textbf{Topic B}}& \multicolumn{1}{c}{\textbf{Topic C}\TopStrut\BotStrut} \\
\rowcolor{white}
\multicolumn{1}{c}{\minicenteredpp{Support for Jean-Luc M\'{e}lenchon \BotStrut}}
 & \multicolumn{1}{c}{\minicenteredpp{Support of abstention\BotStrut}} & \multicolumn{1}{c}{\minicenteredpp{Criticism of  Macron and Le Pen\BotStrut}} \\
\hline
m\'{e}lenchon & twitter.com & macron \TopStrut\\
jean & macron & pen \\
france & faire & marine \\
mai & voter & utm \\
insoumis & vote & emmanuel \\
franceinsoumise & tour & article \\
1er & m\'{e}lenchon & source \\
l\'{e}gislatives & insoumis & fillon \\
1ermai & sansmoile7mai & perdu \\
insoumise & pen & d\'{e}bat \\
tour & faut & politique \\
luc & lepen & discours \\
youtu.be & oui & france \\
rdls26 & jlm & twitter \\
paris & mlp & candidat \\
youtube.com & veut & campaign \\
melenchon & abstention & social \\
legislatives2017 & blanc & medium \\
watch & \'{e}lecteurs & lafarge \\
pr\'{e}sidentielle & voix & bit.ly \BotStrut\\
 \hline
    \end{tabular}
  \end{center}
\end{table}

\begin{table}
\centering
\setlength{\tabcolsep}{0.15em}
\newcommand\TopStrut{\rule{0pt}{2.6ex}}       
\newcommand\BotStrut{\rule[-1.2ex]{0pt}{0pt}} 
\newcolumntype{S}{>{\small}r}
\newcolumntype{T}{>{\small\let\twittersize\footnotesize}l}
\caption{Comparison of impact statistics between accounts on the \twitter{\#MacronLeaks} narrative network:
  tweets (T), total retweets (TRT), most retweeted tweet (MRT), followers (F), first tweet time on
  5~May, PageRank centrality (PR), and causal impact* (CI).}
\label{tab:macronleaks_impact_estimation}
\begin{tabular*}{0.7\linewidth}{T@{\extracolsep{\fill}}SSSSSSS}
\hline
\multicolumn{1}{l}{Screen name} & \multicolumn{1}{c}{\ T} &
  \multicolumn{1}{c}{TRT} & \multicolumn{1}{c}{MRT} &
  \multicolumn{1}{c}{\quad F} &
  \multicolumn{1}{c}{1st time} &
  \multicolumn{1}{c}{\quad PR} &
  \multicolumn{1}{r}{\bf CI*\TopStrut\BotStrut}\\
\hline
\Rowcolor{twitterblue}\twitter{@JackPosobiec} & 95 & 47k & 5k & 261k & 18:49 & 667 &\bf 5.60\TopStrut \\
\twitter{@RedPillDropper} & 32 & 8k & 3k & 8k & 19:33 & 44 &\bf 1.80 \\
\Rowcolor{twitterblue}\twitter{@UserA}\dag & 256 & 59k & 8k & 1k & 19:34 & 7 &\bf 0.05 \\
\twitter{@UserB}\dag & 260 & 54k & 8k & 3k & 20:25 & 424 &\bf 4.84 \\
\Rowcolor{twitterblue}\twitter{@wikileaks} & 25 & 63k & 7k & 5515k & 20:32 & 9294 &\bf 4.18 \\
\twitter{@Pamela\underscore Moore13} & 4 & 4k & 2k & 54k & 21:14 & 294 &\bf 4.16 \\
\Rowcolor{twitterblue}\twitter{@UserC}\dag & 1305 & 51k & 8k & $<{}$1k & 22:16 & 6 &\bf 0.80\BotStrut \\
\hline
\setarstrut{\footnotesize}%
\multicolumn{8}{>{\footnotesize}r}{*Estimate of the causal estimand in {\em\href{\mainpaperDOI}{\color{blue}main paper}}, {\color{blue}Eq.~[\customxrref{eq:individual_impact_causal_estimand}]}\TopStrut} \\
\multicolumn{8}{>{\footnotesize}r}{\dag Anonymized screen names of currently active accounts} \\
\restorearstrut
\end{tabular*}
\end{table}

\newpage

\bibliography{io_detection_pnas}